\documentclass[twocolumn,traditabstract, longauth]{aa}
\DeclareUnicodeCharacter{02BC}{'}
\usepackage{graphicx}
\usepackage[hyperindex,breaklinks=true, colorlinks, citecolor=blue]{hyperref} 
\usepackage{tablefootnote}
\usepackage{txfonts}
\usepackage{xcolor}
\usepackage{multirow}
\usepackage{ulem}
\usepackage{rotating}
\usepackage{orcidlink}
\usepackage{lscape}

\begin{document}
%%%%%%%%%%%%%%%%%%% TITLE PAGE %%%%%%%%%%%%%%%%%%%

% Title of the paper, and the short title which is used in the headers.
% Keep the title short and informative.

\title{ALMAGAL IV. Morphological comparison of molecular and thermal dust emission using the histogram of oriented gradients (HOG) method}
%%%%%%%%%%%%%%%%%%%%%%%%%%%%%%%%%%%%%%%%%%%%%%%%%%%%%%%%%%%%%%%%%%%%%
\author{
C.~Mininni\inst{\ref{rome}}\orcidlink{0000-0002-2974-4703}\and 
S.~Molinari\inst{\ref{rome}}\orcidlink{0000-0002-9826-7525} \and
J.~D.~Soler\inst{\ref{rome}}\orcidlink{0000-0002-0294-4465} \and
\'A.~S\'anchez-Monge\inst{\ref{icecsic}, \ref{ieec}}\orcidlink{0000-0002-3078-9482} \and
A.~Coletta\inst{\ref{rome}, \ref{sapienza}}\orcidlink{0000-0001-8239-8304} \and
M.~Benedettini\inst{\ref{rome}}\orcidlink{0000-0002-0560-3172} \and
A.~Traficante\inst{\ref{rome}}\orcidlink{0000-0003-1665-6402} \and
E.~Schisano\inst{\ref{rome}}\orcidlink{0000-0003-1560-3958} \and
D.~Elia\inst{\ref{rome}}\orcidlink{0000-0002-9120-5890} \and
S.~Pezzuto\inst{\ref{rome}}\orcidlink{0000-0001-7852-1971} \and
A.~Nucara\inst{\ref{rome}, \ref{torvergata}}\orcidlink{0009-0005-9192-5491} \and
P.~Schilke\inst{\ref{unikoeln}}\orcidlink{0000-0002-1730-8832} \and
C.~Battersby\inst{\ref{connecticut}}\orcidlink{0000-0002-6073-9320} \and
P.~T.~P.~Ho\inst{\ref{asiaa}, \ref{hawaii}}\orcidlink{0000-0002-3412-4306} \and
M.~T.~Beltr\'an\inst{\ref{arcetri}}\orcidlink{0000-0003-3315-5626} \and
H.~Beuther\inst{\ref{mpia}}\orcidlink{0000-0002-1700-090X} \and
G.~A.~Fuller\inst{\ref{manchester}, \ref{unikoeln}}\orcidlink{0000-0001-8509-1818} \and
B.~Jones\inst{\ref{unikoeln}}\orcidlink{0000-0002-0675-0078} \and
R.~S.~Klessen\inst{\ref{ita}, \ref{uniheidelberg},\ref{cfa},\ref{radcliffeHav}}\orcidlink{0000-0002-0560-3172} \and
Q.~Zhang\inst{\ref{cfa}}\orcidlink{0000-0003-2384-6589} \and
S.~Walch\inst{\ref{unikoeln}, \ref{datacologne}} \and
Y.~Tang\inst{\ref{asiaa}} \and
A.~Ahmadi\inst{\ref{astron}}\orcidlink{0000-0003-4037-5248} \and
J. Allande\inst{\ref{arcetri},\ref{unifirenze}}\and
A.~Avison\inst{\ref{skaUK}, \ref{manchester}, \ref{almaUK}}\orcidlink{0000-0002-2562-8609} \and
C.~L.~Brogan\inst{\ref{nraoVA}}\orcidlink{0000-0002-6558-7653} \and
F. De Angelis\inst{\ref{rome}}\orcidlink{0009-0002-6765-7413}\and 
F.\ Fontani\inst{\ref{arcetri}, \ref{mpe}, \ref{lerma}} \and
P. Hennebelle\inst{\ref{saclay}}\and
T.~R.~Hunter\inst{\ref{nraoVA}}\orcidlink{0000-0001-6492-0090} \and
K.~G.~Johnston\inst{\ref{lincoln}}\orcidlink{0000-0003-4509-1180} \and
P.~Koch\inst{\ref{asiaa}}\and
R.~Kuiper\inst{\ref{duisburg}}\orcidlink{0000-0003-2309-8963} \and
C.~-Y.~Law\inst{\ref{arcetri}}\and
D.~C.~Lis\inst{\ref{jplcaltech}}\orcidlink{0000-0002-0500-4700} \and
S. Liu\inst{\ref{rome}}\orcidlink{0000-0001-7680-2139}\and
T.~Liu\inst{\ref{shanghai}} \and
S.-Y.~Liu\inst{\ref{asiaa}} \and
L.~Moscadelli\inst{\ref{arcetri}}\orcidlink{0000-0002-8517-8881} \and
T.~M\"oller\inst{\ref{unikoeln}}\orcidlink{0000-0002-9277-8025} \and
A.~J.~Rigby\inst{\ref{leeds}}\orcidlink{0000-0002-3351-2200}\and
K.~L.~J.~Rygl\inst{\ref{bologna}}\orcidlink{0000-0003-4146-9043} \and
P.~Sanhueza\inst{\ref{tokyo},\ref{meguro}, \ref{naoj}}\orcidlink{0000-0002-7125-7685} \and
L. Testi\inst{\ref{u-bo}}\and
Y.-N.~Su\inst{\ref{asiaa}} \and
F.~F.~S.~van der Tak\inst{\ref{sron}, \ref{kapteyn}}\orcidlink{0000-0002-8942-1594} \and
M.~R.~A.~Wells\inst{\ref{mpia}}\orcidlink{0000-0002-3643-5554}\and
L.~Bronfman\inst{\ref{unichile}}\orcidlink{0000-0002-9574-8454}\and
T.~Zhang\inst{\ref{zhejiang}, \ref{unikoeln}}\orcidlink{0000-0002-1466-3484}\and
H.~Zinnecker\inst{\ref{autchile}}
}

\institute{
\label{rome}INAF-Istituto di Astrofisica e Planetologia Spaziale, Via Fosso del Cavaliere 100, I-00133 Roma, Italy
\and
\label{icecsic}Institut de Ci\`encies de l'Espai (ICE), CSIC, Campus UAB, Carrer de Can Magrans s/n, E-08193, Bellaterra (Barcelona), Spain
\and
\label{ieec}Institut d'Estudis Espacials de Catalunya (IEEC), E-08860, Castelldefels (Barcelona), Spain
\and
\label{sapienza}Dipartimento di Fisica, Sapienza Universit\`a di Roma, Piazzale Aldo Moro 2, I-00185, Rome, Italy
\and
\label{torvergata}Dipartimento di Fisica, Universit\`a di Roma Tor Vergata, Via della Ricerca Scientifica 1, I-00133 Roma, Italy
\and
\label{unikoeln}I.\ Physikalisches Institut, Universit\"{a}t zu K\"{o}ln, Z\"{u}lpicher Str.\ 77, D-50937 K\"{o}ln, Germany
\and
\label{connecticut}University of Connecticut, Department of Physics, 2152 Hillside Road, Unit 3046 Storrs, CT 06269, USA
\and
\label{asiaa}Institute of Astronomy and Astrophysics, Academia Sinica, 11F of ASMAB, AS/NTU No.\ 1, Sec.\ 4, Roosevelt Road, Taipei 10617, Taiwan
\and
\label{hawaii}East Asian Observatory, 660 N.\ A'ohoku, Hilo, Hawaii, HI 96720, USA
\and
\label{arcetri}INAF-Osservatorio Astrofisico di Arcetri, Largo E.\ Fermi 5, I-50125 Firenze, Italy
\and
\label{mpia}Max Planck Institute for Astronomy, K\"onigstuhl 17, D-69117 Heidelberg, Germany
\and
\label{manchester}Jodrell Bank Centre for Astrophysics, Oxford Road, The University of Manchester, Manchester M13 9PL, UK
\and
\label{ita}Universit\"at Heidelberg, Zentrum f\"ur Astronomie, Institut f\"ur Theoretische Astrophysik, Heidelberg, Germany
\and
\label{uniheidelberg}Universit\"at Heidelberg, Interdisziplin\"ares Zentrum f\"ur Wissenschaftliches Rechnen, Heidelberg, Germany
\and
\label{cfa}Center for Astrophysics  Harvard \& Smithsonian, 60 Garden St, Cambridge, MA 02138, USA
\and
\label{radcliffeHav}Elizabeth S. and Richard M. Cashin Fellow at the Radcliffe Institute for Advanced Studies at Harvard University, 10 Garden Street, Cambridge, MA 02138, U.S.A.
\and
\label{datacologne}Center for Data and Simulation Science, University of Cologne, Germany
\and
\label{astron}ASTRON, Netherlands Institute for Radio Astronomy, Oude Hoogeveensedijk 4, Dwingeloo, 7991 PD, The Netherlands
\and
\label{unifirenze}Dipartimento di Fisica e Astronomia, Universit\`a degli Studi di Firenze, Via G.\ Sansone 1,I-50019 Sesto Fiorentino, Firenze, Italy
\and
\label{skaUK}SKA Observatory, Jodrell Bank, Lower Withington, Macclesfield, SK11 9FT, UK
\and
\label{almaUK}UK ALMA Regional Centre Node, M13 9PL, UK
\and
\label{nraoVA}National Radio Astronomy Observatory, 520 Edgemont Road, Charlottesville VA 22903, USA
\and
\label{mpe}Max-Planck-Institute for Extraterrestrial Physics (MPE), Garching bei M\"unchen, Germany 
\and
\label{lerma} Laboratory for the study of the Universe and eXtreme phenomena (LUX), Observatoire de Paris, Meudon, France
\and
\label{saclay}Universit\'e Paris-Saclay, Universit\'e Paris-Cit\'e, CEA, CNRS, AIM, 91191 Gif-sur-Yvette, France
\and
\label{lincoln}School of Engineering and Physical Sciences, Isaac Newton Building, University of Lincoln, Brayford Pool, Lincoln, LN6 7TS, United Kingdom
\and
\label{duisburg}Faculty of Physics, University of Duisburg-Essen, Lotharstra{\ss}e 1, D-47057 Duisburg, Germany
\and
\label{jplcaltech}Jet Propulsion Laboratory, California Institute of Technology, 4800 Oak Grove Drive, Pasadena, CA 91109, USA
\and
\label{shanghai}Shanghai Astronomical Observatory, Chinese Academy of Sciences, Shanghai 200030, Peopleʼs Republic of China
\and
\label{leeds}School of Physics and Astronomy, University of Leeds, Leeds LS2 9JT, UK
\and
\label{bologna}INAF-Istituto di Radioastronomia \& Italian ALMA Regional Centre, Via P. Gobetti 101, I-40129 Bologna, Italy
\and
\label{tokyo}Department of Astronomy, School of Science, The University of Tokyo, 7-3-1 Hongo, Bunkyo, Tokyo 113-0033, Japan
\and
\label{meguro}Department of Earth and Planetary Sciences, Institute of Science Tokyo, Meguro, Tokyo, 152-8551, Japan
\and
\label{naoj}National Astronomical Observatory of Japan, National Institutes of Natural Sciences, 2-21-1 Osawa, Mitaka, Tokyo 181-8588, Japan
\and
\label{u-bo}Dipartimento di Fisica e Astronomia, Alma Mater Studiorum - Universit\`a di Bologna
\and
\label{sron}SRON Netherlands Institute for Space Research, Landleven 12, 9747 AD Groningen, The Netherlands
\and
\label{kapteyn}Kapteyn Astronomical Institute, University of Groningen, The Netherlands
\and
\label{unichile}Departamento de Astronomía, Universidad de Chile, Casilla 36-D, Santiago, Chile
\and
\label{zhejiang}Zhejiang Laboratory, Hangzhou 311100, P.R.China
\and
\label{autchile}Universidad Autonoma de Chile, Pedro de
Valdivia 425, 9120000 Santiago de Chile, Chile
}

%%%%%%%%%%%%%%%%%%%%%%%%%%%%%%%%%%%%%%%%%%%%%%%%%%%%%%%%%%%%%%%%%%%%%%
%\include{authorslist}
%\author{C. Mininni\inst{1}\thanks{E-mail: chiara.mininni@inaf.it}, S. Molinari, J. D. Soler, Á. Sanchez-Monge, A. Coletta, M. Benedettini, A. Traficante, E. Schisano, D. Elia, S. Pezzuto, A. Nucara, P. Schilke, M. T. Beltrán, H. Beuther, C. Battersby, P. Ho, B. Jones, D. Lis, K. Johnston, R. Klessen, F. van der Tak, P. Sanhueza, P. Koch, R. Kuiper, L. Moscadelli, K.~L.~J. Rygl, T. Liu, Q. Zhang,  M. R. A. Wells, T. M\"oller,  F. Fontani,
%...  ,  and the ALMAGAL collaboration
%       }

% List of institutions
% These dates will be filled out by the publisher
   \date{Received ; accepted }

% Enter the current year, for the copyright statements etc.
%\pubyear{2021}

% Don't change these lines

  \abstract
  % context heading (optional)
  % {} leave it empty if necessary  
   {The study of molecular line emission is crucial to unveil the kinematics and the physical conditions of gas in star-forming regions. We utilize data from the ALMAGAL survey which provides an unprecedentedly large statistical sample of high-mass star-forming clumps that helps to remove bias and reduce noise (e.g. due to source peculiarities, selection or environmental effects) to determine how well individual molecular species trace continuum emission.  }
  % aims heading (mandatory)
   {Our aim is to quantify the reliability of using individual molecular transitions to derive physical properties of the bulk of the H$_2$ gas, looking at morphological correlations in their overall integrated molecular line emission with the cold dust. For this study we selected transitions  of H$_2$CO, CH$_3$OH, DCN, HC$_3$N, CH$_3$CN, CH$_3$OCHO, SO, and SiO  and compared them with the 1.38\,mm dust continuum emission at different spatial scales in the ALMAGAL sample. We included two transitions of H$_2$CO to understand the validity of the results depending on the excitation condition of the selected transition of a molecular specie. The ALMAGAL project observed more than 1000 candidate high-mass star-forming clumps in ALMA Band 6 at a spatial resolution down to 1000 au. A total of 1013 targets have been analyzed in the present paper, covering all evolutionary stages of the high-mass star-formation process and different conditions of clump fragmentation.%Among the sample of this ALMA large project, 916 sources have been analyzed in this paper, that cover all the evolutionary stages of the high-mass star-formation process and different conditions of clump fragmentation.
   }
  % methods heading (mandatory)
   {We used, for the first time on a large statistical sample, the method of the histogram of oriented gradients (HOG) implemented in the tool \texttt{astroHOG} to compare the morphology of integrated line emission with maps of the 1.38\,mm dust continuum emission. For each clump, we defined two masks: one that covers the extended more diffuse continuum emission and a smaller one that contains only the compact sources. We select these two masks to study if and how the correlation among the selected molecules changes with the spatial scale of the emission, from extended more diffuse gas in the clumps to denser gas in compact fragments (cores). Moreover, we calculated the Spearman's correlation coefficient, and compared it with our astroHOG results.}
  % results heading (mandatory)
   {Among the molecular species analyzed in this paper, only H$_2$CO, CH$_3$OH, and SO show emission on spatial scales comparable with the diffuse 1.38\,mm dust continuum emission. However, from the HOG method, the median correlation of the emission of each of these species with the continuum is only $\sim$24-29\%. In comparison with the dusty dense fragments these molecular species still have low values of correlation, on average below 45\%. The low level of morphological correlation suggests that these molecular lines likely trace on average the clump medium or outer layers around dense fragments (in some cases possibly due to optical depth effects) or are also at this scale tracing inner parts of outflows. On the other hand DCN, HC$_3$N, CH$_3$CN, and CH$_3$OCHO show a good correlation with the dense dust fragments, above 60\%. The worst correlation is seen with SiO, both with the extended continuum emission and with compact sources. Moreover, unlike other outflow tracers, SiO in a large fraction of the sources does not cover well the area of the extended continuum emission. This and the results of the astroHOG analysis reveal that SiO and SO do not trace the same gas, contrary to what was previously thought. From the comparison of the results of the HOG method and the Spearman's correlation coefficient, the HOG method gives much more reliable results than the intensity-based coefficient in estimating the level of similarity of the emission morphology.} 
  % conclusions heading (optional), leave it empty if necessary 
   {}

\titlerunning{ALMAGAL IV}
\authorrunning{Mininni et al.}
    \maketitle
%% From the front matter, we move on to the body of the paper.
%% Sections are demarcated by \section and \subsection, respectively.
%% Observe the use of the LaTeX \label
%% command after the \subsection to give a symbolic KEY to the
%% subsection for cross-referencing in a \ref command.
%% You can use LaTeX's \ref and \label commands to keep track of
%% cross-references to sections, equations, tables, and figures.
%% That way, if you change the order of any elements, LaTeX will
%% automatically renumber them.
%%
%% We recommend that authors also use the natbib \citep
%% and \citet commands to identify citations.  The citations are
%% tied to the reference list via symbolic KEYs. The KEY corresponds
%% to the KEY in the \bibitem in the reference list below. 
% -----------------------------------------------
\section{Introduction} \label{sec:intro}
The study of molecular line emission in star-forming regions is extremely important, not only for understanding the chemical content of the environment in which stars form but also for constraining the physical and dynamic properties of the gas \citep[e.g.][]{Kennicutt_2012, beuther2006, 2018beltran}. 
In fact, from the analysis of line emission, we can infer the average velocity and velocity dispersion of the emitting gas, as well as its column density and temperature, if we can detect multiple transitions of the same molecular species. Moreover, the combined analysis of several molecular species can help constrain quantities such as the column density of H$_2$, the volume density, and the UV radiation field \citep{gratier2017}.
It is known that different molecular species can trace different regions in the star formation scenario (e.g. \citealt{vandischoekblake1998,caselliceccarelli2012,tychoniecnew}), depending on the physical conditions required for the production/release of each molecular species in the gas phase. 
In addition, different transitions of a molecular species can be excited and emit in gas with higher/lower temperatures and densities.
The use of various tracers in star-forming regions is due to the fact that the main component of the interstellar medium (ISM), i.e. the hydrogen molecule (H$_2$), has no permanent dipole moment and the lowest moment of inertia among all the molecular species. 
As a result, there are no allowed transitions of H$_2$ that can be excited under typical conditions in cold and dense of star-forming regions \citep{stahlerpalla}. 
Therefore, an indirect method to trace H$_2$ in star-forming regions on scales below 1 pc, where the CO emission - the typical proxy of H$_2$ - is usually optically thick \citep{Kennicutt_2012}, is to use the dust continuum emission at millimeter wavelengths, assuming a constant ratio between the density of the H$_2$ and that of the dust.
However, much of the crucial information contained in the line emission, such as the kinematics of the material, cannot be obtained from dust emission. 
Therefore, studying 
how well individual molecular species trace the continuum emission is crucial to assess the reliability of the estimates made from the analysis of these species in the bulk of the gas emission under different conditions.\\ \indent
\begin{table*}
\centering

\caption{Summary of the molecular emission lines analyzed in this work.}
\label{table:lines_used}

\begin{tabular}{lccccc}
\hline\hline
transition & $\nu$ & log$_{10}A_{\rm{E}}$ & $E_{\rm{U}}/k_{\rm{B}}$ &$n_{\rm{c}}$(20\,K) &$n_{\rm{c}}$(100\,K) \\% & A$_{ij}$/C$_{ij}$ (20K) & A$_{ij}$/C$_{ij}$ (100K) \\%&$n_{\rm{c}}$(200\,K)\\
 & [GHz] & & [K] & [cm$^{-3}$] & [cm$^{-3}$] \\% &[cm$^{-3}$] &[cm$^{-3}$]  \\%& [cm$^{-3}$] \\
  \hline
  SiO $5 - 4$ &217.104980 &-3.284 & 31.3&$9.8\times10^{5}$ &$7.2\times10^{5}$ \\%& $2.4\times10^{6}$ & $2.6\times10^{6}$\\
   DCN$^{a}$ $3 - 2$ &217.238538 & -3.340 & 20.9 & $5.5\times10^{6}$&$2.1\times10^{6}$ \\%&$3.7\times10^{7}$ & $3.7\times10^{7}$ \\
  H$_2$CO $3_{0,3} - 2_{0,2}$ &218.222192 &  -3.550 & 21.0 & $7.8\times10^{5}$ & $4.7\times10^{5}$\\% & $2.6\times10^{6}$ & $3.4\times10^{6}$\\
  CH$_3$OCHO $17_{3,14}-16_{3,13}\,$E & 218.280900 & -3.822 & 99.7 & -$^{b}$ & -$^{b}$\\% & -$^{b}$ & -$^{b}$\\
 HCCCN $24-23$ &218.324723 &-3.083 &131.0 &$1.3\times10^{6}$ &$7.7\times10^{5}$ \\%& $1.7\times10^{7}$ & $1.6\times10^{7}$\\
 CH$_3$OH $4_{2,3} - 3_{1,2}$ & 218.440063 & -4.329 & 45.5 &$1.3\times10^{5}$ &$8.8\times10^{4}$\\% & $2.0\times10^{7}$ &$7.8\times10^{7}$\\
 H$_2$CO $3_{2,1} - 2_{2,0}$ & 218.760066 &-3.802 &68.1 &$3.1\times10^{5}$ & $2.4\times10^{5}$ \\%& $3.4\times10^{6}$ & $3.2\times10^{6}$\\
SO $6_{5} - 5_{4}$& 219.949442&	-3.874 &35.0 & 4.6$\times10^{5\, c}$&$3.4\times10^{5}$\\%& 2.2$\times10^{6\, c}$&$2.3\times10^{6}$ \\
  CH$_3$CN $12_{1}-11_{1}$ &220.743011 &-3.199 &76.0 &$2.0\times10^{6}$&$8.7\times10^{5}$\\% &$4.7\times10^{6}$&$3.1\times10^{6}$ \\
 CH$_3$CN $12_{0}-11_{0}$ &220.747261 &-3.196 &68.9 &$2.0\times10^{6}$ &$8.7\times10^{5}$\\% &$4.7\times10^{6}$&$3.1\times10^{6}$ \\

 \hline

\end{tabular}
\vspace{1mm}

\begin{flushleft}
\small{ \textbf{Notes:} Col. 1: transition. Col. 2: rest frequency $\nu$. Col. 3: Einstein's coefficient $A_{\rm{E}}$. Col. 4: energy of the upper level $E_{\rm{U}}/k_{\rm{B}}$ in units of K. Col. 5: critical density $n_{\rm{c}}$ calculated at the temperature of 20\,K.  Col. 6: critical density $n_{\rm{c}}$ calculated at the temperature of 100K. The critical densities have been calculated from Eq. (4) of \citet{Shirley2015} using the collisional coefficients $C_{\rm{ij}}$ from LAMDA (Leiden Atomic and Molecular Database).
\textit{a)}: collisional coefficient for DCN not available; thus, we used the collisional coefficients of HCN; \textit{b)}: no collisional coefficient found for the CH$_3$OCHO E-state; \textit{c)}: extrapolated from the values at 100K, 80K, and 60K since collisional coefficients were not available for $T$=20K.  }

\end{flushleft}
\end{table*}
In this framework, we analyze molecular line emission in star-forming regions in the dataset of the ALMA
Evolutionary study of High Mass Protocluster Formation in the Galaxy (ALMAGAL), the largest sample of star-forming regions ever observed at high spatial resolution with the Atacama Large Millimeter/Submillimeter Array (ALMA). ALMAGAL is an ALMA large project (2019.1.00195.L, PIs: Molinari, Schilke, Battersby, Ho; \citealt{ALMAGAL1,ALMAGAL2}) that targeted more than 1000 candidate high-mass star-forming regions with a resolution of $\sim1000\,\mathrm{au}$ covering all the evolutionary stages of the star formation process and different environmental conditions (from the tip of the bar to the outer galaxy) and showing different fragmentation properties (\citealt{ALMAGAL3}, Elia et al. in prep.). 
This will give a robust statistical basis for our study.
\\\indent We  analyzed the emission morphology of eight commonly detected molecular tracers of star-forming regions (H$_2$CO, CH$_3$OH, DCN, HC$_3$N, CH$_3$CN, CH$_3$OCHO, SO, and SiO) in the 1.38\,mm band (ALMA Band 6) and compared it to the dust continuum emission on different spatial scales. \\ \indent To study the morphological correlation of emission from selected molecular species with the dust continuum emission we use the histogram of oriented gradients (HOG) method \citep{soler_astroHOG}.
The HOG method uses the orientation of the spatial gradients of different images (or 3D data cubes) to compare their emission morphology.
Previous applications of the HOG method include the pairing of atomic and molecular emission toward molecular clouds in the Galactic plane \citep{syed2020}, the correlation between neutral atomic hydrogen (H{\sc i}) emission and Faraday synthesis cubes \citep{bracco2020}, the characterization of the line emission from molecular species in three high-mass star formation regions \citep{Liu2020ATOMShog, Gieser2022}, and the comparison between the three-dimensional reconstruction of the interstellar dust density and the H{\sc i} and carbon monoxide (CO) emission \citep{soler2023}.
Studies of the correlation between the emission in high-resolution observations of high-mass star-forming regions have been mostly carried out using intensity-based statistical correlators, such as cross-correlation, Pearson's coefficient or Spearman's coefficient, to study the correlation between different molecular species \citep[e.g.][]{guzman2018ApJScrosscorrelation,li2022crosscorrelation}. \\ \indent
In this paper, we present, for the first time on such a large sample, a more sophisticated approach that compares the total morphology of the emission and not only the intensity pixel by pixel, as usually done so far. In particular, we compare the morphology of the integrated intensity maps of several molecular species with the morphology of continuum emission from the cold dust in the largest sample of star-forming regions using the HOG method.

This paper is organized as follows.
In Section~2, we present the ALMAGAL sample and the molecular species analyzed.
In Section~3, we describe the observations and the data reduction process. 
In Section~4, we describe the derivation of the integrated line intensity maps (moment-0 maps) and the tool used to investigate the morphological correlation of the molecular species with the dust continuum emission: \texttt{astroHOG}.
Section~5 presents a discussion of the results where we show that emitting in the same regions of the maps does not imply for all the species that the emission morphology is well correlated, and demonstrate the importance of using a dedicated tool to study the morphology.
Finally, in Section~6, we present our conclusions.

\begin{figure*}[t!]
    \centering
    \includegraphics[width=17.5cm]{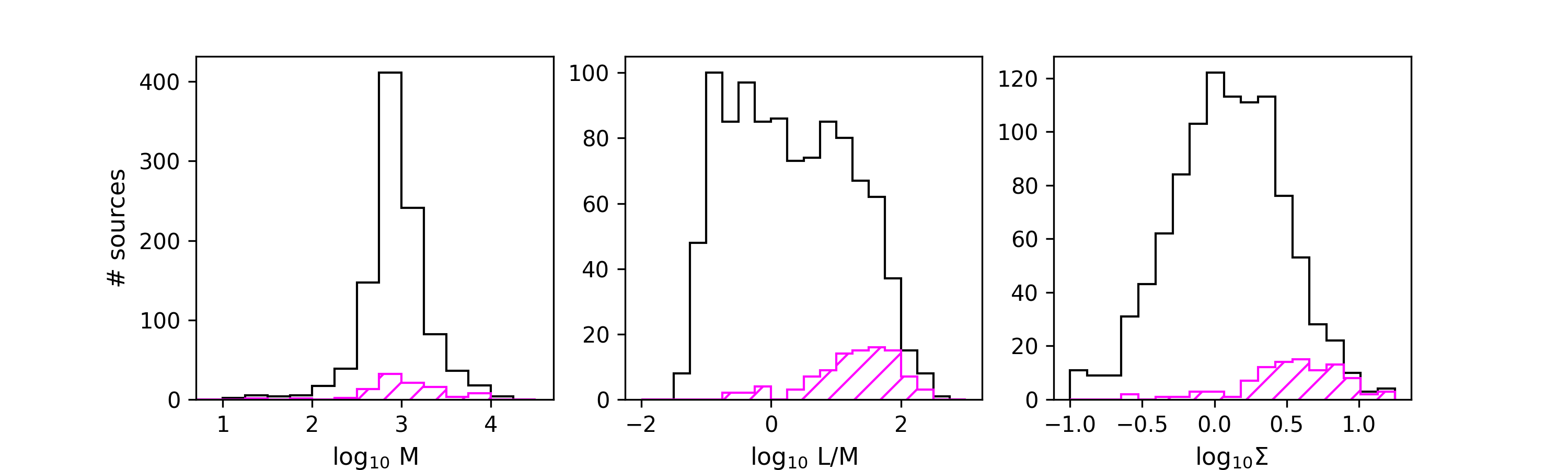}
    \caption{Physical properties of the clumps in the ALMAGAL %full
    sample, shown in black. The subsample for which we created moment-0 maps using a fixed velocity range of $\pm5\,$km/s around the clump systemic velocity (see Section 4.1) is shown in magenta with shading. From left to right: mass of the clumps $M$ in units of M$_{\odot}$, $L/M$ ratio of the clumps in units of $\mathrm{L_{\odot}/M_{\odot}}$, and surface density of the clumps $\Sigma$ in units of g/cm$^{2}$.}
    \label{fig:subsampleclumpsproperties}
\end{figure*}
\section{Sources and Molecules selection}
\subsection{The sample}

The ALMAGAL sample includes 1017 high-mass clumps with masses $M$\,$>$\,500\,$M_{\odot}$, surface densities $\Sigma$\,$>$\,0.1\,g\,cm$^{-2}$, located at heliocentric distances up to 7.5\,kpc, and covering evolutionary stages from starless clumps to H{\sc ii} regions. Due to the close proximity (below the FOV size) of 3 pairs of sources and the exclusion of Eta Carinae, the sample analyzed is of 1013 sources (see \citealt{ALMAGAL1} for further details).
The sample spans four orders of magnitude in the luminosity-to-mass ratio ($L/M$), a distance-independent indicator of the evolutionary stage of a star-forming region \citep{molinari2008,cutexmolinari2016,molinari2019, giannetti2017, elia2017, elia2021, konig2017}.
During the first stages of star formation, $L/M$ increases due to the increase of the accretion rate with time and therefore accretion luminosity, while the central object is still very embedded in its envelope so that the total mass $M$ does not vary significantly.

Using this parameter, it is possible to roughly divide starless sources, having $L/M<1\,L_{\odot}/M_{\odot}$, from protostellar objects, with $1<L/M<10\,L_{\odot}/M_{\odot}$, and from candidate ultra-compact HII (UC HIIs) regions, with $L/M>10\,L_{\odot}/M_{\odot}$ \citep{cesaroni2015uchii, elia2017, Elia2020uchii, elia2021}.

Within the ALMAGAL sample, the clump masses are in the range between $10^{2}\,M_{\odot}$ and $10^{4}\,M_{\odot}$, and the number of fragments varies from 1 up to 50, with a median value over the whole sample of 5 \citep{ALMAGAL3}. 
We give an overview of the properties of the ALMAGAL sample in Figure \ref{fig:subsampleclumpsproperties}, where the pink histogram highlights the properties of a subsample of sources for which the automated routine for the creation of moment-0 maps failed (see Section 4.1). For these sources, moment-0 maps have been created integrating over a fixed velocity range.\\ \indent
The ALMAGAL survey and its dataset have already been described by \citet{ALMAGAL1} and \citet{ALMAGAL2}, including a description of the reduction pipeline and full characterization of the clumps. Moreover, \citet{ALMAGAL3} performed a detailed study of the fragmentation properties in each clump, and \cite{molly2024} studied the accretion flows of gas from the lines of H$_2$CO in a sub-sample of ALMAGAL sources.
\subsection{Molecular transitions analyzed}
We  analyzed the emission morphology of eight commonly detected molecular tracers of star-forming regions (H$_2$CO, CH$_3$OH, DCN, HC$_3$N, CH$_3$CN, CH$_3$OCHO, SO, and SiO; the selected transitions are listed in Table 1) in the 1.38\,mm band (ALMA Band 6) and compared it to the dust continuum emission on different spatial scales.
To investigate the dependence of the results on the selected transition for each molecular species, and on the upper level energy and critical density, we selected two of the three available transitions for H$_2$CO, which is expected to be the most commonly detected molecule in the sample. \\\indent The species selected include both simple and complex molecules (i.e. molecules made of six or more atoms) as well as well-known shock tracers such as SiO and SO. 
Even though we expect the emission of those molecular species to not correlate with the extended dust continuum emission, we do not know how those molecular species correlate with continuum emission at cores scale, and with this study we will be able to quantify it. This is especially true for SO, since other studies have underlined that SO can also trace a hot envelope around protostellar sources  \citep[e.g.][]{tychoniecnew}. The spectral setup of ALMAGAL includes also lines of $^{13}$CO and C$^{18}$O. However, we excluded from the analysis these two tracers since in most cases their emission covers the entire FOV of the observations, and extend even further, unlike all other species analyzed in this paper, which raises problems in the cleaning of those spectral channel, leading to possible artefacts and filtering.\\ \indent In the following paragraphs we give a brief review of the main formation pathways for all the molecular species analyzed in the paper.\\\indent
H$_2$CO forms on grains from hydrogenation of CO \citep{charnley1997, watanabe2004, garrod2008}, and is released in gas phase when the temperature raises above 40\,K \citep{garrod2008}, but also through gas-phase reactions \citep{leteuff2000, garrod2006, fuchs2009, cachon2019}. Methanol is quite exclusively formed on dust-grains, mainly from the hydrogenation of CO, after the formation of H$_2$CO (\citealt{charnley1997, watanabe2004}; CO $\rightarrow$ HCO $\rightarrow$ H$_2$CO $\rightarrow$ CH$_2$CO, CH$_3$O $\rightarrow$ CH$_3$OH), or through a secondary route by CH$_3$OH+H$_2$CO $\rightarrow$ CH$_3$OH + HCO, on dust grains \citep{simons2020, santos2022}. It is then released in gas phase mainly due to thermal desorption from T$\sim$150\,K \citep{luna2017}, but non-thermal desorption mechanisms are at play to explain its detection in cold environment. These two molecules can also be released due to grain-sputtering and therefore probe outflows, as their emission shows high-velocity wings and their column densities have been found to be well correlated with the one of SiO \citep{li2022}. Indeed, also in our ALMAGAL sample, we found evidence of contribution from outflowing gas in the emission of these two molecular species.\\ \indent HCCCN can form through gas phase reactions starting mainly from the  release in gas-phase of CH$_4$ above 25\,K, that reacts with C$^{+}$ to form C$_2$H$_2$, followed by C$_2$H$_2$+CN$\rightarrow$HCCCN+H, and increase its production at T>55\,K when also C$_2$H$_2$ thermally desorb from dust grains \citep{hassel2008}. However, the HCCCN formed in gas-phase is partly depleted onto dust-grains, and the highest abundances are reached when it desorb at T above $\sim 90\,$K \citep{taniguchi2019}. Therefore it is expected to be detected only in the more warm regions.\\ \indent The main pathway of formation for CH$_3$CN is on the surface of dust grains from the reactions of CH$_3$ with CN \citep{garrod2008}. Another grain surface pathways for its formation is the multiple hydrogenation of C$_2$N, and modest abundances of this molecules are also produced in gas-phase following the desorption of HCN from the grains \citep{garrod2008}.\\ \indent CH$_3$OCHO formation is mainly the result of the reactions of HCO and CH$_3$OH on grains. This reaction can occur already at T$\sim 10\,$K, but higher temperature around 20-40\,K increase the methyl formate production thanks to the higher mobility of the reactants. The majority of CH$_3$OCHO is released in gas-phase at temperature above 100\,K \citep{burke2015}, so also this line is expected in the hot regions around protostars.\\ \indent The reactions that leads to the formation of DCN are gas-phase reactions, with both cold and warm/high temperature reactions. However, unlike other deuterated species, the predominant part of the production of DCN takes place in warm gas up to and even above 70\,K 
\citep{millar1989, turner2001, roueff2007, albertsson2013}.  \\ \indent SiO is a well known shock tracer  and its abundance is enhanced up to six order of magnitudes in outflow regions \citep{martinpintado1992}. In those regions grain sputtering releases in gas-phase a significant amount of atomic Si or Si-bearing molecules, that leads to the formation of SiO \citep{schilke1997, jimenezserra2008}. However, a narrow SiO emission has been observed, that could be due to low-velocity schocks, or cloud-cloud collisions \citep{jimenezserra2008, louvet2016, cosentino2022, duartecabral2014A&A}. Lastly, S atoms freeze on dust grains during the cold collapse phase, where H$_2$S is mainly formed. After its release in gas phase at high temperature (T$>100\,$K) rapid hot gas-phase reactions lead to the formation of SO \citep{charnley1997, wakelam2004, wakelam2011}. However, \citet{esplugues2013} observed an increase of 3 orders of magnitude in SO abundances in regions affected by shocks, and \citet{fontani2023} observed that SO can trace both quiescient and turbulent material depending on the source, thus the sputtering of dust grains is likely increasing the SO formation in outflow regions, as in the case of SiO.
% OBSERVATIONS ===============================================================
\section{Observations and data reduction}

The ALMAGAL program (project
code: 2019.1.00195.L in Cycle 7) was observed between October 2019 and July 2022. To achieve the requested linear resolution of $\sim1000\,$au without losing information on diffuse material at the clump scale, each target was observed with the 7-m array (hereafter 7M) and with two configurations of the main ALMA array. To resolve almost the same spatial scale of about 1000 au for all the sources that are at different distances, the sample was divided in two bins of distance. The \textit{near} sources (i.e. sources with d$\,<4.7$ kpc) were observed in configurations C-2 and C-5 of the main array, while the \textit{far} sources (i.e. sources with d$\,>4.7$ kpc) were observed in configurations C-3 and C-6. 
Using the standard pipeline, we calibrated the data of the single arrays (hereafter \textit{\sc 7m}; \textit{\sc tm2}, C-2 and C-3, intermediate resolution configuration of the main array; \textit{\sc tm1}, C-5 and C-6, highest resolution configuration of the main array). The ALMAGAL team developed a dedicated pipeline to obtain the final cubes using joint deconvolution of the various configurations observed that includes self-calibration for the most intense sources. Moreover, an iterative procedure starting from the standard ALMA-pipeline function \texttt{findcont} was implemented, to refine the determination of the continuum channels used to create the continuum images and the line-only cubes. A detailed description of this routine, together with a quality assessment of the final products, is presented in \citet{ALMAGAL2}. For the purpose of this work, we decided to analyze the cubes and continuum maps obtained by the joint deconvolution of \textit{\sc 7m} and \textit{\sc tm2} data only (hereafter \textit{\sc 7m+tm2}), that reach a linear resolution of $\sim 5000\,$au, because they recover with a better signal-to-noise ratio the diffuse emission on large scales compared to the \textit{\sc 7m+tm2+tm1} cubes obtained from the joint deconvolution of all the available configurations that reach the aforementioned maximum resolution of $\sim1000\,$au.\\ \indent
The spectral setup of the observations covers four spectral windows in total: two spectral windows with 
a 1.9 GHz bandwidth and spectral resolution of 0.98 MHz ($\sim1.3\,$km\,s$^{-1}$) centered at 217.8 GHz and 220.0 GHz, and two higher resolution spectral windows with 0.47 GHz bandwidth and spectral resolution of 0.24 MHz ($\sim0.3\,$km/s) centered at 218.3 GHz and 220.6 GHz, respectively.

% METHODOLOGY ===============================================================
\section{Methodology/Analysis}

\subsection{Moment maps creation and detection}

\begin{figure*}
    
    \centering
    \includegraphics[trim={0.6cm 0 0 0}, clip=true]{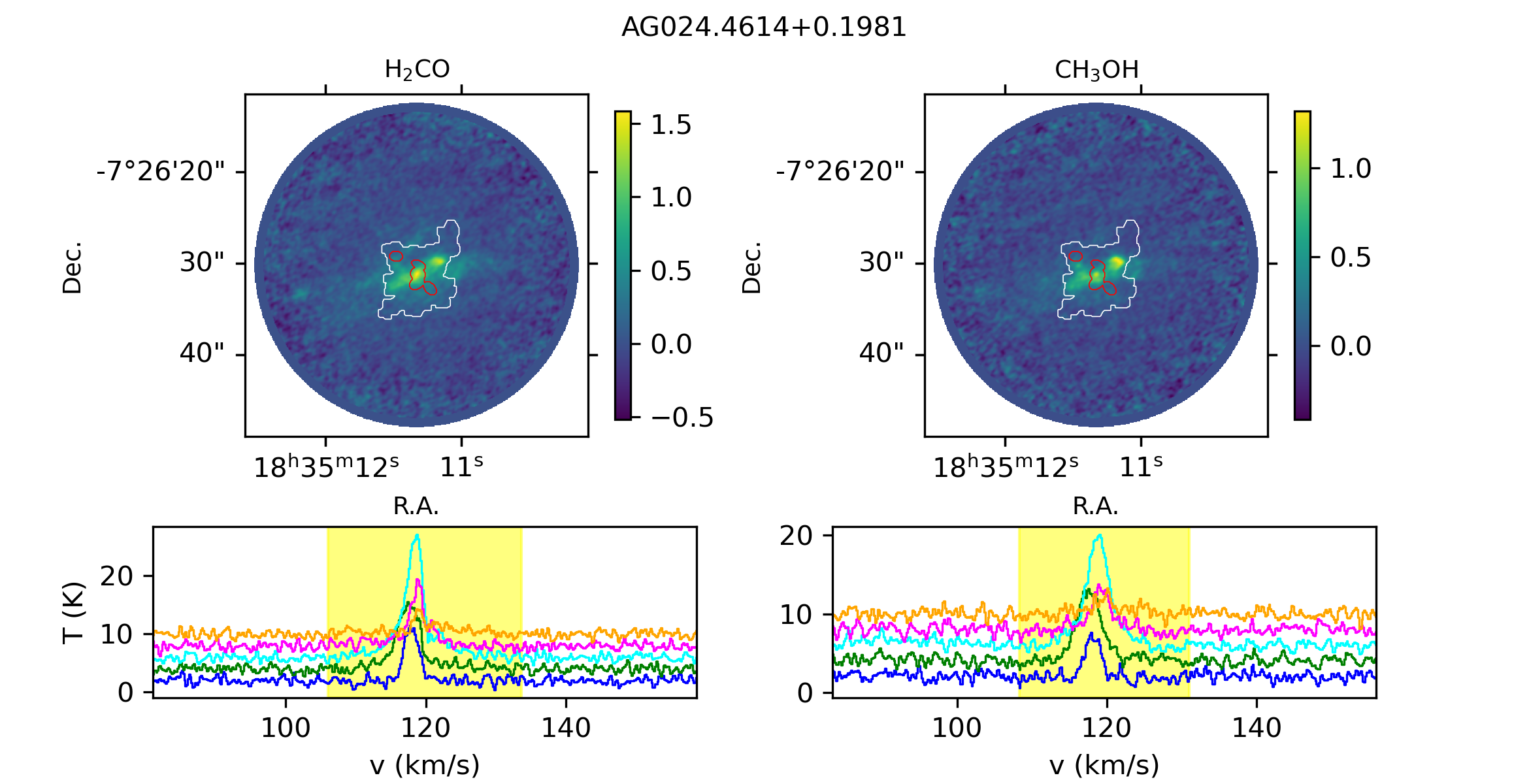}
    \caption{Moment-0 maps and spectral range created with the automated routine described in Sect. 4.1. \textit{Upper row:} moment-0 map of H$_2$CO $3_{0,3}-2_{0,2}$ (left) and CH$_3$OH $2_{2,3}-3_{1,2}$ (right) for the source AG024.4614+0.1981. 
    The color scales are in units of Jy/beam km/s. 
    \textit{Lower row:} spectra averaged over the twelve continuum cores detected in the source (ellipse with semi-axis the HWHM = 1.177$\sigma$ from the 2D Gaussian fit). Each core spectra has an y-axis offset to better show the spectra.
    The yellow area indicates the velocity integration range to obtain the moment-0 maps shown in the upper panel. The red contours are the contours of mask$_{\rm{com.}}$, while the white contour is the contour of mask$_{\rm{ext.}}$, defined from the continuum emission (see Section 4.4 for details).
    For this source, the continuum emission is shown in Fig. \ref{fig:example3typesofcontinuum}.}
    \label{fig:mom0example}
\end{figure*}

\begin{figure*}
    \centering
    \includegraphics[width=18.2cm, trim={0cm 1.6cm 0.7cm 0cm}, clip=true]{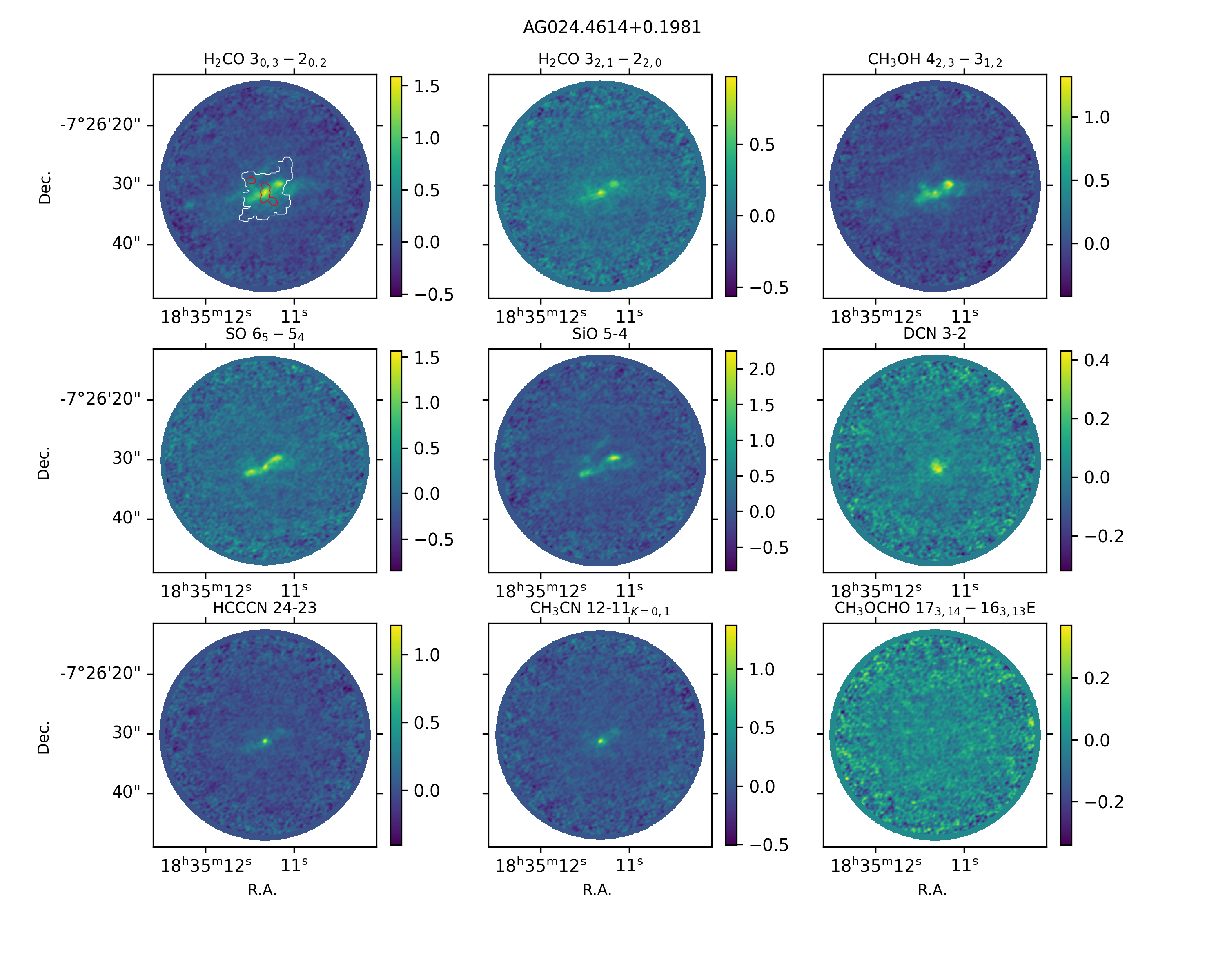}
    \caption{Moment-0 maps of all the transitions analysed in this paper for the source AG024.4614+0.1981. The colorscales are in units Jy\,beam$^{-1}$\,km\,s$^{-1}$. In the upper-left panel we show the continuum emission masks for reference; red contour: mask of all the compact sources inside the source, white contour: mask of the continuum emission above 3 SNR.}
    \label{fig:allmoments}
\end{figure*}

The generation of the moment-0 maps for the sources in our sample was not feasible individually, given the large number of fields. 
We automated the procedure to estimate, for each molecular line in each source, the extremes of integration of the line in the velocity space, and then we created the moment-0 maps integrating the spectral cube inside that range for all the pixels. The details of this automated routine are given in Appendix B. \\ \indent  
The CH$_{3}$CN $12_{0}-11_{0}$ and $12_{1}-11_{1}$ lines are heavily blended in several sources. 
Therefore, for these two transitions, we created a single moment-0 map that includes the emission of both transitions.\\ \indent All the maps for each field and transition were then visually inspected to validate them. This was done by looking at the line observed in the spectra averaged on each continuum compact source detected by the source extraction algorithm CuTEx (CUrvature Thresholding EXtractor \citealt{cutexMolinari+11, cutexmolinari2016} see Sect. 4.4; following the same methodology used in \citet{ALMAGAL3} for \textit{\sc 7m+tm2+tm1} data) in the clump, and checking that the velocity extremes used for the integration agree with the velocity ranges of the lines in these spectra.
The employed  method gives good extremes of integration, i.e. extremes that include the whole line for all the cores in the clump, in the majority of the sources. Two example moment 0 maps with spectra averaged over the area of each cores in the field is shown in Fig \ref{fig:mom0example}, while Fig \ref{fig:allmoments} shows the moment-0 maps of all the transitions for one source. The number of sources for which the automated routine gave integration extremes that encompass all the line, without line blending problems, is 916, while for the remaining 97 sources we created the maps using a fixed velocity range.
In fact, a fraction of the sample has, in at least one transition, moment-0 maps for which the extremes of integration automatically derived do not encompass the whole line or partially encompass other transitions. 
This is especially true for extremely line-rich sources where most lines are blended with transitions of other molecular species. For those clumps, the moment0-maps have been recreated integrating the cube over a fixed velocity range of $\pm 5\,$km/s around the systemic velocity of the clump for all the pixels. This range is a compromise between including the majority of the line, while excluding blendings in line-rich cores. These maps, in some spatial pixels, do not include the total velocity range of the line emission, since they clearly exclude possible high-velocity wings and a portion of the line if it is larger than 10km/s.

\begin{figure*}
    \centering
    \includegraphics[width=18.2cm, trim={0 1.8cm 0 2.3cm}, clip=true]{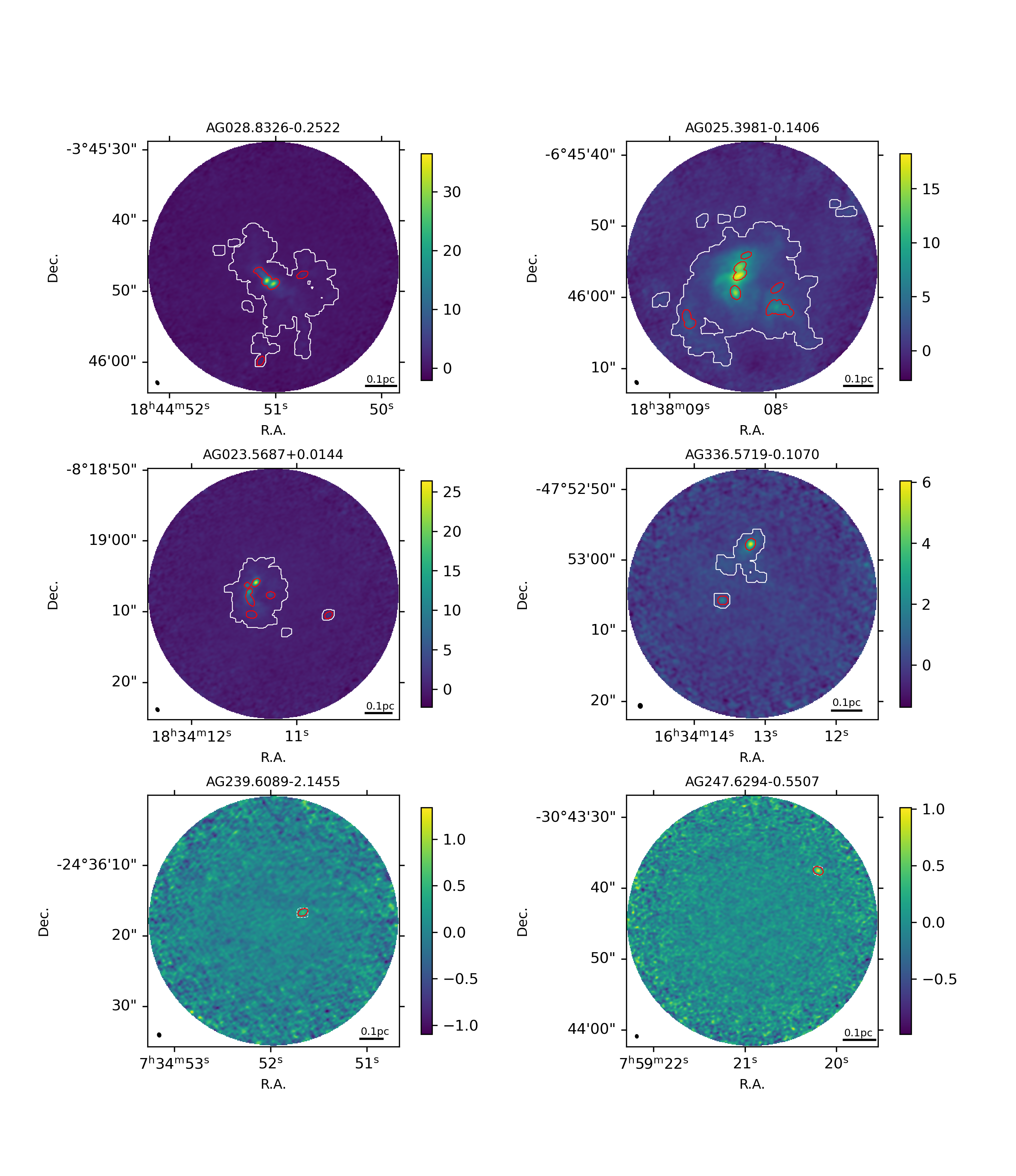}
    \caption{ Examples of continuum emission %inside the 
    for six sources in the
    ALMAGAL sample.
    \textit{Upper row:} continuum maps of two example sources with large diffuse emission outside the cores. \textit{Middle row:} continuum maps of two example sources for which the diffuse emission is present but more restricted in area. \textit{Lower row:} continuum maps of two example sources in which the continuum emission is well described by the mask of all compact sources only (the mask$_{\rm{ext,}}$ is plotted in dashed white for reference). Red contour: mask$_{\rm{comp.}}$, white contour: mask$_{\rm{ext.}}$ (see Section 4.4). The continuum image color scales are in Jy/beam.}
    \label{fig:example3typesofcontinuum}
\end{figure*}

\subsection{Detection and mask on moment-0 maps}\label{sec:mom0mask}

To determine if a selected transition is detected in a target and to compare the emission of the different molecular species with the continuum emission, we created the masks of the line emission from the moment-0 maps. 
Given a moment-0 map of a selected transition of a specific field, we created a mask of all the pixels with SNR$\,>3$, excluding noise fluctuation. More details on the refinement of the moment-0 mask can be found in Appendix A. 
We required that at least one of the regions remaining inside the mask has a number of pixels $>N_{\rm{beam}}$, since this would be the size of emission of a point-like source. 
If this condition is met, we considered the transition as detected in the field.
\\ \indent For methyl-formate (CH$_3$OCHO), the molecule with lowest SNR and least detected (detected in 139 sources), we visually inspected the spectra extracted towards the cores for the detected sources to confirm all the detections, since false detections could possibly alter the results on such a small sample of detections (see Section 5). 
We revealed only 29 false detections that were introduces into the sample where CH$_3$OCHO was detected and we excluded these sources for the following analysis, leaving 110 sources with a real detection. 
This is a very small number of false detections out of the sample of 903 sources without emission of CH$_3$OCHO (3\%), and it corroborates the robustness of the automatic method applied. 
We can expect a similar number of false detections for the other molecular species. 
However, these would not influence the detection statistics because they are negligible compared to the total number of detections for all the other species. Moreover, those false detections would not enter the astroHOG analysis because the regions of false emission are at the edge of the FOV and likely do not intersect with the masks defined for the continuum (see Sections 4.3 and 4.4).

\subsection{Morphological analysis with astroHOG}

We performed the comparison of the continuum and integrated intensity (moment-0) maps of the molecular transitions listed in Table \ref{table:lines_used} using the HOG method presented in \citet{soler_astroHOG}.
The HOG method is a machine vision tool based on the use of the intensity gradient orientation to characterize the similarities between two images. It is at the core of many algorithms used in object detection and classifications in image processing \citep[see, for example,][]{leonardis2006}.
The implementation of the HOG method introduced by \citet{soler_astroHOG}, {\tt astroHOG}\footnote{code available through GitHub at\\ https://github.com/solerjuan/astroHOG}, uses tools from circular statistics to quantify the significance of the accordance of the orientation between the intensity gradients.

The basic analysis in {\tt astroHOG} for two given 2D intensity maps $I_{ij}^{\rm{A}}$ and $I_{ij}^{\rm{B}}$ is as follows.
The gradient of the two images $\nabla I_{ij}^{\rm{A}}$ and $\nabla I_{ij}^{\rm{B}}$ are calculated using Gaussian derivatives, which are the results of convolution of the images with the derivative of a Gaussian with characteristic FWHM $\Omega$ \citep{soler2013}.
This procedure is equivalent to calculating the derivative using weighted differences over points within a diameter $\Omega$.
By definition, the orientation between the gradients in a certain position $ij$ of the maps, which is also the orientation between the iso-intensity contours, is
\begin{equation}
\phi_{ij} =   \arctan \biggl(\frac{ \bigl(\nabla I_{ij}^{\rm{A}}\times\nabla I_{ij}^{\rm{B}}\bigr)\cdot \hat{z}}{\nabla I_{ij}^{\rm{A}}\cdot\nabla I_{ij}^{\rm{B}}}\biggr)\,,
\end{equation}
where $\hat{z}$ is the unit vector in the direction perpendicular to the plane of R.A. and Dec.
If the two maps are identical, all of the angles $\phi_{ij}$ are equal to 0$^{\circ}$.
If the intensity in the two maps is entirely uncorrelated, the distribution of the angles $\phi_{ij}$ should be uniform.
Thus, quantifying the similarity in the intensity distribution in the two images is related to evaluating if the distribution of angles peaks around 0$^{\circ}$. \\ \indent {\tt astroHOG} tests whether the distribution of the relative orientations $\phi_{ij}$ is uniformly distributed or has a preferential alignment at $\phi$\,$=$0$^{\circ}$, using the projected Rayleigh statistic \citep[PRS][]{durand1958, Jow2018}
\begin{equation}
\label{eq:Vraileigh}
V = \frac{\sum_{ij} w_{ij}\,\rm{cos}\,(2\phi_{\mathrm{ij}})}{\bigl(\sum_{ij}\,w_{ij}^{2}/2\bigr)^{1/2}}\,,
\end{equation}
\begin{table*}

\caption{Number of clumps with detection from moment-0 maps}
\label{table:detection}
\begin{center}
\begin{tabular}{lr|lrrrrr}
\hline\hline
& all &$L/M$: &$<0.1$ & $0.1-1$ & $1-10$ & $10-100$ & $>100$\\
\hline
    tot. sources &1013 & & 55 & 368 &316 &244 &24 \\
    \hline
     SiO $5-4$                          & 582   (57.5\%) & &19 (34.6\%)& 176  (47.8\%) & 187  (59.2\%) & 178  (73.0\%) & 21  (87.5\%)\\
     DCN $3-2$                          & 423   (41.8\%) & &4  (7.3 \%)& 58   (15.8\%) & 145  (45.9\%) & 191  (78.3\%) & 22  (91.7\%)\\
    H$_2$CO $3_{0,3}-2_{0,2}$           & 772	(76.2\%) & &30 (54.6\%)& 242  (65.8\%) & 241  (76.3\%) & 234  (95.9\%) & 23	(95.8\%) \\
    CH$_3$OCHO $17_{3,14}-16_{3,13}\,$E & 110	(10.9\%) & &0  (0.0 \%)& 0    (0.0 \%) & 25   (7.9 \%) &  71  (29.1\%) & 13	(54.2\%)\\
      HCCCN $24-23$                     & 371	  (36.6\%) & &1  (1.8\%) & 47   (12.8\%)  & 128  (40.5\%) & 172(  70.5\%) & 21  (87.5\%) \\
    CH$_3$OH $4_{2,3}-3_{1,2}$          & 607	(59.9\%) & &19 (34.6\%)& 169  (45.9\%) & 195  (61.7\%) & 200  (82.0\%) & 21	(87.5\%)\\
     H$_2$CO $3_{2,1} - 2_{2,0}$        & 569   (56.2\%) & &17 (30.9\%)& 140  (38.0\%) & 185  (58.5\%) & 202  (82.8\%) & 22  (91.7\%)\\
    SO $6_5-5_4$                        & 650	(64.2\%) & &13 (23.6\%)& 164  (44.6\%) & 225  (71.2\%) & 222  (91.0\%) & 23	(95.8\%)\\
    CH$_3$CN $12-11$ \tiny{K=0,1}       & 353	(34.9\%) & &4  (7.3 \%)& 43   (11.7\%) & 122  (38.6\%) & 162  (66.4\%) & 21	(87.5\%)\\

    \hline
\end{tabular}
\end{center}
\vspace{1mm}
\small{ \textbf{Notes:} for 6 sources  $L/M$ is  not available, due to a lack of source counterparts in Herschel bands that make it not possible to derive physical parameters as for the majority of the sample that was selected from the Hi-GAL survey (see Section 3.1.1 of \citealt{ALMAGAL1}); thus they are only included in the ‘‘all" column. These six sources are: AG012.7862-0.1784, AG014.9963-0.6733, AG030.8695-0.1031, AG305.2021+0.2073, AG322.1735+0.6443, AG331.5166-0.0937. The ranges in $L/M$ are in units L$_{\odot}$/M$_{\odot}$. }
\end{table*}
where $w_{ij}$ are the statistical weights corresponding to $\phi_{ij}$. Therefore, it is also independent of the intensity of the gradients, but is defined only from the relative orientations of gradients in the two maps. 
In our application of the method, we use the statistical weights to account for the fact that different values of $\phi_{ij}$ inside a beam are not independent.
Thus, we chose weight $w_{ij}$\,$=$\,$w$\,$=$\,$\delta/\Delta$, where $\delta$ is the size of the pixels of the image and $\Delta$ is the size of the beam of the observations. Since, in our case, the weights are the same for all the pixels, Eq. \ref{eq:Vraileigh} becomes:
\begin{equation}
\label{eq:Vwithconstantweights}
V = \Bigl(\frac{2}{N}\Bigr)^{1/2} \sum_{ij}\rm{cos}(\,2\phi_{\mathrm{ij}})\,.
\end{equation}
If the angles $\phi_{\rm{ij}}$ are randomly distributed, as in the case of two completely unrelated images or an image of noise with a real image, the sum of $\cos\,(2\phi_{\mathrm{ij}})$ will give values close to zero. 
The parameter $V$ is thus our estimate of how well the two images are morphologically correlated. 
In the case of identical images, in which all $\phi_{ij}=0^{\circ}$, we obtain the maximum possible value of $V$:
\begin{equation}
\label{eq:Vmax}
V_{\mathrm{max}}\,=\,(2\,N)^{1/2},
\end{equation}
where $N$ is the number of angles $\phi_{ij}$, which is equal to the number of pixels of the image (or inside the mask selected).
Since we aim to compare a large fraction of the ALMAGAL sources, which maps not always have the same numbers of pixels and whose emission covers different portions of the map, we introduced a normalized projected Rayleigh statistic
\begin{equation}
    V_{\rm{N}} = V/V_{\rm max}\,,
    \label{eq:normalizedV}
\end{equation}
where $V_{\mathrm{max}}$ is defined for each clump and mask used (see Section 4.4). Following the definitions, $V_{\rm{N}}$ can only take values between 1 and -1, where negative values are reached when the gradients are more likely to be oriented perpendicularly rather than parallel in the two maps.
The value of $V_{\rm{N}}$ can be interpreted as the percentage of the two maps that shows parallel gradients (if $V_{\rm{N}}>$0), thus the percentage of the maps that is morphologically similar. It has to be noted, that since this method relies only on the relative orientation of the gradients, it is not sensitive to investigate if two tracers peak in the same position or if their emission (assumed to be Gaussian and centered in the same pixel) has the same size (i.e. FWHM), since these quantities are defined on the intensity of the emission.\\ \indent 
We estimated the uncertainties of $V$ and $V_{\rm{N}}$ using the Monte Carlo (MC) method implemented inside the {\tt astroHOG} package, which works as follows. 
Giving as input the standard deviations, $\sigma_{\rm{A}}$ and $\sigma_{\rm{B}}$, of the two images $I^{\rm{A}}$ and $I^{\rm{B}}$ and the number of MC iterations, $N_{\rm{MC}}$, the procedure creates $N_{\rm{MC}}$ images $I^{\rm{A}}_{i} = I^{\rm{A}}+I^{\rm{A}}_{noise}$ adding to the original image an image of random noise with standard deviation $\sigma_{\rm{A}}$, and $N_{\rm{MC}}$ analogous images $I^{\rm{B}}_{j}$. 
An estimate of $V$ is then computed for all the $N_{\rm{MC}}^2$ possible couples  ($I^{\rm{A}}_{i}$\,,\,$I^{\rm{B}}_{j}$), from which it derives the best estimate of $V$ and its standard deviation. 

% --------------------------------------------------------------
\subsection{Details of the application of astroHOG}\label{sec:detailsonASTROHOG}
In this paper, we investigate how the emission of different species is correlated with the dust continuum emission. Moreover, it is interesting to explore if and how this correlation changes at different spatial scales. In particular, we want to investigate the level of correlation in the clump medium, which is traced by the extended diffuse continuum emission, and in the denser fragments which are traced by the compact dust emission.  
For this reason,  we run astroHOG both on the mask of the extended emission (all pixels with the emission of continuum above 3 SNR, mask$_{\rm{ext.}}$) and on the mask of all the compact structures that we identified with the CuTEx algorithm (mask$_{\rm{com.}}$) for each target. More details on the creation of these masks are given in Appendix A. 
 The continuum emission, with the masks defined, are shown in Fig.~\ref{fig:example3typesofcontinuum} for six sources as examples of the variety of emission morphologies in the sample. In some cases, especially for fainter sources, part of the extended emission might not be recovered due to the sensitivity limitation.
The size of mask$_{\rm{ext.}}$ with respect to mask$_{\rm{com.}}$ varies from source to source, with 6 extreme cases in which $N^{mask}_{\rm{ext.}}<2\,N^{mask}_{\rm{com.}}$, where $N^{mask}_{\rm{ext.}}$ and  $N^{mask}_{\rm{com.}}$ are the number of pixels in the two masks (see bottom row of Fig. 4). For these 6 sources we  considered only the mask$_{\rm{com.}}$ for the analysis.
\\ \indent
To give some reference numbers, the median value of the area covered by mask$_{\rm{com.}}$ is of 0.004\,pc$^{2}$, while the median for mask$_{\rm{ext.}}$ is of 0.04\,pc$^{2}$. Both masks possibly include regions with outflows in the molecular emission. This is more relevant for mask$_{\rm{ext.}}$, but depending on the geometry of the outflows and how close from the compact source the shock is present, may also affect mask$_{\rm{com.}}$.  A full disentanglement of the two masks from the regions with outflow emission (i.e. masks of extended/compact emission not co-spatial with outflows, and a mask only covering the outflow regions) would require a prior dedicated study on outflows to obtain the region covered by it, which is beyond the scope of this paper.\\\indent
We expect that, in some cases, the continuum emission of high-mass star forming regions at 1.38\,mm is not only due to thermal dust emission, but may have a contribution from free-free emission. Therefore, we checked for available ALMAGAL counterparts in the CORNISH survey at $5$ GHz (\citealt{Purcell+08,Hoare+12}). We identified 112 matched clumps, mostly evolved clumps (i.e., $L/M>10\,\mathrm{L_{\odot}/M_{\odot}}$), compatible with the potential presence of HII regions. We do not exclude those clumps from the main analysis presented in the paper, since we do expect that the free-free emission will affect the emission of a small region inside each clump. However, we present in Appendix E the same results from the HOG analysis excluding the sources associated with a CORNISH counterparts, to show that the main conclusions of this work are not affected. We show in the same Appendix that also the inclusion of the 97 sources with moment-0 maps created over the fixed $+/-5\,$km/s range (which do not include the totality of the emission of the transition for the pixel of the line-rich cores and in the pixels with high-velocity wings) do not affect the results.
\\\indent
We also expect that in some clumps the line emission in a portion of the map could be optically thick. However, this still provides information about how well the line is tracing the continuum, therefore we do not exclude sources due to optical depth consideration. \\\indent
\begin{figure*}[t!]
    \centering
    \includegraphics[width=18cm, trim={2.4cm 0 2.4cm 0}]{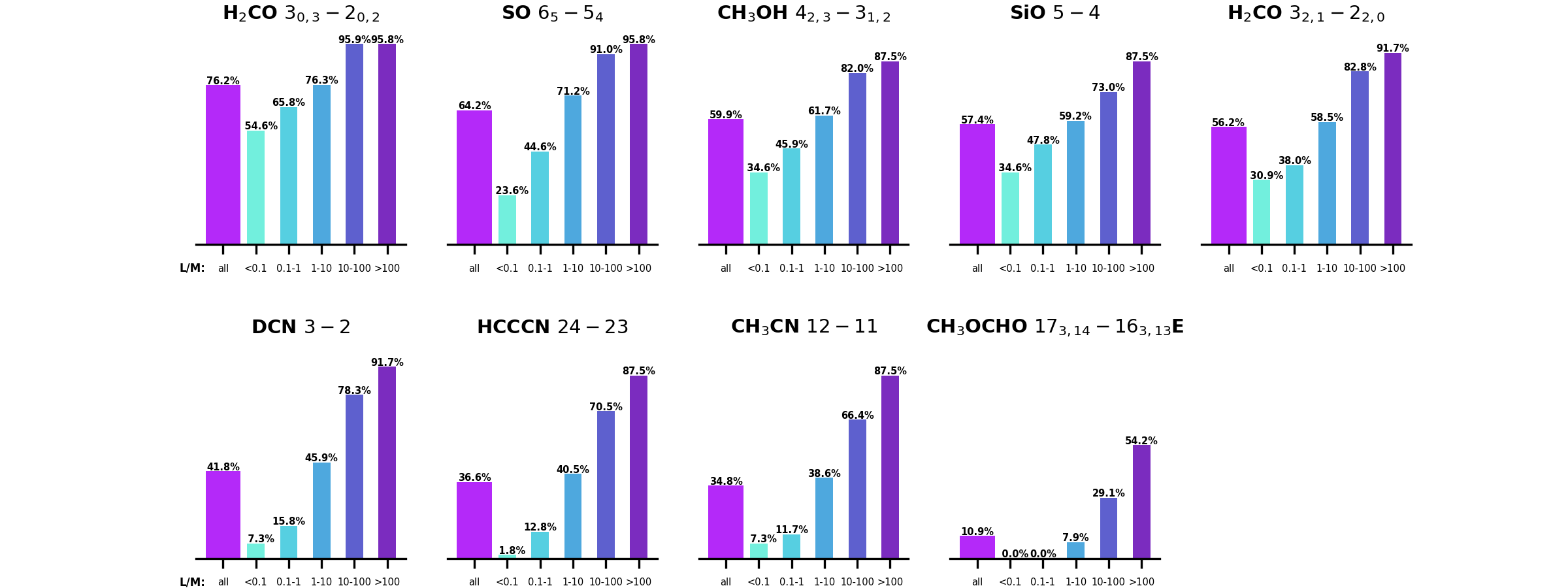}
    \caption{Percentage of detection per clump of the molecular transitions analyzed in this work on the whole sample and divided in ranges of $L/M$. 
    The molecular transitions are sorted by the rate of detection over the whole sample.}
    \label{fig:detectionpie}
\end{figure*}
To compare the morphological emission of the line tracers considered in this work with the continuum emission, we derive the $V_{\rm{N}}$ parameters from {\tt astroHOG} and the Spearman's correlation coefficient, $\rho_{\rm{s}}$. The latter is a statistical correlation coefficient that determines if two sets of data, x$_{\rm{i}}$ and y$_{\rm{i}}$, are well correlated from their intensity values.\\ \indent We compute the correlations of the moment-0 maps and the continuum maps only on the intersection mask between the moment-0 emission masks defined in Sect. \ref{sec:mom0mask} and the continuum emission masks defined here and Appendix A, for both the HOG method and the Spearman's coefficient correlation. %and in detail in Appendix \ref{appendix:continuummasks}. 
This choice was made considering that in determining $V_{\rm{N}}$, the area where only one of the two maps shows emission, while the other is only noise, should contribute with a value close to 0. 
Thus, the $V_{\rm{N}}$ value would be lowered by the percentage of the area of the map where only one of the tracers is detected.\\\indent 
To have robust and reliable results, for the intersection with mask$_{\rm{com.}}$ we took all the compact sources that have at least 60\% of their area covered also by line emission, and checked that the pixels in the masks cover at least three times the area of the beam; for mask$_{\rm{ext.}}$ we performed the analysis only if the intersection with the mask of moment-0 covers at least 5 times the area of the beam and if the number of pixels of the intersection is at least two times the number of pixels of the mask derived from the intersection of mask$_{\rm{com.}}$.
Lastly, to obtain an error estimate on $V_{\rm{N}}$, we used the MC method calculating 100 times $V_{\rm{N}}$ to derive its error (see Section 4.3). The noise maps considered in the MC method are not correlated below the beam size, unlike the noise of interferometric maps. However, the value of correlation of the real noise in our dataset is below 3$\%$ (see Appendix B), and the correlation in the noise is not relevant in the regions with S/N>3 considered inside our masks.

% RESULTS ===============================================================

\section{Results}

\subsection{Statistics of detections on moment-0 maps}\label{sec:resultdetectionstat}
\begin{figure}[h!]
    \centering
    \includegraphics[width=8.5cm, trim={0.cm 0cm 0 1cm}, clip=True]{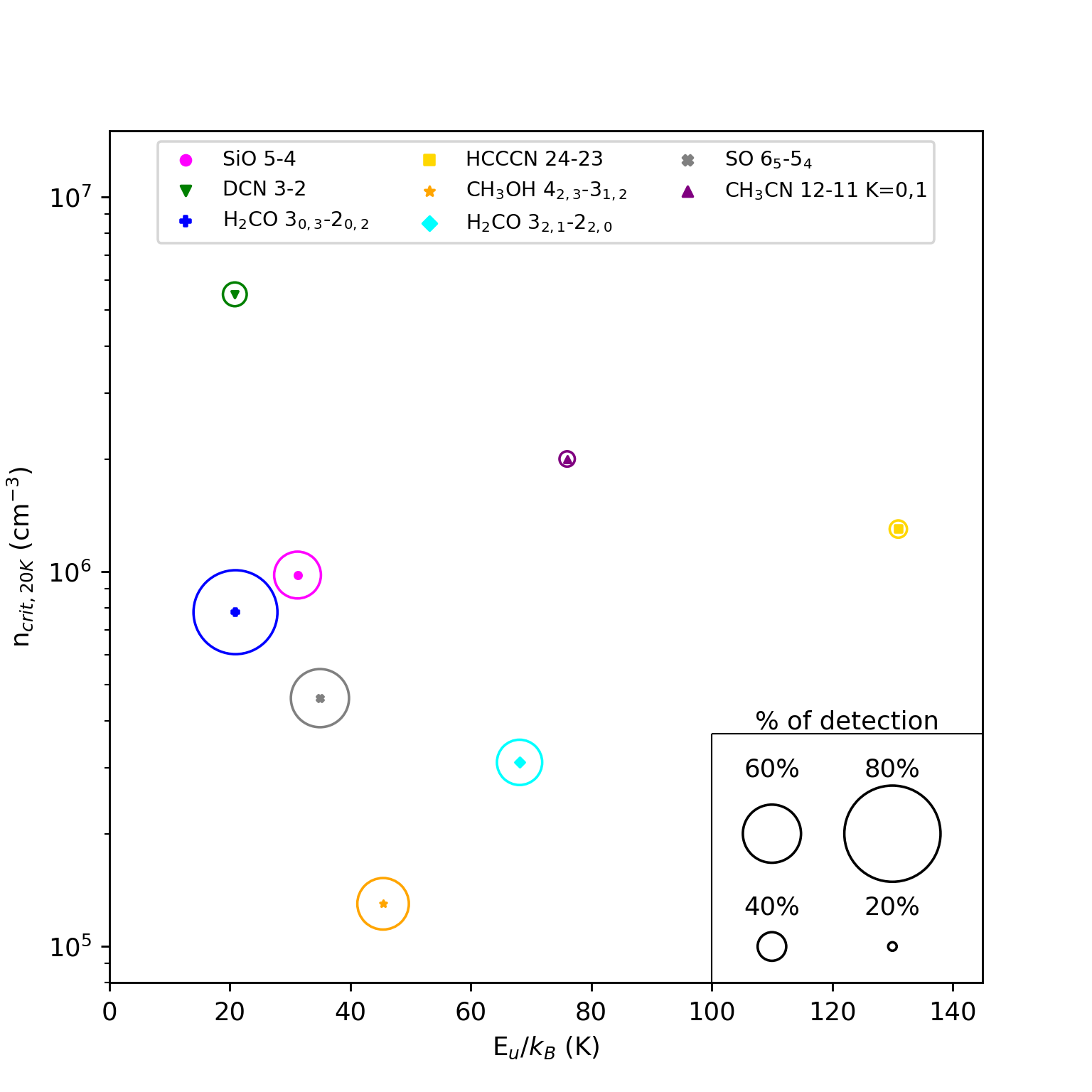}\\
    \caption{
    Percentage of detection on the whole sample of the molecular transitions, shown as the diameter of the circles, in a scatter plot of the upper state energy transition and of the transition's critical density at 20\,K.
    }
    \label{fig:detectionratewithncrit}
\end{figure}

\begin{figure}
    \centering
    \includegraphics[width=0.95\linewidth, trim={1.cm 18.6cm 1cm 1cm}, clip=True]{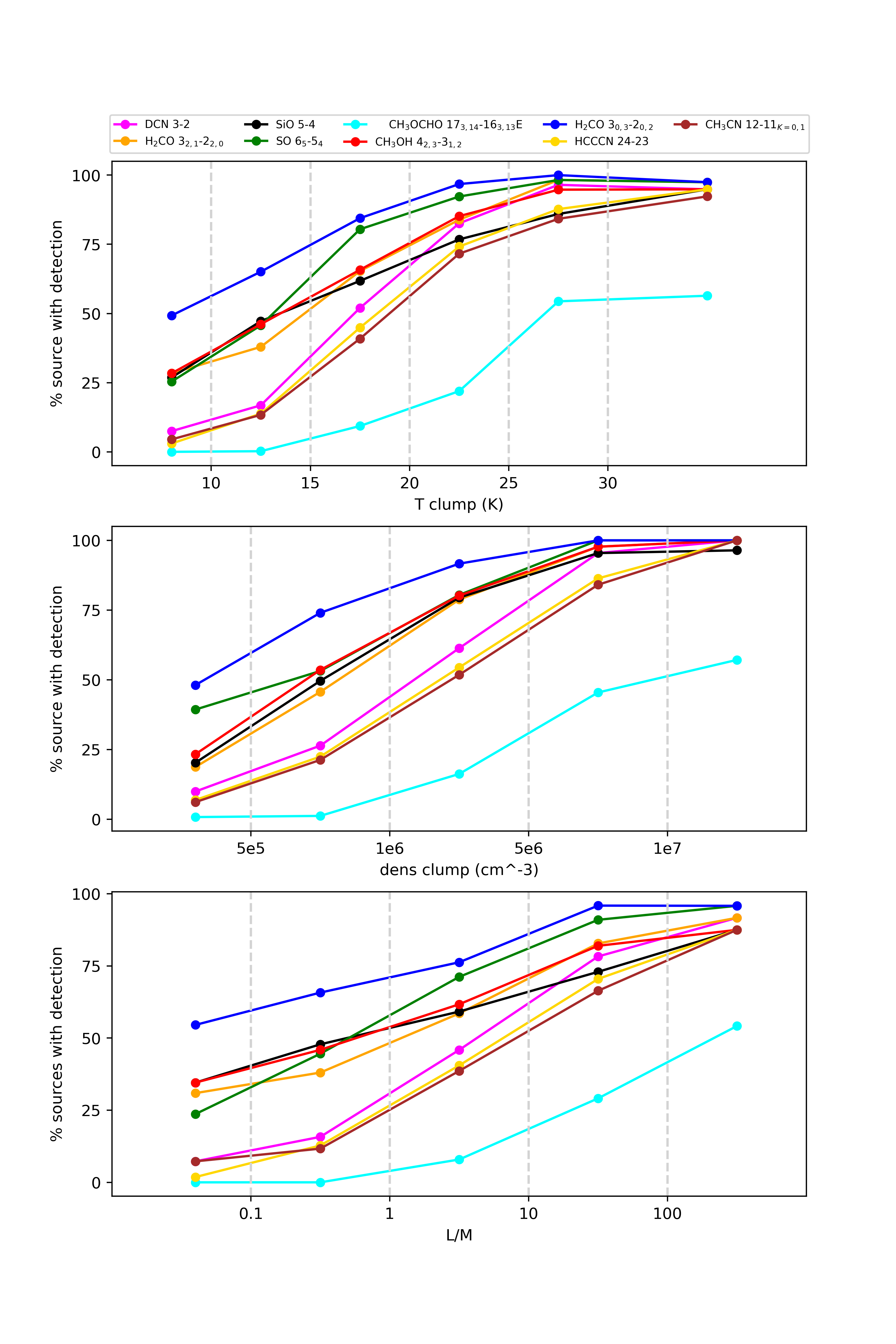}\\
    \includegraphics[width=0.95\linewidth, trim={1.cm 1cm 1cm 20cm}, clip=True]{figures/grid_properties_detection.png}\\
    \caption{Increase of the detection statistic of the molecular transitions analyzed, divided in bins of temperature of the clump (upper panel) and evolutionary stage of the clump L/M (lower panel).}
    \label{fig:grid_prop}
\end{figure}

In this section, we analyze the statistics of the detection of the different molecular transitions, starting from the emission of the moment-0 maps. 
The number of clumps where a specific tracer is detected is given in Table \ref{table:detection}  together with the corresponding percentage of the total sample, which is also shown in Fig. \ref{fig:detectionpie}. 
Moreover, Table \ref{table:detection} and Fig. \ref{fig:detectionpie} show the percentage of detections when dividing the sample into 5 ranges of luminosity-to-mass ratio ($L/M$), a distance independent evolutionary indicator \citep{molinari2008,cutexmolinari2016,molinari2019, giannetti2017, elia2017, elia2021, konig2017}. Sources with $L/M<1\,\rm{L_{\odot}/M_{\odot}}$ are considered starless candidates. Considering the large number of sources in the ALMAGAL sample, we divided them into 5 ranges, to explore also very young sources ($L/M<0.1\,\mathrm{L_{\odot}/M_{\odot}}$) and the most evolved ($L/M>100\,\mathrm{L_{\odot}/M_{\odot}}$) ones.
\\\indent Part of the differences in the detection statistics, and of the increase of the detection rate with L/M, originates from the chemical pathways that lead to the formation of these molecular species, which we reviewed in Section 2.2. \\\indent Together with the chemistry taking place, to interpret the difference in the detection rate for different molecular transitions, we have to consider that the critical density,  $n_{\rm{crit}}$,  and upper state energy, $E_{\rm{U}}/\kappa_{B}$ (where $\kappa_{B}$ it the Boltzmann's constant), of the selected transitions play a role in their detectability. These two parameters are listed in Table \ref{table:lines_used}, except for the critical density of CH$_3$OCHO since there are no collisional cross sections available for the E-conformer. We can see that the molecule with the highest detection rate is  $\rm{H_2CO}$  with 76\% in the  $\rm{H_2CO}\, 3_{0,3}-2_{0,2}$ transition and 56\% in the $3_{2,1}-2_{2,0}$, which is the molecular species that form both in gas-phase and on grains and that can be released from the grains at the lowest temperature, and also in shock regions thanks to sputtering of the grains. The highest detected transition has the lowest upper energy of all the transitions analyzed in the sample (20\,K), while $\rm{H_2CO}$ $3_{2,1}-2_{2,0}$ has an upper energy of 68\,K. After H$_2$CO, the most detected transitions are those of SO, CH$_3$OH, and SiO with a detection rate of $\sim60\%$. The formation of those molecular species happens on grains, or through rapid gas-phase reactions after the release of some precursor atoms/species, and are then released thanks to sputtering or thermal desorption. The specific transitions analyzed in this paper, also have similar upper energy transitions between 30\,K and 45\,K. DCN $3-2$ has a detection rate of 42\% over the entire sample, despite having one of the lowest $E_{\rm{U}}/\kappa_{B}$. However, the abundance of this molecular specie is enhanced in warm/hot gas through gas-phase reactions, it is a deuterated species, and the specific transition has the highest critical density among the transitions analyzed. The three least detected molecular transitions are those of HCCCN, CH$_3$CN, and CH$_3$OCHO, respectively. The abundances of these molecular species are mainly enhanced in hot gas thanks to their release from grains at temperature above 90\,K, therefore they are mostly detected only on the more evolved sources in the sample. CH$_3$OCHO is the most complex molecular species in the sample, which is also the least detected. The detection rates of the different transitions as a function of $E_{\rm{U}}/\kappa_{B}$ and $n_{\rm{crit}}$, calculated for a gas temperature of 20\,K, are given in Fig. \ref{fig:detectionratewithncrit}. From the image, the percentages of detection within the total sample show a clear decreasing trend with $E_{\rm{U}}/\kappa_{B}$ for values up to 70\,K, together with a decreasing trend with $n_{\rm{crit}}$, as highlighted by DCN $3-2$.
\\ \indent The percentage of detections as a function of $L/M$ varies noticeably for each molecular species. The rate of detection increases for all species from the lower value of $L/M$ to the highest ones, where all of the molecular transitions except of CH$_3$OCHO  show at least a detection rate of $\sim87$\%.\\
\indent
From the lower panel of Fig. \ref{fig:grid_prop}, we note that H$_2$CO (both transitions), CH$_3$OH, SiO, and SO have already detection rates above $\sim25$\% in the less evolved sources ($L/M<0.1\,L_{\odot}/M_{\odot}$), with H$_2$CO 3$_{0,3}-2_{0,2}$ reaching already more than 50\%. This high detection is expected for H$_2$CO that can desorb at $\sim40\,$K from grains and have gas-phase creation pathways, and CH$_3$OH which has already been found in cold environment, despite a high desorption temperature. The $24\%$ and $34\%$ detection rate of SO and SiO can could be related to the higher efficiency of outflow observations to reveal the star-formation process from its very beginning compared to the infrared emission \citep[e.g.][]{motte2007SiOinCygnusX, LiOutflows2020, TownerAlmaIMFOutflow} or to the presence of a narrow Gaussian component of SiO and SO in star-forming regions \citep[e.g.][]{motte2007SiOinCygnusX, SanhuezaSiOnarrow, duartecabral2014A&A, csengeri2016atlasgal2}, whose origin is caused by low-velocity shocks from cloud-cloud collisions, but not fully understood yet. The presence of outflow in very early stages could have also contributed to part of the high detection rate detected in H$_2$CO and CH$_3$OH. However, disentangling the two components as well as the identification of outflows is beyond the scope of this paper and will be explored in forthcoming papers of the ALMAGAL collaboration. \\ \indent
On the other hand DCN, HCCCN, CH$_3$CN, and CH$_3$OCHO have a statistics of detection lower than 8\% for L/M below 0.1\,$\mathrm{L_\odot/M_\odot}$, since their chemical pathways require high temperature for their formation/release in gas-phase. The increase of the detection rate of these molecular species is thus more sensible (the trend is steeper) to the evolutionary stage and also the mean temperature of the clump (see upper panel of Fig. \ref{fig:grid_prop}), as derived by \citet{elia2017, elia2021} from the SED fitting of the 160-850$\,\mu$m continuum emission.  
\begin{figure*}[h]
    \centering
    \includegraphics[height=2.9cm, trim={0.6cm, 0, 1.7cm, 0}, clip=True]{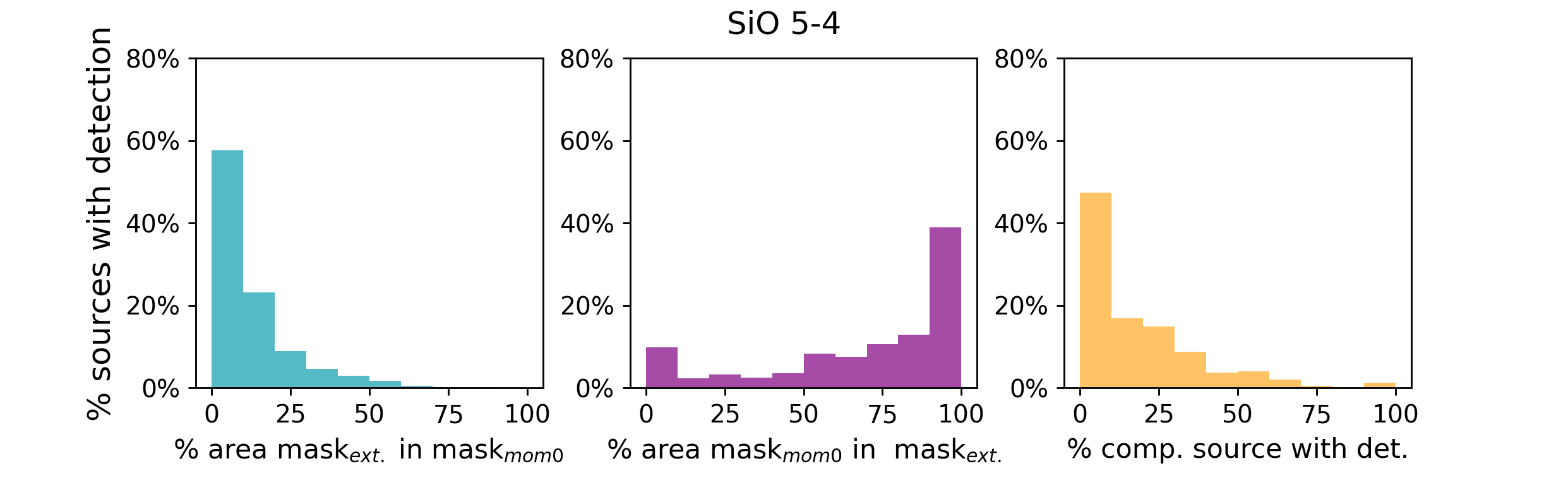}
    \includegraphics[height=2.9cm, trim={0.89cm, 0, 1.7cm, 0}, clip=True]{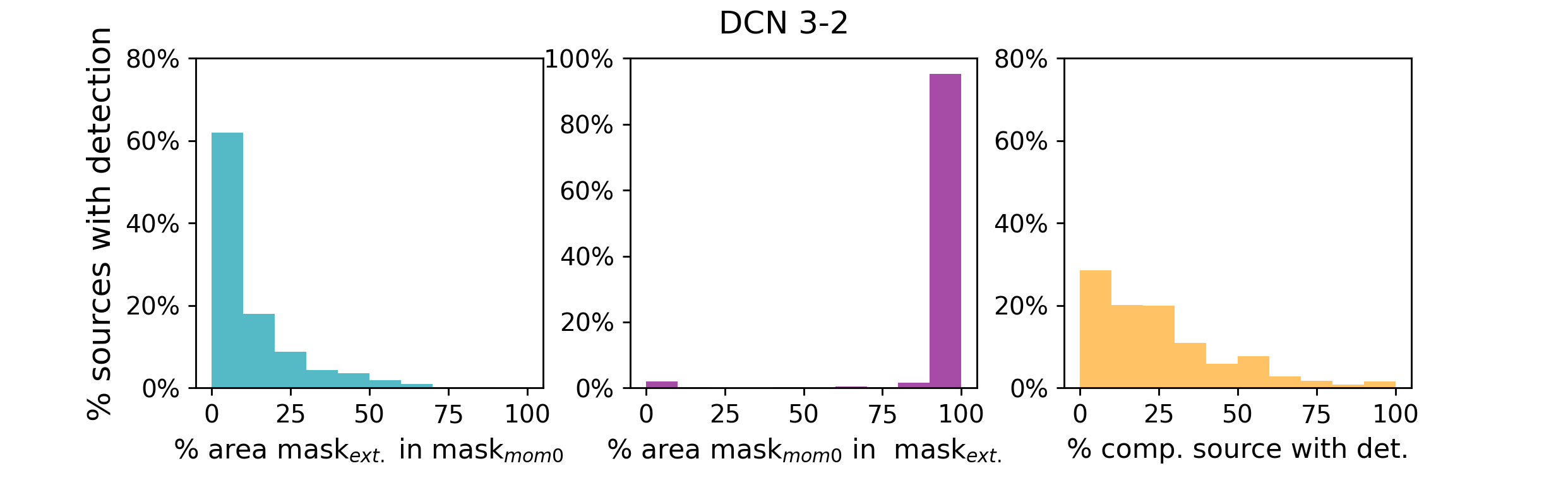}\\
    \includegraphics[height=2.9cm, trim={0.89cm, 0, 1.7cm, 0}, clip=True]{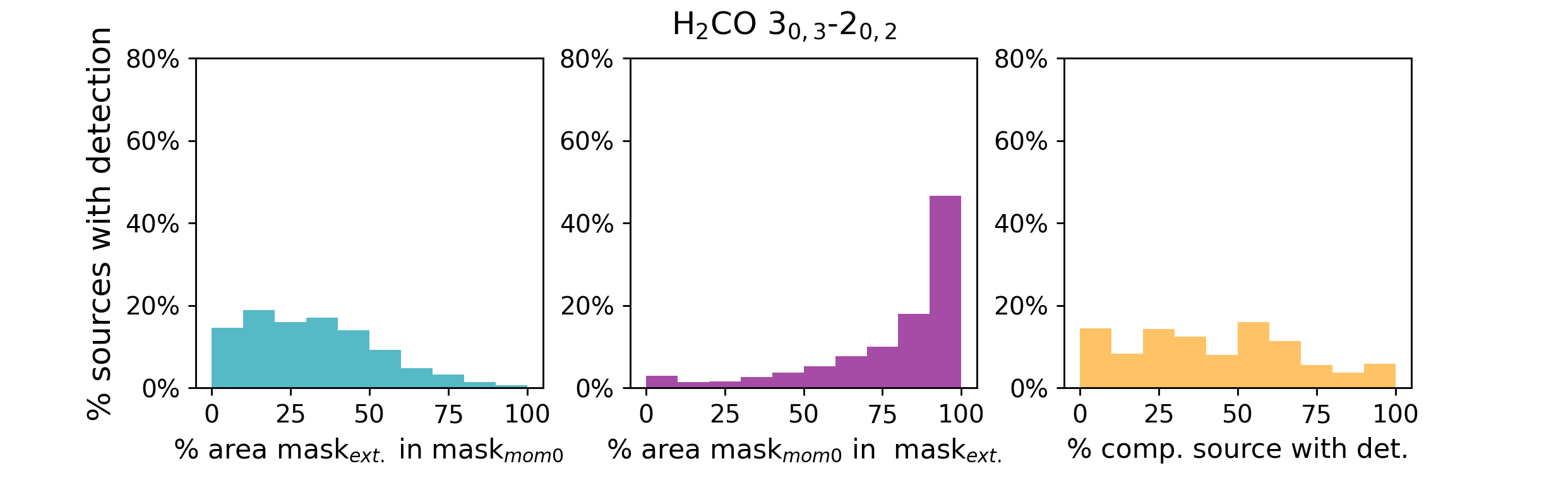}
    \includegraphics[height=2.9cm, trim={0.6cm, 0, 1.7cm, 0}, clip=True]{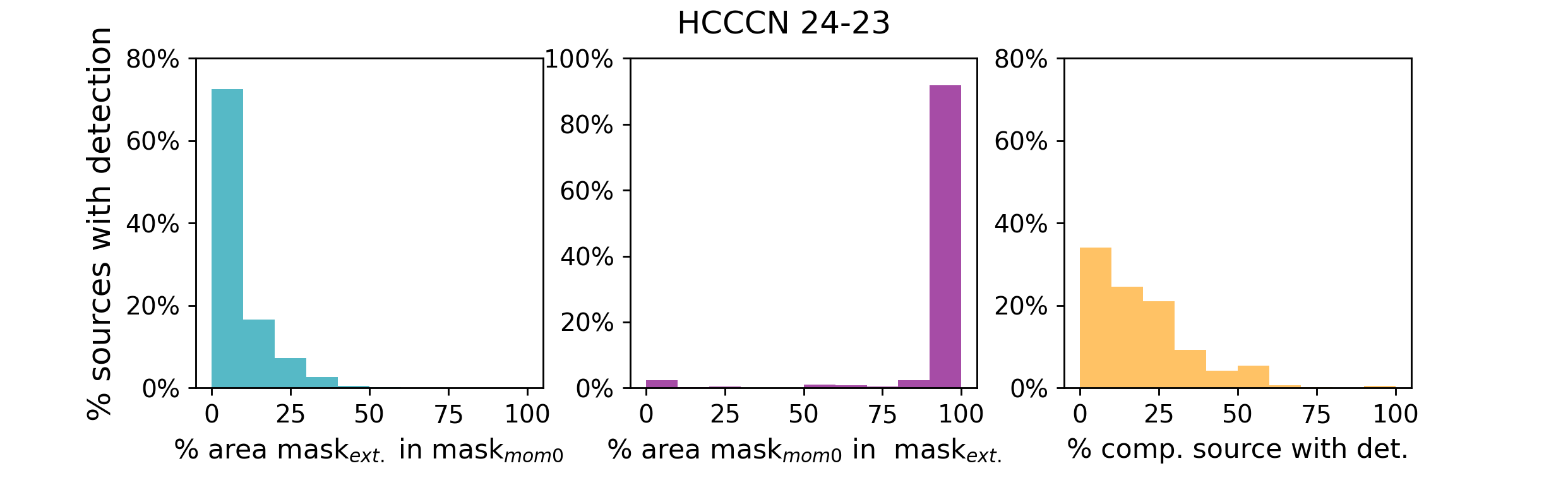}\\
     \includegraphics[height=2.9cm, trim={0.89cm, 0, 1.7cm, 0}, clip=True]{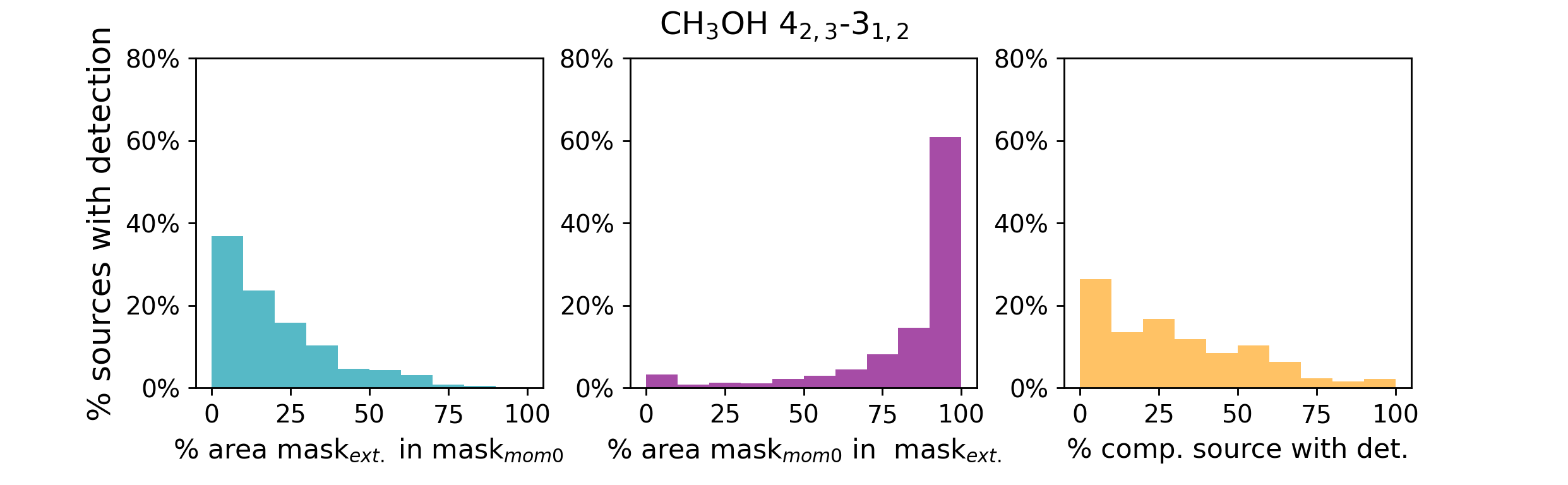}
    \includegraphics[height=2.9cm, trim={0.89cm, 0, 1.7cm, 0}, clip=True]{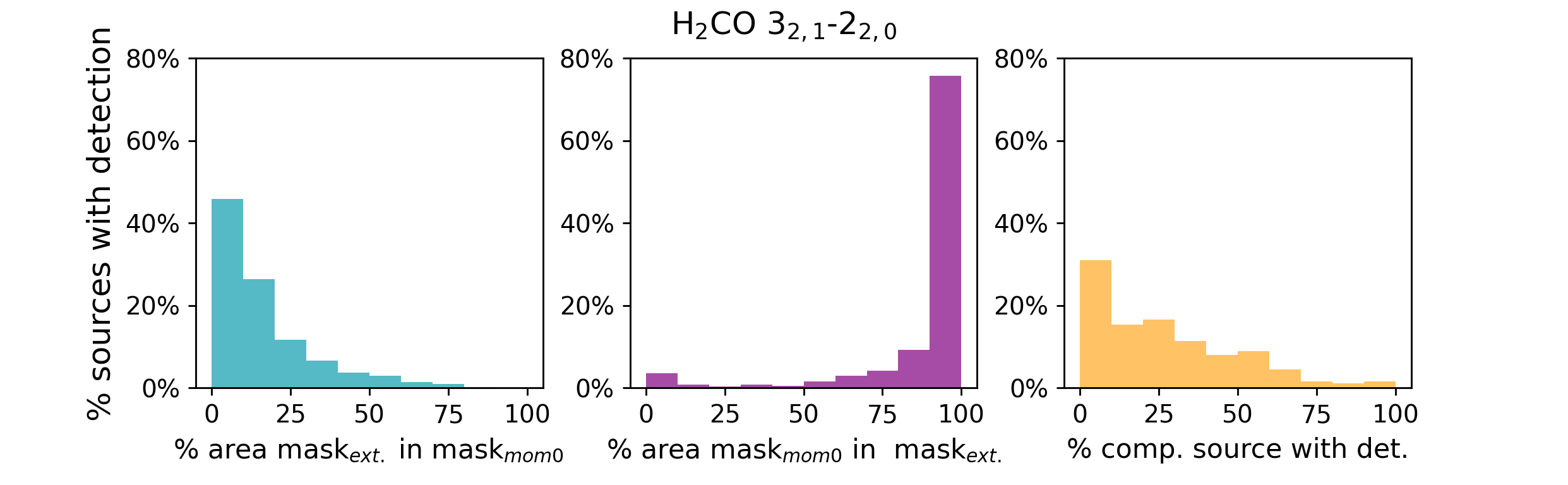}\\
    \includegraphics[height=2.9cm, trim={0.6cm, 0, 1.7cm, 0}, clip=True]{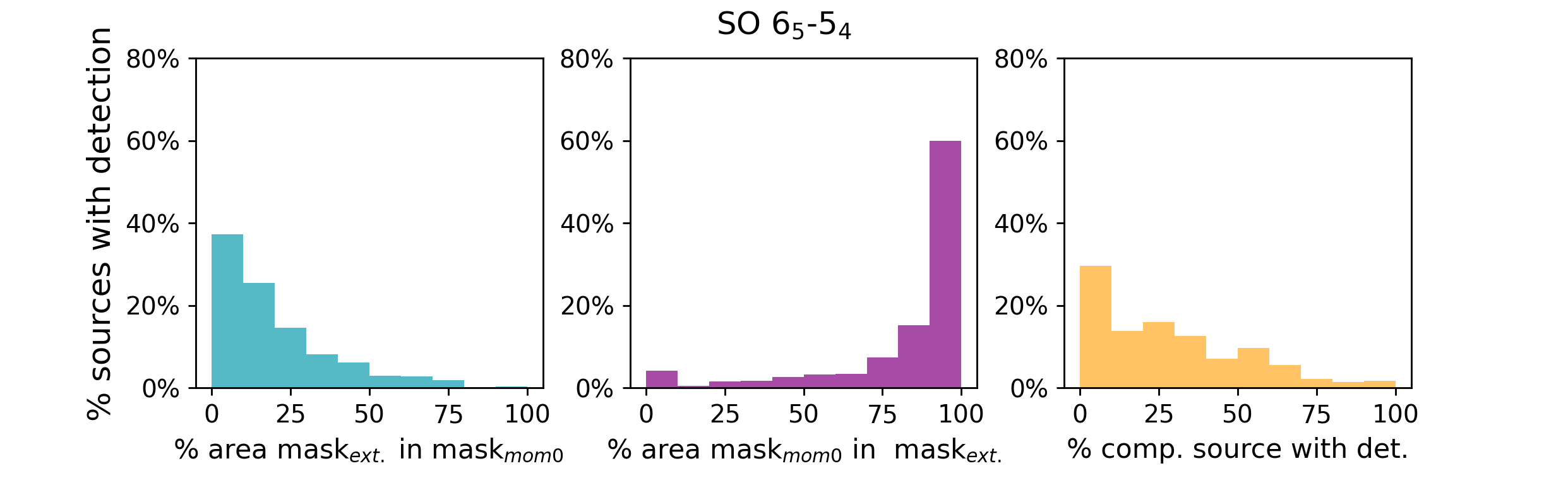} 
    \includegraphics[height=2.9cm, trim={0.89cm, 0, 1.7cm, 0}, clip=True]{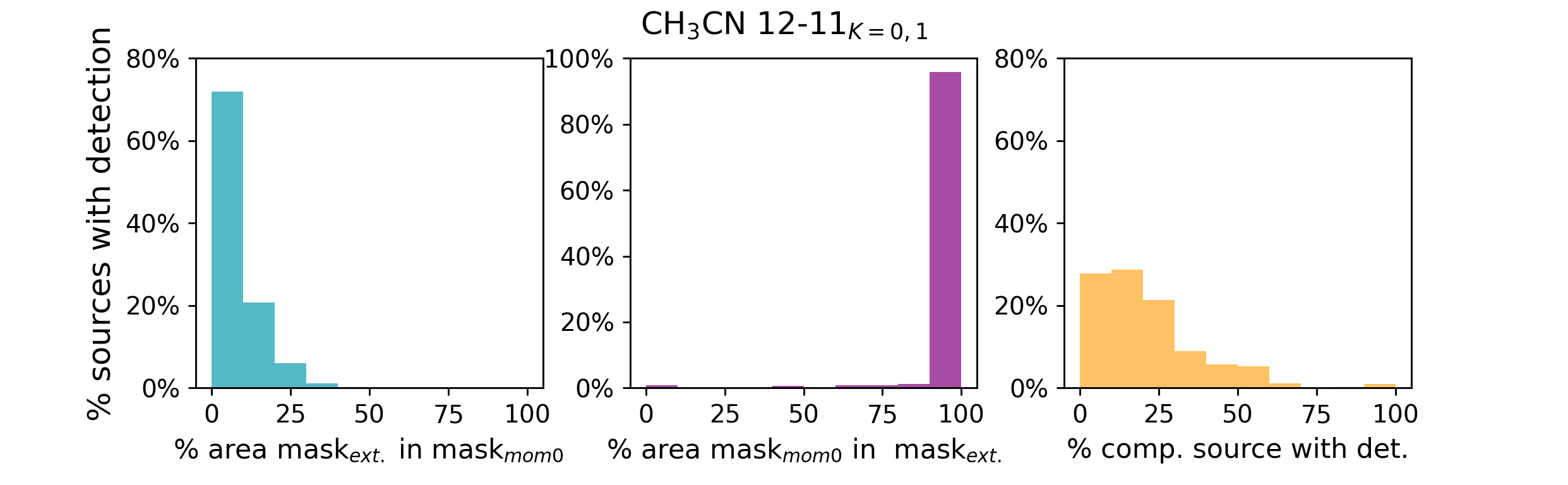}\\
    \includegraphics[height=2.9cm, trim={0.89cm, 0, 1.7cm, 0}, clip=True]{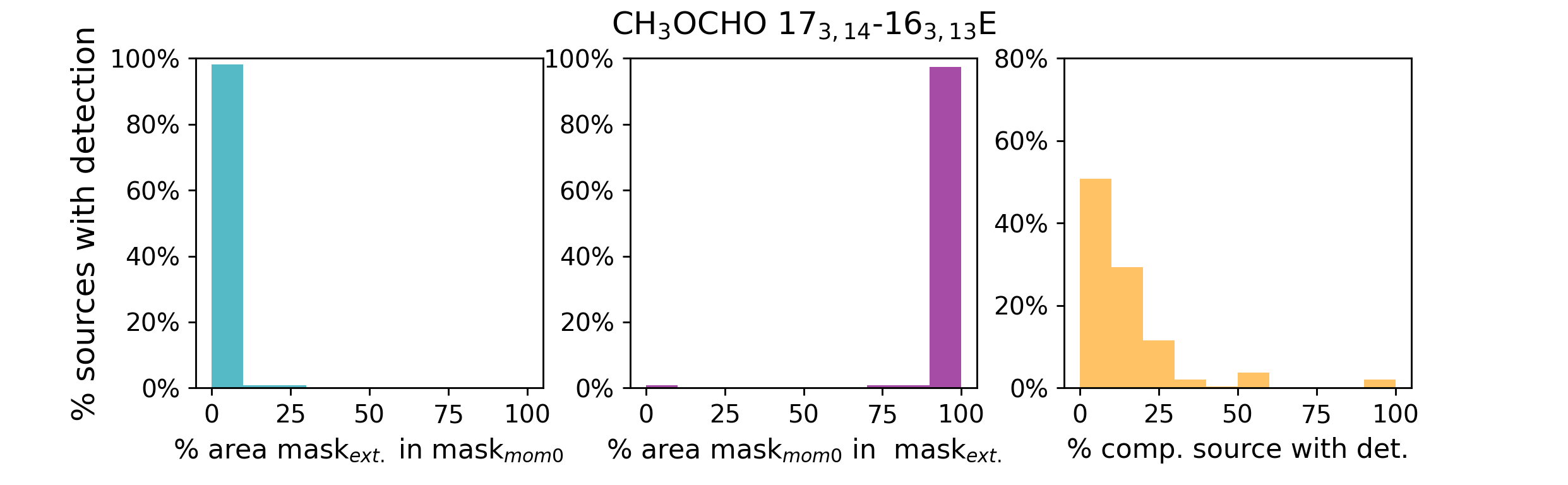}
    \caption{For each molecular transition, three histograms are presented: \textit{blue histogram}: percentage of mask$_{\rm{ext.}}$ covered by the moment-0 mask of the selected transition;  \textit{purple histogram}: percentage of the moment-0 mask of the selected transition covered by the mask$_{\rm{ext.}}$;  \textit{yellow histogram}: percentage of compact sources in which the transition is detected, i.e. compact sources with at least 60\% of the area covered by the moment-0 mask of the selected transition. %Notice that for the blue histogram of CH$_3$OCHO $17_{3,14}-16_{3,13}$E the scale on the y-axis is different from all the other panels, going up to 100\%.}
    }
    \label{fig:step_detection_H2COtoSiO}
\end{figure*}
\subsection{Area of line emission vs area of continuum emission}

In this section, we will discuss the area of emission of the selected molecular lines with respect to the area of the emission of the continuum at the two levels  (extended continuum and emission of the compact sources) defined by the mask presented in Sect. \ref{sec:detailsonASTROHOG}. 
In Fig.  \ref{fig:step_detection_H2COtoSiO}, we display for each molecular transition the histogram of the percentage of the area of mask$_{\rm{ext.}}$ covered also by the emission in the moment-0 map (in blue), the percentage of the mask of the emission of the moment-0 map covered by the mask$_{\rm{ext.}}$, and the percentage of dense cores detected in the molecular line (i.e. cores with
at least 60\% of the area covered by the moment-0 mask of the selected transition, in yellow).
Only H$_2$CO, CH$_3$OH, and SO, already all known to be tracing also outflows, have at least 25\% of sources with detection for which the molecular line emission covers more than 25\% of the extended continuum mask. In particular the percentage is of 66\%, 40\%, 38\%, and 27\% for H$_2$CO 3$_{0,3}-2_{0,2}$, SO, CH$_3$OH, and H$_2$CO 3$_{2,1}-2_{2,0}$, respectively. 
On the other hand, HCCCN, CH$_3$CN, and CH$_3$OCHO only have 11\%, 7\%, and 1\% respectively of sources that cover more than 25\% of the extended continuum mask. DCN and SiO have an intermediate number of sources around $\sim20\%$ for which their emission cover at least 25\% of the extended continuum mask. The fact that DCN can, in some sources, trace not only the dense emission, but also be present in diffuse gas, and also be easily detected at an advanced stage of the star-formation process and trace outflows has also been shown by \cite{SakaiASHES_DCNoutflow}, \cite{cunningham2023DCN}, and \cite{gieser2021}, likely due to warm gas phase reactions occurring \citep{roueff2007}. \\ \indent
Also, in the case of SiO emission, only 20\% of the sources cover more than 25\% of the area of mask$_{\rm{ext.}}$. Thus, SiO shows a different overall emitting region than SO, even though both are outflow tracers. The reasons behind the low percentage of SiO coverage of the extended continuum emission mask do not imply that SiO does not have extended emission, but that in most cases this is mainly not in the region overlapping with the continuum emission. In fact the purple histogram of SiO shows the largest percentage of sources having less than 50\% of the moment-0 map covered by the mask of the extended continuum emission, with respect to all the other molecular species, including the other outflow tracers SO, H$_2$CO, and CH$_3$OH.

\begin{figure*}
    \includegraphics[width=20cm, trim={1.2cm, 0.3cm, 0, 0}, clip=True]{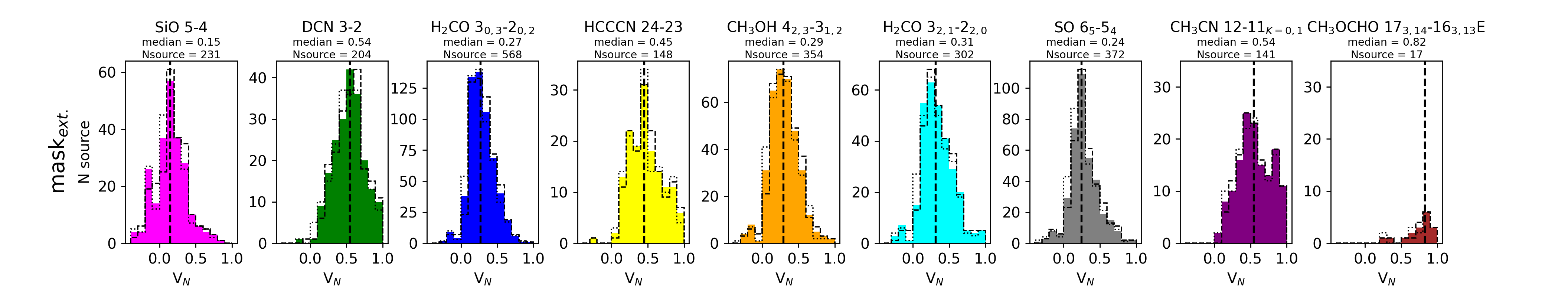}\\
    \includegraphics[width=20cm, trim={1.2cm, 0, 0, 0.7cm}, clip=True]{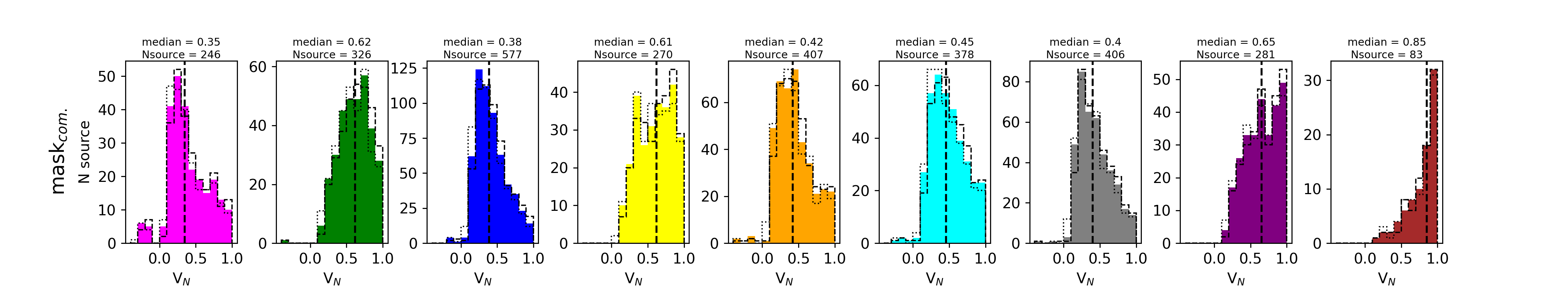}
    \caption{
    Morphological correlation between the continuum emission and the line emission (moment-0) for the indicated transition, as quantified by $V_{\rm{N}}$ parameter from the \texttt{astroHOG} method.
    Values of $V_{\rm{N}}$\,$\approx$\,0 and $V_{\rm{N}}$\,$\approx$\,1  correspond to no correlation and perfect correlation in the distribution of the two emissions, respectively.
    The upper panels correspond to the correlation on the intersection of the moment-0 line emission mask and the mask$_{\rm{ext.}}$.
    The lower panels correspond to a mask selecting only the compact sources where the line is detected, starting from mask$_{\mathrm{com.}}$.
    The dotted and dashed histograms represent how the distribution would change if all the values were replaced by V$_{\mathrm{N}}-\,$error (the error on the value of $V_{\mathrm{N}}$ is derived using an MC method, see Sect. 4.3) and $V_{\mathrm{N}}+\,$error, respectively. }
    \label{fig:resultsastroHOG}
\end{figure*}
\subsection{astroHOG results}
\begin{figure}[t!]
    \includegraphics[width=10cm, trim={0.5cm, 0, 0, 0.5cm}, clip=True]{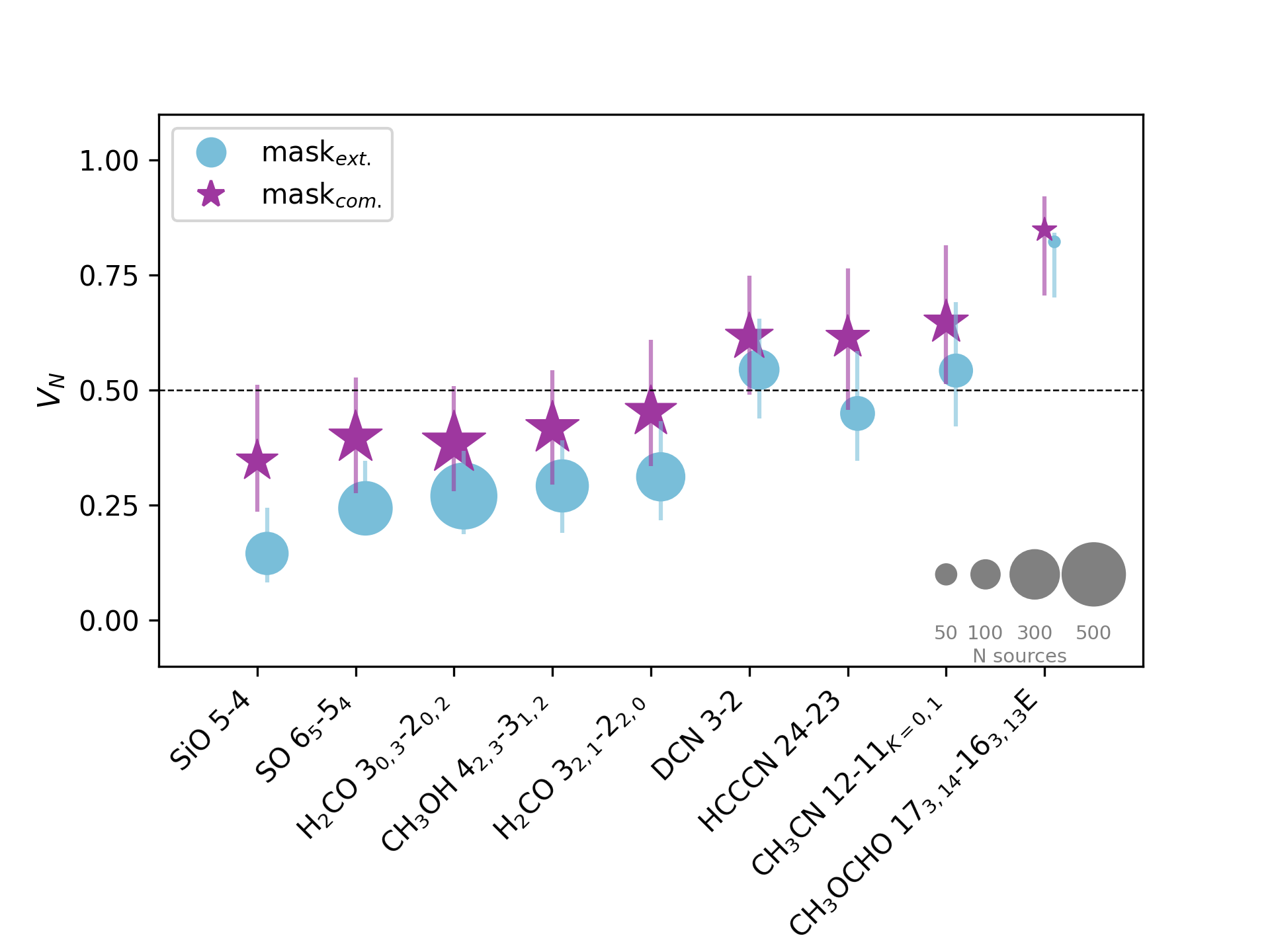}
    \caption{Median value of $V_{\mathrm{N}}$ histograms for the mask$_{\rm{ext.}}$ (light-blue points) and for the mask$_{\mathrm{com.}}$ (purple stars) for the different molecular transitions studied in this paper. The horizontal dashed line represent the values of $V_{\mathrm{N}}$ for which $\sim50\%$ of the area has a good agreement in the two maps (continuum and moment-0).}
    \label{recapVmedian}
\end{figure}

The results of the morphological comparison between the continuum emission and the emission of the molecular lines selected in this paper with \texttt{astroHOG} are shown in Fig. \ref{fig:resultsastroHOG} and are tabulated in Table D.1 and D.2, for the first 40 sources for the lines of H$_2$CO on the mask$_{\rm{ext.}}$ and mask$_{\rm{com.}}$, respectively. The full versions of Table D.1 and D.2 and the analogous Tables for all the other molecular lines are available through Zenodo. The top row of Fig. \ref{fig:resultsastroHOG} shows the results from the comparison with the diffuse emission (mask$_{\rm{ext.}}$), while the bottom row shows the same results with the dense emission (mask$_{\rm{com.}}$). Moreover, in Fig. \ref{recapVmedian} we plot only the median values of the histograms presented in Fig. \ref{fig:resultsastroHOG}. Values of $V_{\mathrm{N}}$ above 0.02 deviate from the statistical correlation that can be found running astroHOG on noise maps (see Appendix C), and are considered significant. The same plot excluding sources with CORNISH counterparts, i.e. excluding possible free-free contamination in the continuum emission, or with only the sources for which moment-0 maps have been created selecting a fixed range of +/-5km/s range are presented in Fig. \ref{fig:NORADIO}, and shows no significant variation from the main results of Fig. \ref{recapVmedian}.  \\ \indent As shown in the previous section, only H$_2$CO (both transitions), CH$_3$OH, and SO cover in more than the 25\% of the sources at least 25\% of the area of mask$_{\rm{ext.}}$. The two transitions of H$_2$CO and CH$_3$OH have similar median values, 0.27-0.29 (27-29\%), while the median morphological correlation of SO with the continuum at the diffuse scale is 0.24 (24\%).  
These percentages highlight that on average these transitions do not trace well the continuum emission arising from the diffuse thermal dust emission. 
In fact, these numbers can be interpreted as saying that, on average, only 24 to 29\% of the area compared is morphologically well correlated. We report the results also for the rest of the molecular transitions, but the results of the previous section should be kept in mind, i.e. that SiO and DCN have only 20\% of the sources covering more than 25\% of the area of mask$_{\rm{ext.}}$, while for the rest of the molecular species the percentage is below 10\%. Therefore the results of the astroHOG on mask$_{\rm{ext.}}$ are in general for this transitions limited to a smaller area. SiO shows the lowest value of correlation, with a median value of 0.15 (15\%). This value is the lowest in the entire sample of transitions analyzed. On the other hand, the results for HCCCN, DCN, and CH$_3$CN are $\sim0.5(50\%)$, while for CH$_3$OCHO (for which we have only 17 sources, while in the other case we always have more than $\sim150$  following the selection criteria on the masks detailed in section 4.4) we found a value of 0.84 (84\%). 
These considerations should be taken statistically over the source sample, since for all species we find some specific sources in which $V_{N}$ is very close either to 0 or to 1 (100\%).\\ \indent For the comparison on compact continuum emission, the molecular transition with the lowest value of morphological correlation is again SiO, with a median correlation of  0.35 (35\%). The other molecular species can be divided in two groups. The first group consists of H$_2$CO (both transitions), CH$_3$OH, and SO. For this group, we found median 
correlation values of around 40\%, with an increase between the values found for the same molecular transitions at the diffuse scale. However, these molecules, even at the scales and densities of the cores, do not present the same morphology of the continuum emission.\\ \indent 
The second group of molecules, which includes HCCCN, CH$_3$CN, DCN, and CH$_3$OCHO, show a very high correlation with the morphology of the continuum at the dense cores scales. For these molecules, the value of the normalized projected Rayleigh statistic is in the range 62\%-65\%, with the exception of CH$_3$OCHO that has a median value of 87\%. CH$_3$OCHO and CH$_3$CN (respectively an O-bearing and an N-bearing molecule) are the two tracers with the best agreement in morphology with the compact continuum. Most of the transitions considered for these molecules have values of $E_{\rm{U}}/\kappa_{B}$ at the higher end of the range covered by all the transitions analyzed in this paper, with the exception of the transition of DCN (from Table 1 the DCN 3-2 line has $E_{\rm{U}}/\kappa_{B}\sim21\,$K), which still shows a high degree of morphological correlation. Moreover, in the first group of molecules we also find the transition of H$_2$CO$\,3_{2,1}-2_{2,0}$ which has a value of $E_{\rm{U}}/\kappa_{B}$ similar to those of CH$_3$CN, but still has a low correlation with the continuum morphology (below 50\%). This leads us to conclude that the difference between the two groups of molecular transitions is related to the molecular species. However, some bias due to the selected transition is present especially on the compact mask.\\ 
\begin{figure}[h!]
    \includegraphics[width=10cm, trim={0.5cm, 3.5cm, 0, 0.5cm}, clip=True]{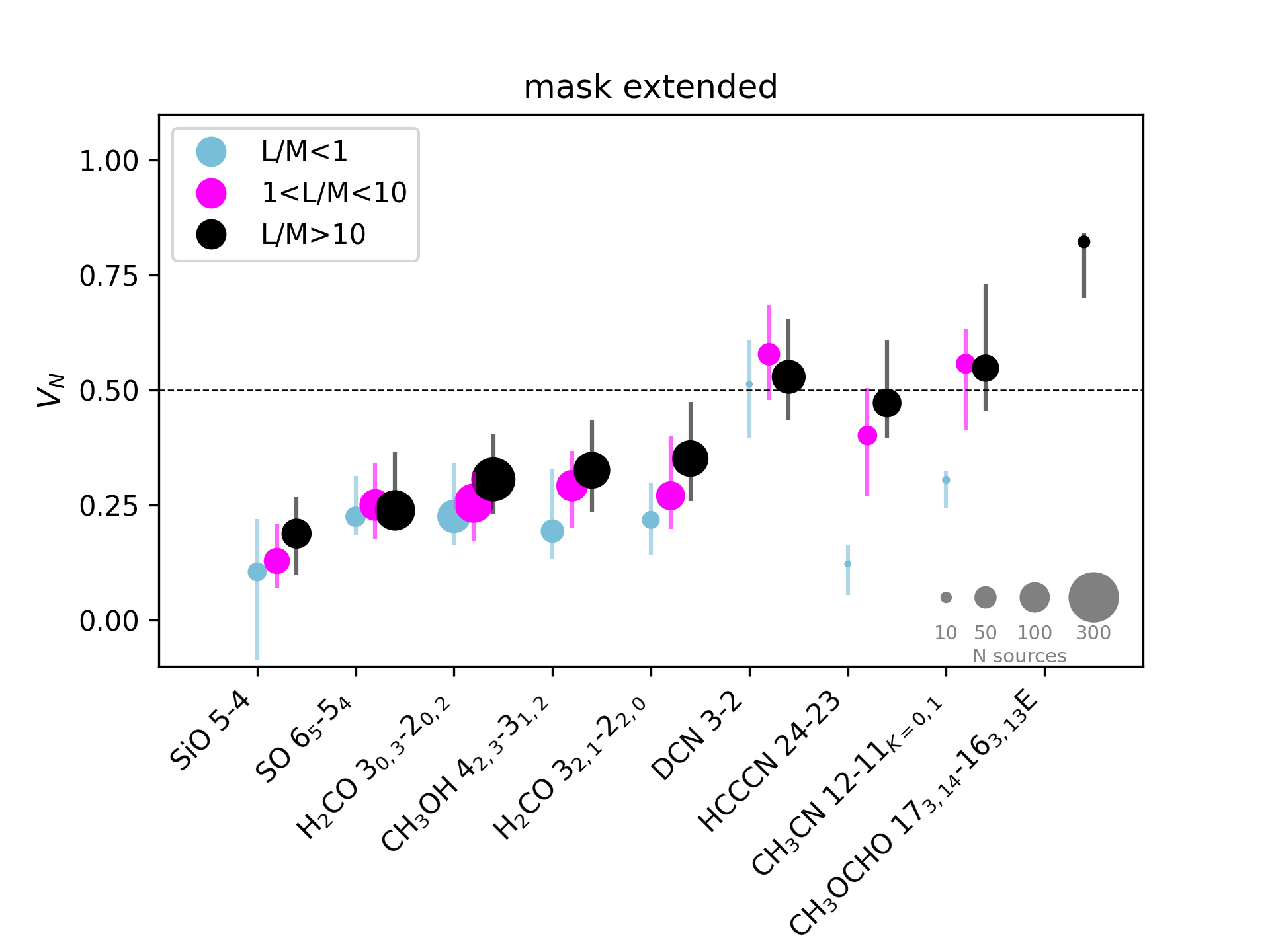}\\
    \includegraphics[width=10cm, trim={0.5cm, 0, 0cm, 0.5cm}, clip=True]{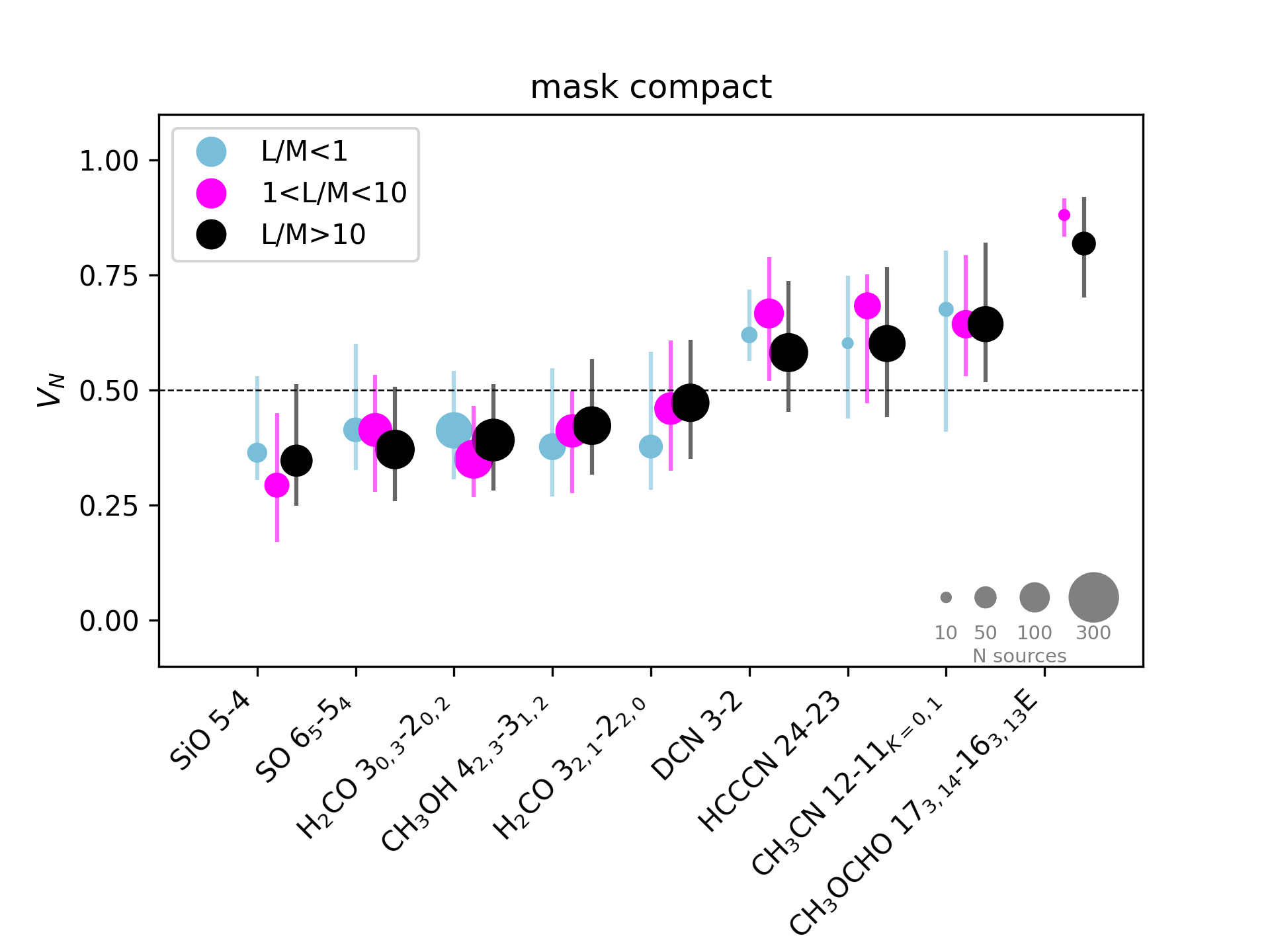}
    \caption{Upper panel: Median value of $V_{\mathrm{N}}$ histograms for the mask$_{\rm{ext.}}$ dividing the sample in three evolutionary stages; lower panel: Median value of $V_{\mathrm{N}}$ histograms for the mask$_{\mathrm{com.}}$ dividing the sample in three evolutionary stages. The horizontal dashed line represent the values of $V_{\mathrm{N}}$ for which $\sim50\%$ of the area has a good agreement in the two maps (continuum and moment-0).}
    \label{recapVmedianLsuM}
\end{figure}
\indent In fact, we analyzed two lines of H$_2$CO, with two different $E_{\rm{U}}/\kappa_{B}$ ($21\,$K and $68\,$K, respectively), and from the results we can see that there is a mild improvement in the morphological correlation with the continuum using the transition with higher  $E_{\rm{U}}/\kappa_{B}$ at the extended scale, but at this scale it is quite consistent with the other transition, while a larger increase (from 31\% to 45\%) is seen at the scale of the compact sources. We can see  that a dependence on the selected transition is present. Therefore, the results of this paper can be considered valid also for other transitions of the same molecular specie, in a reasonable range of $E_{\rm{U}}/\kappa_{B}$ around the transitions presented here.\\\indent
The difference between the two groups of molecules can be traced back to their mechanism of formation (see section 5.1), together with the conditions of excitation of the selected transition. SiO, whose formation pathway is strictly linked to sputtering of dust grains in shock/outflows regions, register the lowest morphological correlation with the continuum emission. All the other three molecular species that show a low level of morphological correlation are species that have been shown to be associated to shocks. At the scale of compact emission these molecular species have a higher degree of correlation, but still below 50\%. In this case, the main emission could likely come not only from the cores but also from an outer envelope surrounding them, since the conditions of the release of these molecular species and/or the low/intermediate excitation temperature allow their emission. Still, a contribution to outflows close to protostellar objects can be present also at core scales. The comparison of SO and SiO on both masks shows a difference between the two molecules, as already highlighted by the results of Section 5.2. In fact, SO has in general values closer to those of H$_2$CO and CH$_3$OH. This is likely due to the importance of the formation route of SO after the thermal desorption of H$_2$S from dust grains due to thermal desorption. On the other hand, DCN, HCCCN, CH$_3$CN and CH$_3$OCHO have pathways of formation that require high temperature for their formation/release in gas phase. The higher values of the upper energy of the transitions (with the exception of DCN) also contribute to the fact that they probe the warm/hot gas linked to the evolved cores, without contribution from their colder outer envelope, thus showing a good morphological correlation with the continuum over the mask$_{\rm{com.}}$. These molecular species shows less extended emission, as seen in Sect. 5.2, but the area of extended continuum emission that is in some cases covered by their emission show an intermediate level of correlation $\sim50\%$. 
\\
\indent We also explored the possible dependence of the results of the morphological correlation of the selected transitions with the dust continuum emission with the evolutionary stage of the clumps,  dividing the sample in three range of L/M: L/M\,<\,1\,L$_{\odot}$/M$_{\odot}$, 1\,L$_{\odot}$/M$_{\odot}$ \,<\,L/M\,<\,10\,L$_{\odot}$/M$_{\odot}$, and L/M >\,10\,L$_{\odot}$/M$_{\odot}$. The results are shown in Fig.  \ref{recapVmedianLsuM}. In general, the distributions of the morphological correlation of almost all the molecular lines with the continuum for the three evolutionary stages, are consistent within the errors. The only clear differences are for HCCCN and CH$_3$CN for L/M<1 in the mask$_{\rm{ext.}}$. However, the statistics for these two points are based on 10 or less sources. The fact that we found no correlation with the evolutionary stage of the clump, indicated by the L/M value, could not exclude a possible dependence with the evolutionary stage of the single fragments forming inside the clump itself. In fact, a clump can can contain fragments in different evolutionary stages (especially for sources with high values of L/M). An in depth analysis on the evolutionary stages of the cores, looking at the temperature of the cores, or at the presence of outflows originating from them, is not available yet. Therefore, the mask$_{\rm{comp}}$ that include all the cores, can mix together cores in different evolutionary stages, including both startless and protostellar cores. This is true also for mask$_{\rm{ext}}$ that can contain emission from the starless cores and their surroundings together with the emission from more evolved and dynamical regions. For this reasons, a possible dependence with the evolutionary stage of the individual cores might  be undetectable, due to mixing of the two in the same masks.
 %Therefore, the two masks of the continuum we defined in this paper for each clump can contain emission from both starless and protostellar objects, making a possible dependence with the evolutionary stage of the individual cores undetectable, due to mixing of the two in the same masks.
For some of the molecular species, such as CH$_3$CN and CH$_3$OCHO there could also be no difference due to a selection effect, i.e. the cores with emission of these molecular species in all the range of L/M are in the same phase, i.e. already protostellar objects with high temperature at their center, therefore the different L/M bins, that probe different evolutionary stages at clump level, in these cases might still be probing cores at the same stage of evolution.  

\subsection{Comparison with Spearman's correlation coefficient}
\label{sec:pearsoncoeff}
\begin{figure*}
\includegraphics[height=4.5cm, trim={2cm, 0.0cm, 0, 0}, clip=True]
{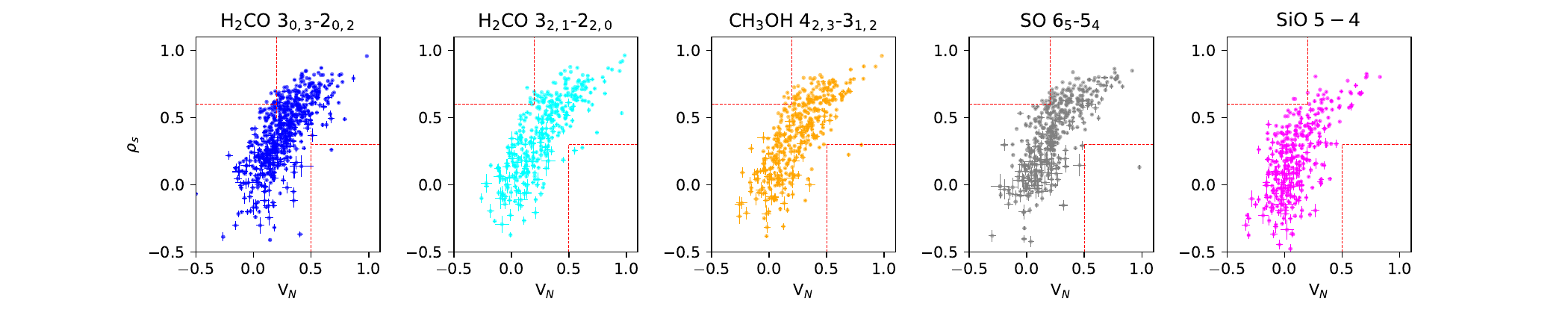 }\\
\includegraphics[height=4.5cm, trim={0, 0, 0, 0}, clip=True ]{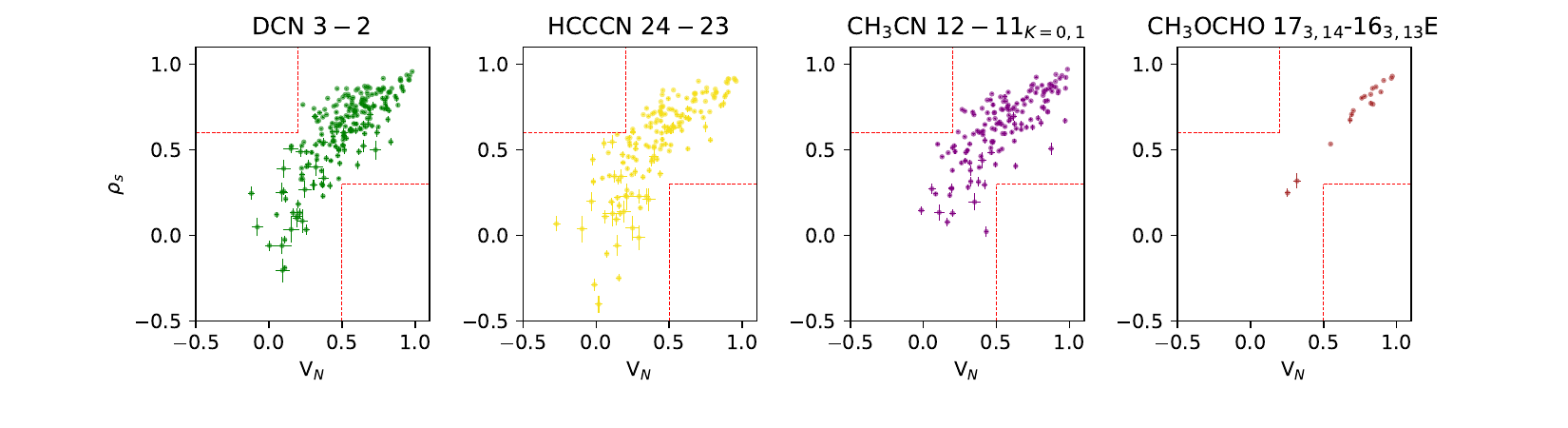}
\caption{Comparison between Spearman's correlation coefficient and the $V_{N}$ parameter from the HOG method in the comparison between the continuum and the indicated species using mask$_{\rm{ext.}}$. The two boxes delimited by red dotted-dashed lines are the regions in the plot where the two estimators give the more contrasting results. }
\label{fig:spearman1}
\end{figure*}

\begin{figure*}
\centering
\includegraphics[height=4.5cm, trim={2cm, 0.0cm, 0, 0}, clip=True]
{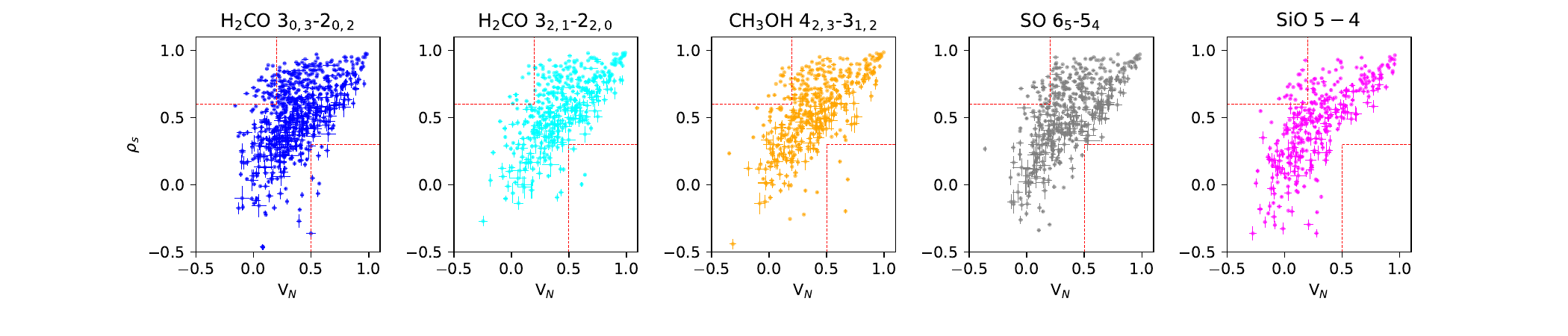}\\
\includegraphics[height=4.5cm, trim={0, 0, 0, 0}, clip=True ]{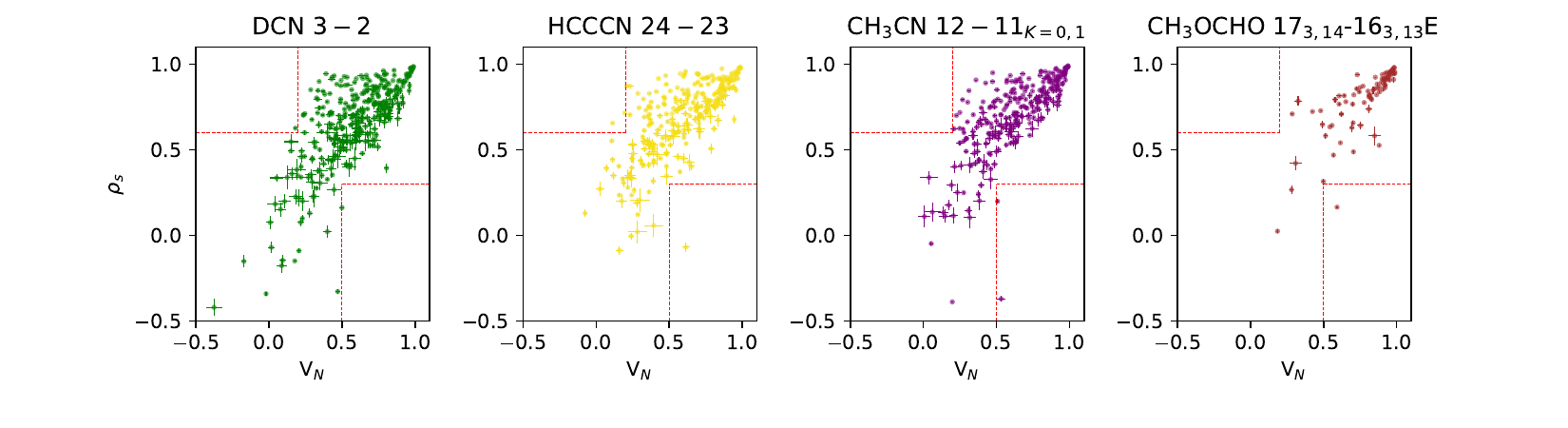}
\caption{Same as Fig.~\ref{fig:spearman1}, but for mask$_{\rm{com.}}$.
}
\label{fig:spearman2}
\end{figure*}

\begin{figure*}
\centering
    \includegraphics[width=13cm,trim={0.2cm, 5cm, 1.0cm, 5.0cm}, clip=True]{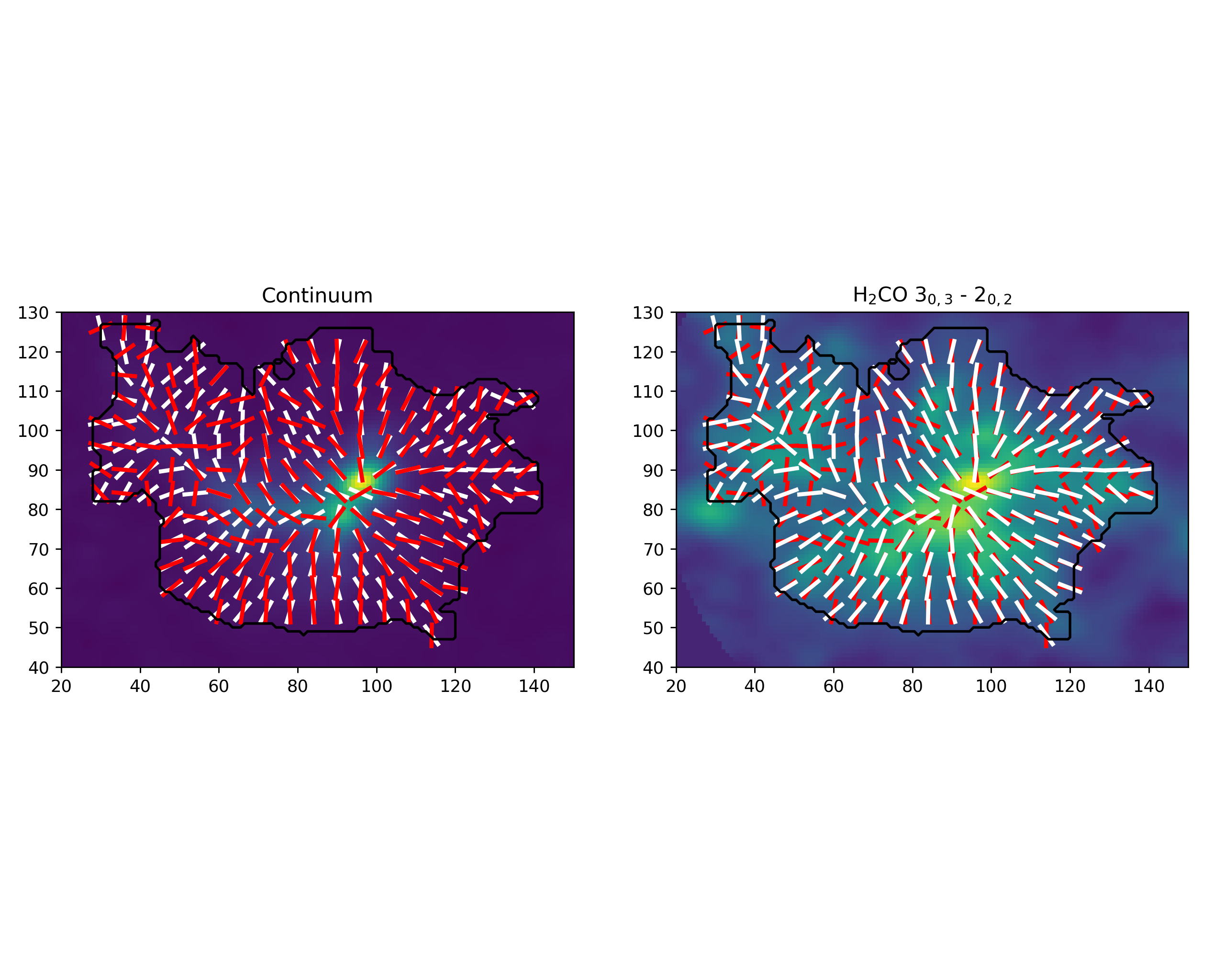}\\
    \includegraphics[width=11cm, trim={0cm, 0cm, 0cm, 0.3cm}, clip=True]{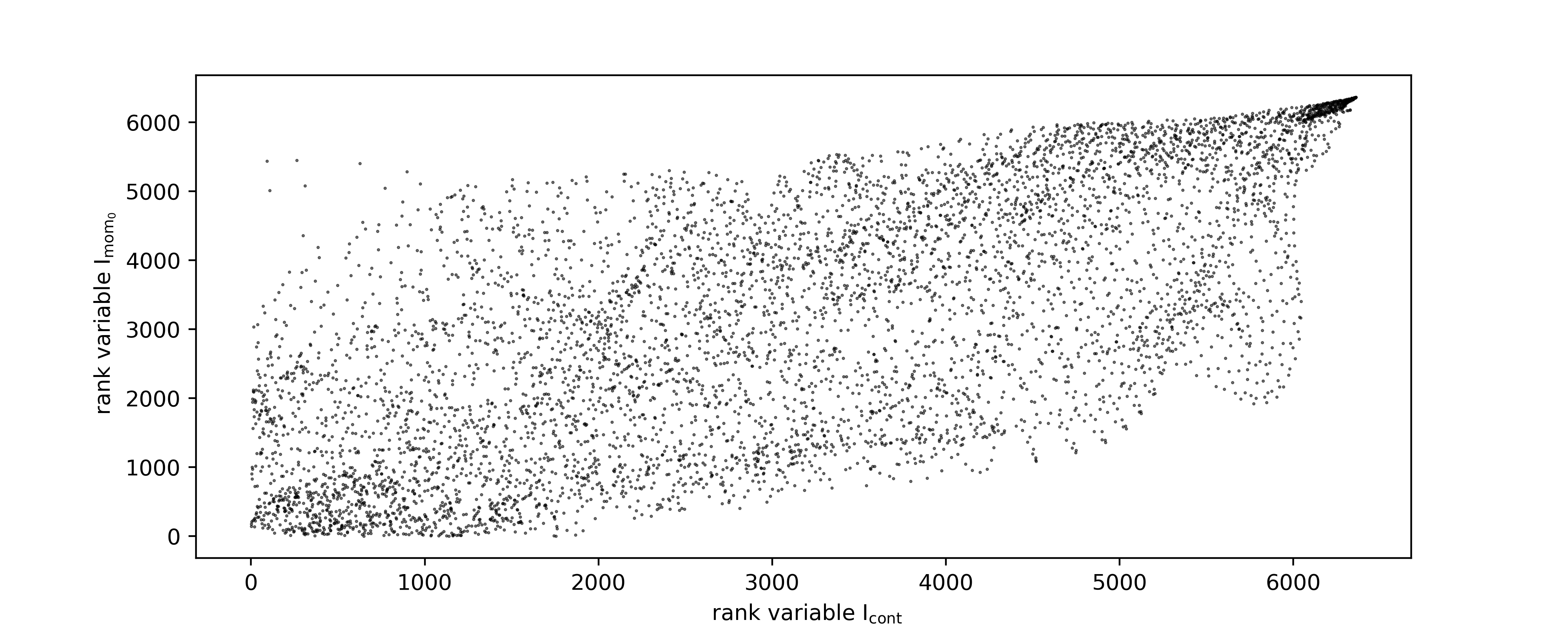}
   
    \caption{Example of a source in which the HOG method 
   indicates low morphological correlation and the 
    Pearson's correlation points out high correlation.
    The source is AG019.8843-0.5337 and in the comparison of the continuum and of the moment-0 of H$_2$CO $3_{0,3}-2_{0,2}$ in the mask$_{\rm{ext.}}$ $V_{\rm{N}}$=0.20, while  $\rho_{\rm{s}}=0.67$. The two panels in the upper row show the continuum and the moment-0 map of H$_2$CO $3_{0,3}-2_{0,2}$ with superimposed the directions of the gradients in the two maps (red: moment-0, white: continuum, respectively). The panel in the lower row shows the scatter plot of the pixel-to-pixel rank variables in the two maps inside the same mask (in black in the images of the two maps) where both the Spearman's correlation and the Projected Rayleigh Statistic, $V_{\rm{N}}$, have been computed.}
     \label{fig:goodpearsonbadV}
\end{figure*}

\begin{figure*}
\centering
    \includegraphics[width=14.5cm,trim={0.2cm, 6cm, 1.0cm, 6.0cm}, clip=True]{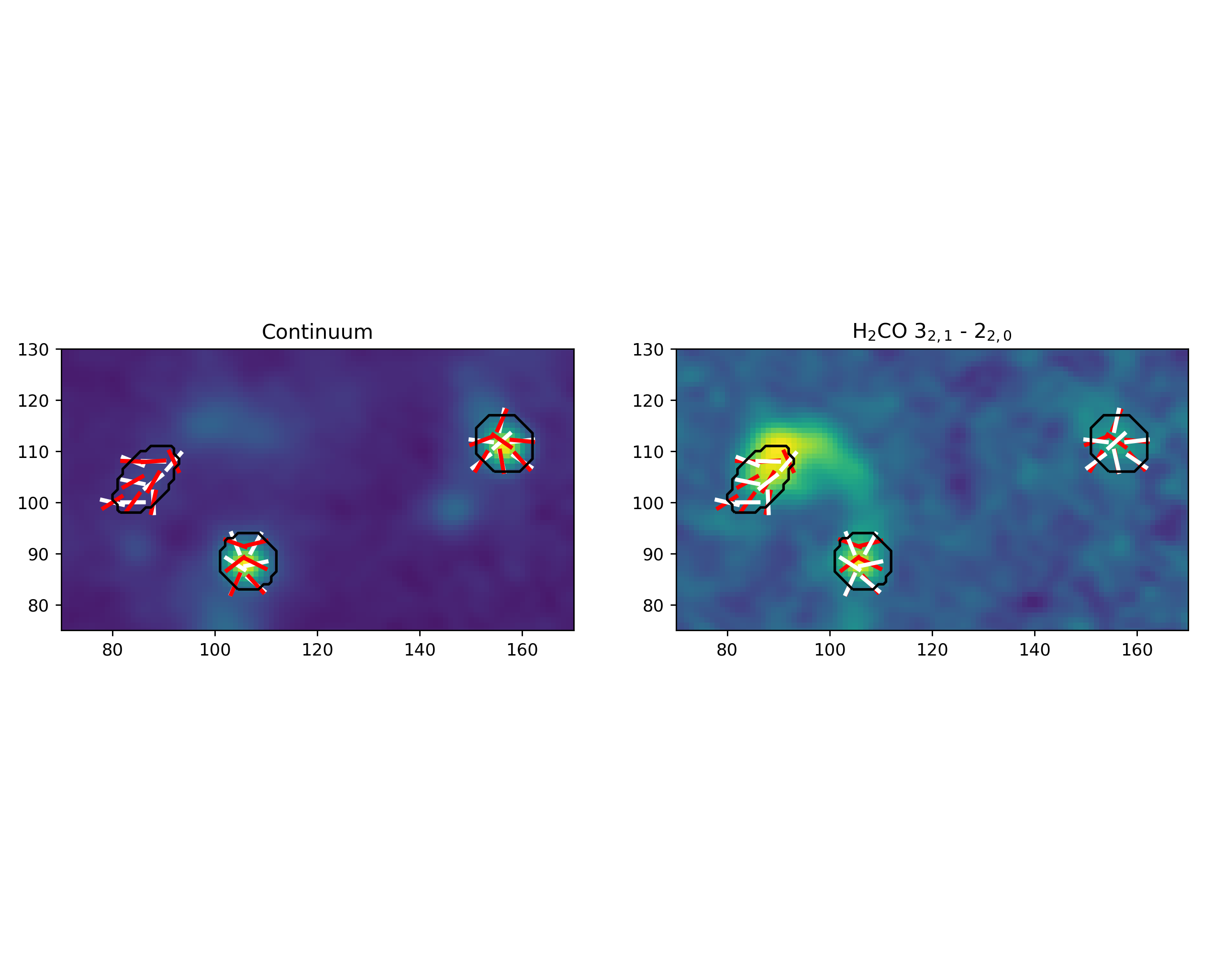}
    \\
    \includegraphics[width=11cm, trim={0cm, 0cm, 0cm, 0.3cm}, clip=True]{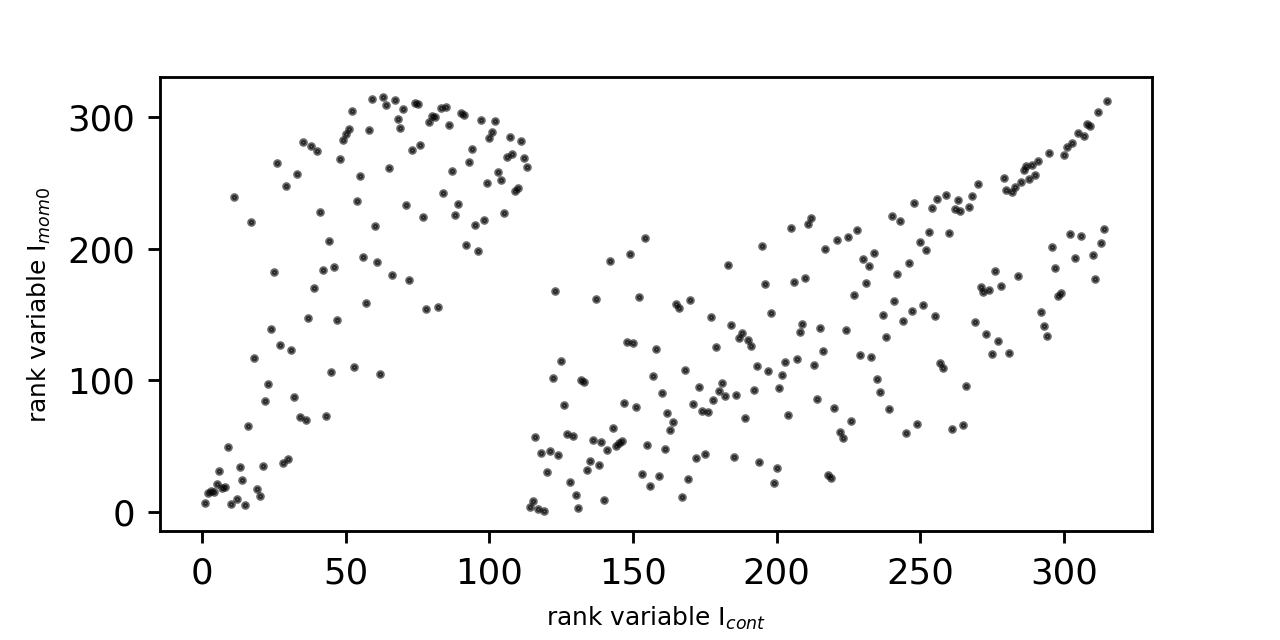}
    \caption{Example of a source in which the HOG method 
    indicates high morphological correlation
    %gives a good correlation 
    and the Spearman's correlation coefficient %gives a bad correlation.
    points out low correlation.
    The source is AG332.7016-0.5870 and in the comparison of the continuum and of the moment-0 of H$_2$CO $3_{2,1}-2_{2,0}$ in the mask$_{\rm{com.}}$ $V_{\rm{N}}$=0.64, while $\rho_{\rm{s}}=0.07$. The two panels in the upper row show the continuum and the moment-0 map of H$_2$CO $3_{2,1}-2_{2,0}$ with superimposed the directions of the gradients in the two maps (red: moment-0, white: continuum, respectively). The panel in the lower row shows the scatter plot of the pixel-to-pixel rank variables in the two maps inside the same mask (in black in the images of the two maps) where both the Spearman's correlation and the Projected Rayleigh Statistic, $V_{\rm{N}}$, have been computed.}
     \label{fig:goodVbadpearson}
\end{figure*}

In this section we compare the results obtained with the HOG method with one of the most widely used tool in the literature for computing the correlation between two tracers: the Spearman's correlation coefficient, $\rho_{\rm{s}}$. The definition of this coefficient starts from those of the Pearson's correlation coefficient,  $\rho_{\rm{p}}$, that determines if two data sets, x$_{i}$ and y$_{i}$, are well correlated through a linear relation. The parameter $\rho_{\rm{p}}$ is defined as:
\begin{equation}
\rho_{\rm{p}} = \frac{\sum_{i}(x_{i}-\hat{x})(y_{i}-\hat{y})}{\sqrt{\sum_{i}(x_{i}-\hat{x})^2}\sqrt{\sum_{i}(y_{i}-\hat{y})^2}}\,,
\end{equation}
where in our case the datasets x$_{i}$ and y$_{i}$ are the pixels-to-pixels intensities of the moment-0 map selected and of the continuum map of the source, and $\hat{x}$ and $\hat{y}$ are their mean values. Therefore, this statistical correlation coefficient is intensity based, unlike the HOG method, that is gradient based, and imply an underlying linear correlation between the two datased.\\
\indent In the case of the Spearman's coefficient, no underlying linear relation is assumed, and it is defined as the Pearson's correlation coefficient between the rank variables, i.e. the variables that report the rank in order or brightness for each pixel of the two maps.\\\indent We present here the Spearman's coefficient results only, which is the most suitable to use, but we tested also the Pearson's correlation coefficient, and the values are given in Tables D.1 and D.2 and on Zenodo\footnote{https://doi.org/10.5281/zenodo.15236491}. In general, the Pearson's correlation coefficient gives fairly similar results to those of the Spearman's correlation coefficient in comparison with astroHOG.\\ \indent We compute  $\rho_{\rm{s}}$ using the same masks used for the HOG method, and in Figure \ref{fig:spearman1} we can see the comparison of the results of this method with the HOG method on the mask$_{\rm{ext.}}$ for all the species and in Figure \ref{fig:spearman2} for all the species on the mask$_{\rm{com.}}$. We can see that the results of the statistical correlation coefficient do not differ much compared to the results of the HOG method, and in general there is an underlying positive correlation among the results of them with the HOG method, but a high degree of scatter is found. %Between the Pearson's correlation coefficient and the Spearman's correlation coefficient, the latter show a mildly reduced scatter with the HOG method. Moreover, $\rho_{\rm{s}}$ reach in general lower values with respect to $\rho_{\rm{p}}$ on mask$_{\rm{ext}}$, since it is less affected by high-dynamical range effects due to the fact that correlates the array of the ranks. 
For high values of  $\rho_{\rm{s}}\sim0.7$, there is a large spread in $V_{\rm{N}}$ with some values below 0.5. We can find points where the two methods even give clearly contrasting results (see the two areas in the upper-left and lower-right corner of each plot in Fig.  \ref{fig:spearman1} and  \ref{fig:spearman2}: $V_{\rm{N}}>$0.5 and $\rho_{\rm{s}}<0.3$ and $V_{\rm{N}}<$0.2 and $\rho_{\rm{s}}>0.6$ ). The majority of the sources with highly contrasting results between the two methods fall in the upper-left corner of the plots, i.e. in the region where from the Spearman's correlation coefficient there is a moderate/high correlation, but a very low correlation based on the HOG results, which means that even if there is a good level of correlation in the intensity of the continuum and the molecular species, the two emissions do not have a similar morphology. An example is given in Fig. \ref{fig:goodpearsonbadV}, where we show both the continuum map and the moment-0 map of H$_2$CO $3_{0,3}-2_{0,2}$ for AG019.8843-0.5337 with the direction of the gradients, together with the scatter plot of the intensity from which the Spearman's correlation coefficient is calculated. The gradients do not in general match well as indicated by the value of $V_{\rm{N}}$, and also the majority of the points in the scatter plot do not show a linear trend. The trend that results in $\rho_{\rm{s}}\sim0.67$ is driven by a small number of points that correspond to the brightest source in the continuum, which is also the brightest in the moment-0 map. The $\rho_{\rm{s}}$ is therefore mostly biased by a small number of good matching points due to the dynamic range of the values of intensity. This effect is relevant also for the Pearson's correlation coefficient, having a value of 0.69 in this case. \\ \indent
In Figure \ref{fig:goodVbadpearson} we can see an example of one of the clearest opposite (less common) cases, where $V_{\rm{N}}$ has  high values, while $\rho_{\rm{s}}$ is very low. It is possible to see that in the clump AG332.7016-0.5870 the morphology of H$_2$CO $3_{2,1}-2_{2,0}$ and the continuum on the cores with H$_2$CO $3_{2,1}-2_{2,0}$ detected is matching well (looking at the gradients, $V_{\rm{N}}=0.64$), but the brightest source in the continuum does not correspond to the brightest source in H$_2$CO $3_{2,1}-2_{2,0}$ emission. Therefore, the value of the intensity-based correlator $\rho_{s}$ is $\sim0.07$ in this case. From this example we can understand that the two types of statistical correlators give different results, and are suited to answer different questions. If the main goal is to compare the morphology of two tracers, the HOG method is to be preferred to intensity-based correlator. On the other hand, if the goal of a study is to determine if two tracers peak at the same positions the latter is the most suitable. Moreover, the values of intensity-based correlator should be carefully checked in case of maps with large dynamical range.
%\begin{figure*}
%\includegraphics[height=4.5cm, trim={2cm, 0.0cm, 0, 0}, clip=True]
%{figures/pears_vs_V_maskdiffuse_part1_all.pdf}\\
%\includegraphics[height=4.5cm, trim={0, 0, 0, 0}, clip=True ]{figures/pears_vs_V_maskdiffuse_part2_all.pdf}
%\caption{Comparison between Pearson's correlation coefficient and the $V_{N}$ parameter from the HOG method in the comparison between the continuum and the indicated species using mask$_{\rm{ext.}}$. The two boxes delimited by red dotted-dashed lines are the regions in the plot where the two estimators give the more contrasting results. }
%\label{fig:pearson1}
%\end{figure*}

%\begin{figure*}
%\centering
%\includegraphics[height=4.5cm, trim={2cm, 0.0cm, 0, 0}, clip=True]
%{figures/pears_vs_V_maskcores_part1_all.pdf}\\
%\includegraphics[height=4.5cm, trim={0, 0, 0, 0}, clip=True ]{figures/pears_vs_V_maskcores_part2_all.pdf}
%\caption{Same as Fig.~\ref{fig:pearson1}, but for mask$_{\rm{com.}}$.
%}
%\label{fig:pearson2}
%\end{figure*}

% CONCLUSIONS ===============================================================
\section{Conclusions}
In this paper we analyzed the integrated emission of lines from 8 of the most commonly detected molecular species (i.e. H$_2$CO, CH$_3$OH, DCN, HCCCN, CH$_3$CN, CH$_3$OCHO, SO, and SiO) in the largest survey of high-mass star-forming clumps ever observed at high angular resolution, ALMAGAL. %In order 
To evaluate how well these species trace 
%also 
the cold dust, we used an innovative method, the histogram of oriented gradients (HOG, with the package astroHOG), that quantifies the level of similarity in the morphology of two images. The method only uses the angle between the gradients in the two images to derive the level of correlation, and is completely independent of the intensity values of the two images as well as the intensity of the gradients. We compared the morphology of the moment-0 maps of these tracers with the dust continuum emission at 1.38\,mm using the HOG method, on two spatial scales of the extended emission and of the compact sources only. We also compared the results of this method with those obtained using  the Spearman's correlation coefficient, which is an intensity-based correlator. 
We summarize the main results:
\begin{itemize}
    \item All molecular species show a detection rate increasing with the $L/M$ ratio of the sources, reaching a percentage of detection close to or above 87\% for sources with $L/M>100\,\rm{L_{\odot}/M_{\odot}}$ in all cases but CH$_3$OCHO, which is only detected towards 110 sources in the whole sample of this paper.
    \item Among the analyzed molecular species, only the emission of H$_2$CO, CH$_3$OH, and SO covers the extended continuum (mask$_{\rm{ext.}}$) in a large fraction of sources. The smallest fraction is observed for SiO, where the extended emission, if present, is mostly not co-spatial with the continuum emission, as expected since SiO is mainly emitted in the outflow lobes, where continuum from cold dust is not observable. This highlights a difference between SO and SiO, where the former is to be considered not only as a shock tracer. We can also see that in some sources (with the exact number possibly limited by the sensitivity of the data) DCN is detected not only toward the dense cores but also shows diffuse emission.
    \item Even though H$_2$CO, CH$_3$OH, and SO cover the extended continuum emission, the morphological comparison using the HOG method (\texttt{astroHOG}) shows a low level of correlation with only 24-29\% of the area where both emissions are detected in good morphological agreement.
    \item The analysis of the morphology on the dense fragments shows that DCN, HCCCN, CH$_3$CN and CH$_3$OCHO have a strong morphological correlation with the compact continuum emission with values above 60\%. On the other hand, SiO has the lowest correlation with the compact continuum emission, followed by H$_2$CO, CH$_3$OH, and SO that has still low/intermediate values of correlation (below 45\%).
    \item The results of astroHOG divided in three bins of the evolutionary indicator L/M do not show clear differences in the morphological correlation with the evolution of the clump. This result can be biased by the fact that inside each clump we can have cores in prestellar phase and protostellar phase mixed together.  This effect is not observed for some of the molecular species, for which the emission is only arising from more evolved cores/regions inside the clump, namely HCCCN, CH$_3$CN, and CH$_3$OCHO. For these three molecular species the similarity of the HOG results in all the evolutionary stages of the clump, might be due to a selection effect that include only the more evolved cores inside the analysis in all the bins of L/M.
    \item We see a positive trend between the HOG result ($V_{\mathrm{N}}$) that correlates the general morphology of the emission and the  Spearman's coefficients ($\rho_{\rm{s}}$), that measure only the correlation of the intensity. However, there is a large scatter and there are sources in which the two type of methods give clearly contrasting results. Due to their intrinsic nature, the two type of methods are best suited to answer different questions, with the HOG method better performing in comparison of morphologies.
\end{itemize}

From these results, we can roughly divide the molecular transition analyzed into two classes. 
The first is composed of H$_2$CO, CH$_3$OH, SO, and SiO transitions, that do not trace the same morphology as the continuum both on the extended and compact emission scales. Moreover, SiO emission differs from that of the other outflow tracers. This is confirmed both from the percentage of overlap of SiO emission and continuum emission, in comparison with the same values for the other tracers, and from the results of astroHOG. However, a direct comparison using astroHOG of the moment-0 maps of different molecular species, that will better investigate also this aspect, will be presented in a forthcoming paper. \\\indent
The second group of molecular transitions contains lines of DCN, HCCCN, CH$_3$CN and CH$_3$OCHO, which do not trace in the majority of the sources the most diffuse part of the continuum emission while have a good morphological correlation with the compact dust continuum emission. The exception in the latter category is DCN, whose emission traces part of the diffuse emission toward $\sim20\%$ of sources compared to less than 10\% in the other molecular lines in this group.
\\ \indent
These results are to be considered valid in a reasonable range of $E_{\rm{U}}/\kappa_{B}$ values around the ones of the transitions presented here, since we have seen a dependence with the excitation temperature of the transitions for the two lines of H$_2$CO, especially on compact sources. \\\indent The total emission detected that is co-spatial with the densest fragments for the species with a low level of correlation on the fragment mask indicated that these molecular transitions have a large contribution of emission from the intraclump medium or from outer layers around the fragments (in some cases also potentially due to optical depth effects). On the contrary, the total emission towards densest fragments of species with strong morphological correlation is dominated by the emission from the fragments themself.

%% IMPORTANT! The old "\acknowledgment" command has been depreciated. It was
%% not robust enough to handle our new dual anonymous review requirements and
%% thus been replaced with the acknowledgment environment. If you try to 
%% compile with \acknowledgment, you will get an error print to the screen
%% and in the compiled pdf.
%% 
%% Also note that the acknowledgment environment does not support long amounts of text. If you have a lot of people and institutions to acknowledge, do not use this command. Instead, create a new \section{Acknowledgments}.
\section*{Acknowledgements}
C.M. and the INAF-IAPS team acknowledge funding from the European
Research Council (ERC) under the European Union's Horizon 2020 program
through the ECOGAL Synergy grant (ID 855130). R.K. acknowledges financial support via the Heisenberg Research Grant funded by the Deutsche Forschungsgemeinschaft (DFG, German Research Foundation) under grant no.~KU 2849/9, project no.~445783058. P.S. was partially supported by a Grant-in-Aid for Scientific Research (KAKENHI Number JP22H01271 and JP23H01221) of JSPS. P.S. was supported by Yoshinori Ohsumi Fund (Yoshinori Ohsumi Award for Fundamental Research). C.\ B.  gratefully  acknowledges  funding  from  National  Science  Foundation  under  Award Nos. 2108938, 2206510, and CAREER 2145689, as well as from the National Aeronautics and Space Administration through the Astrophysics Data Analysis Program under Award ``3-D MC: Mapping Circumnuclear Molecular Clouds from X-ray to Radio,” Grant No. 80NSSC22K1125. 
A.S.~M. acknowledges support from the RyC2021-032892-I grant funded by MCIN/AEI/10.13039/501100011033 and by the European Union `Next GenerationEU’/PRTR, as well as the program Unidad de Excelencia María de Maeztu CEX2020-001058-M, and support from the PID2023-146675NB-I00 (MCI-AEI-FEDER, UE). RSK acknowledges financial support from the ERC via Synergy Grant ECOGAL (project ID 855130),  from the German Excellence Strategy via the Heidelberg Cluster "STRUCTURES" (EXC 2181 - 390900948), and from the German Ministry for Economic Affairs and Climate Action in project "MAINN" (funding ID 50OO2206).  RSK also thanks the 2024/25 Class of Radcliffe Fellows for highly interesting and stimulating discussions.  Part of this research from D.C.L. was carried out at the Jet Propulsion Laboratory, California Institute of Technology, under a contract with the National Aeronautics and Space Administration (80NM0018D0004). LB gratefully acknowledges support by the ANID BASAL project FB210003.\\
This paper makes use of the following ALMA data:
project
code 2019.1.00195.L. ALMA is a partnership of ESO (representing
its member states), NSF (USA) and NINS (Japan), together with NRC (Canada), MOST and ASIAA (Taiwan), and KASI (Republic of Korea), in cooperation with the Republic of Chile. 
The Joint ALMA Observatory is operated by ESO,
AUI/NRAO and NAOJ.

%% To help institutions obtain information on the effectiveness of their 
%% telescopes the AAS Journals has created a group of keywords for telescope 
%% facilities.
%
%% Following the acknowledgments section, use the following syntax and the
%% \facility{} or \facilities{} macros to list the keywords of facilities used 
%% in the research for the paper.  Each keyword is checked against the master 
%% list during copy editing.  Individual instruments can be provided in 
%% parentheses, after the keyword, but they are not verified.

%% Similar to \facility{}, there is the optional \software command to allow 
%% authors a place to specify which programs were used during the creation of 
%% the manuscript. Authors should list each code and include either a
%% citation or URL to the code inside ()s when available.

%% Appendix material should be preceded with a single \appendix command.
%% There should be a \section command for each appendix. Mark appendix
%% subsections with the same markup you use in the main body of the paper.

%% Each Appendix (indicated with \section) will be lettered A, B, C, etc.
%% The equation counter will reset when it encounters the \appendix
%% command and will number appendix equations (A1), (A2), etc. The
%% Figure and Table counter will not reset.
\bibliographystyle{aa}
\bibliography{sample631.bib}

\begin{thebibliography}{74}
\expandafter\ifx\csname natexlab\endcsname\relax\def\natexlab#1{#1}\fi

\bibitem[{{Albertsson} {et~al.}(2013){Albertsson}, {Semenov}, {Vasyunin}, {Henning}, \& {Herbst}}]{albertsson2013}
{Albertsson}, T., {Semenov}, D.~A., {Vasyunin}, A.~I., {Henning}, T., \& {Herbst}, E. 2013, \apjs, 207, 27

\bibitem[{{Beltr{\'a}n} {et~al.}(2018){Beltr{\'a}n}, {Cesaroni}, {Rivilla}, {S{\'a}nchez-Monge}, {Moscadelli}, {Ahmadi}, {Allen}, {Beuther}, {Etoka}, {Galli}, {Galv{\'a}n-Madrid}, {Goddi}, {Johnston}, {Klaassen}, {K{\"o}lligan}, {Kuiper}, {Kumar}, {Maud}, {Mottram}, {Peters}, {Schilke}, {Testi}, {van der Tak}, \& {Walmsley}}]{2018beltran}
{Beltr{\'a}n}, M.~T., {Cesaroni}, R., {Rivilla}, V.~M., {et~al.} 2018, \aap, 615, A141

\bibitem[{{Beuther} {et~al.}(2006){Beuther}, {Zhang}, {Sridharan}, {Lee}, \& {Zapata}}]{beuther2006}
{Beuther}, H., {Zhang}, Q., {Sridharan}, T.~K., {Lee}, C.~F., \& {Zapata}, L.~A. 2006, \aap, 454, 221

\bibitem[{{Bracco} {et~al.}(2020){Bracco}, {Jeli{\'c}}, {Marchal}, {Turi{\'c}}, {Erceg}, {Miville-Desch{\^e}nes}, \& {Bellomi}}]{bracco2020}
{Bracco}, A., {Jeli{\'c}}, V., {Marchal}, A., {et~al.} 2020, \aap, 644, L3

\bibitem[{{Burke} {et~al.}(2015){Burke}, {Puletti}, {Brown}, {Woods}, {Viti}, \& {Slater}}]{burke2015}
{Burke}, D.~J., {Puletti}, F., {Brown}, W.~A., {et~al.} 2015, \mnras, 447, 1444

\bibitem[{{Caselli} \& {Ceccarelli}(2012)}]{caselliceccarelli2012}
{Caselli}, P. \& {Ceccarelli}, C. 2012, \aapr, 20, 56

\bibitem[{{Cesaroni} {et~al.}(2015){Cesaroni}, {Pestalozzi}, {Beltr{\'a}n}, {Hoare}, {Molinari}, {Olmi}, {Smith}, {Stringfellow}, {Testi}, \& {Thompson}}]{cesaroni2015uchii}
{Cesaroni}, R., {Pestalozzi}, M., {Beltr{\'a}n}, M.~T., {et~al.} 2015, \aap, 579, A71

\bibitem[{{Chac{\'o}n-Tanarro} {et~al.}(2019){Chac{\'o}n-Tanarro}, {Caselli}, {Bizzocchi}, {Pineda}, {Sipil{\"a}}, {Vasyunin}, {Spezzano}, {Punanova}, {Giuliano}, \& {Lattanzi}}]{cachon2019}
{Chac{\'o}n-Tanarro}, A., {Caselli}, P., {Bizzocchi}, L., {et~al.} 2019, \aap, 622, A141

\bibitem[{{Charnley} {et~al.}(1997){Charnley}, {Tielens}, \& {Rodgers}}]{charnley1997}
{Charnley}, S.~B., {Tielens}, A.~G.~G.~M., \& {Rodgers}, S.~D. 1997, \apjl, 482, L203

\bibitem[{{Coletta} {et~al.}(2025){Coletta}, {Molinari}, {Schisano}, {Traficante}, {Elia}, {Benedettini}, {Mininni}, {Soler}, {S{\'a}nchez-Monge}, {Schilke}, {Battersby}, {Fuller}, {Beuther}, {Zhang}, {Beltr{\'a}n}, {Jones}, {Klessen}, {Walch}, {Fontani}, {Avison}, {Brogan}, {Clarke}, {Hatchfield}, {Hennebelle}, {Ho}, {Hunter}, {Johnston}, {Klaassen}, {Koch}, {Kuiper}, {Lis}, {Liu}, {Lumsden}, {Maruccia}, {M{\"o}ller}, {Moscadelli}, {Nucara}, {Rigby}, {Rygl}, {Sanhueza}, {van der Tak}, {Wells}, {Wyrowski}, {De Angelis}, {Liu}, {Ahmadi}, {Bronfman}, {Liu}, {Su}, {Tang}, {Testi}, \& {Zinnecker}}]{ALMAGAL3}
{Coletta}, A., {Molinari}, S., {Schisano}, E., {et~al.} 2025, \aap, 696, A151

\bibitem[{{Cosentino} {et~al.}(2022){Cosentino}, {Jim{\'e}nez-Serra}, {Tan}, {Henshaw}, {Barnes}, {Law}, {Zeng}, {Fontani}, {Caselli}, {Viti}, {Zahorecz}, {Rico-Villas}, {Meg{\'\i}as}, {Miceli}, {Orlando}, {Ustamujic}, {Greco}, {Peres}, {Bocchino}, {Fedriani}, {Gorai}, {Testi}, \& {Mart{\'\i}n-Pintado}}]{cosentino2022}
{Cosentino}, G., {Jim{\'e}nez-Serra}, I., {Tan}, J.~C., {et~al.} 2022, \mnras, 511, 953

\bibitem[{{Csengeri} {et~al.}(2016){Csengeri}, {Leurini}, {Wyrowski}, {Urquhart}, {Menten}, {Walmsley}, {Bontemps}, {Wienen}, {Beuther}, {Motte}, {Nguyen-Luong}, {Schilke}, {Schuller}, {Zavagno}, \& {Sanna}}]{csengeri2016atlasgal2}
{Csengeri}, T., {Leurini}, S., {Wyrowski}, F., {et~al.} 2016, \aap, 586, A149

\bibitem[{{Cunningham} {et~al.}(2023){Cunningham}, {Ginsburg}, {Galv{\'a}n-Madrid}, {Motte}, {Csengeri}, {Stutz}, {Fern{\'a}ndez-L{\'o}pez}, {{\'A}lvarez-Guti{\'e}rrez}, {Armante}, {Baug}, {Bonfand}, {Bontemps}, {Braine}, {Brouillet}, {Busquet}, {D{\'\i}az-Gonz{\'a}lez}, {Di Francesco}, {Gusdorf}, {Herpin}, {Liu}, {L{\'o}pez-Sepulcre}, {Louvet}, {Lu}, {Maud}, {Nony}, {Olguin}, {Pouteau}, {Rivera-Soto}, {Sandoval-Garrido}, {Sanhueza}, {Tatematsu}, {Towner}, \& {Valeille-Manet}}]{cunningham2023DCN}
{Cunningham}, N., {Ginsburg}, A., {Galv{\'a}n-Madrid}, R., {et~al.} 2023, \aap, 678, A194

\bibitem[{{Dame}(2011)}]{dame2011}
{Dame}, T.~M. 2011, arXiv e-prints, arXiv:1101.1499

\bibitem[{{Duarte-Cabral} {et~al.}(2014){Duarte-Cabral}, {Bontemps}, {Motte}, {Gusdorf}, {Csengeri}, {Schneider}, \& {Louvet}}]{duartecabral2014A&A}
{Duarte-Cabral}, A., {Bontemps}, S., {Motte}, F., {et~al.} 2014, \aap, 570, A1

\bibitem[{{Durand} \& {Greenwood}(1958)}]{durand1958}
{Durand}, D. \& {Greenwood}, J.~A. 1958, Journal of Geology, 66, 229

\bibitem[{{Elia}(2020)}]{Elia2020uchii}
{Elia}, D. 2020, Communications of the Byurakan Astrophysical Observatory, 67, 241

\bibitem[{{Elia} {et~al.}(2021){Elia}, {Merello}, {Molinari}, {Schisano}, {Zavagno}, {Russeil}, {M{\`e}ge}, {Martin}, {Olmi}, {Pestalozzi}, {Plume}, {Ragan}, {Benedettini}, {Eden}, {Moore}, {Noriega-Crespo}, {Paladini}, {Palmeirim}, {Pezzuto}, {Pilbratt}, {Rygl}, {Schilke}, {Strafella}, {Tan}, {Traficante}, {Baldeschi}, {Bally}, {di Giorgio}, {Fiorellino}, {Liu}, {Piazzo}, \& {Polychroni}}]{elia2021}
{Elia}, D., {Merello}, M., {Molinari}, S., {et~al.} 2021, \mnras, 504, 2742

\bibitem[{{Elia} {et~al.}(2017){Elia}, {Molinari}, {Schisano}, {Pestalozzi}, {Pezzuto}, {Merello}, {Noriega-Crespo}, {Moore}, {Russeil}, {Mottram}, {Paladini}, {Strafella}, {Benedettini}, {Bernard}, {Di Giorgio}, {Eden}, {Fukui}, {Plume}, {Bally}, {Martin}, {Ragan}, {Jaffa}, {Motte}, {Olmi}, {Schneider}, {Testi}, {Wyrowski}, {Zavagno}, {Calzoletti}, {Faustini}, {Natoli}, {Palmeirim}, {Piacentini}, {Piazzo}, {Pilbratt}, {Polychroni}, {Baldeschi}, {Beltr{\'a}n}, {Billot}, {Cambr{\'e}sy}, {Cesaroni}, {Garc{\'\i}a-Lario}, {Hoare}, {Huang}, {Joncas}, {Liu}, {Maiolo}, {Marsh}, {Maruccia}, {M{\`e}ge}, {Peretto}, {Rygl}, {Schilke}, {Thompson}, {Traficante}, {Umana}, {Veneziani}, {Ward-Thompson}, {Whitworth}, {Arab}, {Band ieramonte}, {Becciani}, {Brescia}, {Buemi}, {Bufano}, {Butora}, {Cavuoti}, {Costa}, {Fiorellino}, {Hajnal}, {Hayakawa}, {Kacsuk}, {Leto}, {Li Causi}, {Marchili}, {Martinavarro-Armengol}, {Mercurio}, {Molinaro}, {Riccio}, {Sano}, {Sciacca}, {Tachihara}, {Torii}, {Trigilio}, {Vitello}, \&
  {Yamamoto}}]{elia2017}
{Elia}, D., {Molinari}, S., {Schisano}, E., {et~al.} 2017, \mnras, 471, 100

\bibitem[{{Esplugues} {et~al.}(2013){Esplugues}, {Tercero}, {Cernicharo}, {Goicoechea}, {Palau}, {Marcelino}, \& {Bell}}]{esplugues2013}
{Esplugues}, G.~B., {Tercero}, B., {Cernicharo}, J., {et~al.} 2013, \aap, 556, A143

\bibitem[{{Fontani} {et~al.}(2023){Fontani}, {Roueff}, {Colzi}, \& {Caselli}}]{fontani2023}
{Fontani}, F., {Roueff}, E., {Colzi}, L., \& {Caselli}, P. 2023, \aap, 680, A58

\bibitem[{{Fuchs} {et~al.}(2009){Fuchs}, {Cuppen}, {Ioppolo}, {Romanzin}, {Bisschop}, {Andersson}, {van Dishoeck}, \& {Linnartz}}]{fuchs2009}
{Fuchs}, G.~W., {Cuppen}, H.~M., {Ioppolo}, S., {et~al.} 2009, \aap, 505, 629

\bibitem[{{Garrod} \& {Herbst}(2006)}]{garrod2006}
{Garrod}, R.~T. \& {Herbst}, E. 2006, \aap, 457, 927

\bibitem[{{Garrod} {et~al.}(2008){Garrod}, {Widicus Weaver}, \& {Herbst}}]{garrod2008}
{Garrod}, R.~T., {Widicus Weaver}, S.~L., \& {Herbst}, E. 2008, \apj, 682, 283

\bibitem[{{Giannetti} {et~al.}(2017){Giannetti}, {Leurini}, {Wyrowski}, {Urquhart}, {Csengeri}, {Menten}, {K{\"o}nig}, \& {G{\"u}sten}}]{giannetti2017}
{Giannetti}, A., {Leurini}, S., {Wyrowski}, F., {et~al.} 2017, \aap, 603, A33

\bibitem[{{Gieser} {et~al.}(2021){Gieser}, {Beuther}, {Semenov}, {Ahmadi}, {Suri}, {M{\"o}ller}, {Beltr{\'a}n}, {Klaassen}, {Zhang}, {Urquhart}, {Henning}, {Feng}, {Galv{\'a}n-Madrid}, {de Souza Magalh{\~a}es}, {Moscadelli}, {Longmore}, {Leurini}, {Kuiper}, {Peters}, {Menten}, {Csengeri}, {Fuller}, {Wyrowski}, {Lumsden}, {S{\'a}nchez-Monge}, {Maud}, {Linz}, {Palau}, {Schilke}, {Pety}, {Pudritz}, {Winters}, \& {Pi{\'e}tu}}]{gieser2021}
{Gieser}, C., {Beuther}, H., {Semenov}, D., {et~al.} 2021, \aap, 648, A66

\bibitem[{{Gieser} {et~al.}(2022){Gieser}, {Beuther}, {Semenov}, {Suri}, {Soler}, {Linz}, {Syed}, {Henning}, {Feng}, {M{\"o}ller}, {Palau}, {Winters}, {Beltr{\'a}n}, {Kuiper}, {Moscadelli}, {Klaassen}, {Urquhart}, {Peters}, {Longmore}, {S{\'a}nchez-Monge}, {Galv{\'a}n-Madrid}, {Pudritz}, \& {Johnston}}]{Gieser2022}
{Gieser}, C., {Beuther}, H., {Semenov}, D., {et~al.} 2022, \aap, 657, A3

\bibitem[{{Gratier} {et~al.}(2017){Gratier}, {Bron}, {Gerin}, {Pety}, {Guzman}, {Orkisz}, {Bardeau}, {Goicoechea}, {Le Petit}, {Liszt}, {{\"O}berg}, {Peretto}, {Roueff}, {Sievers}, \& {Tremblin}}]{gratier2017}
{Gratier}, P., {Bron}, E., {Gerin}, M., {et~al.} 2017, \aap, 599, A100

\bibitem[{{Guzm{\'a}n} {et~al.}(2018){Guzm{\'a}n}, {Guzm{\'a}n}, {Garay}, {Bronfman}, \& {Hechenleitner}}]{guzman2018ApJScrosscorrelation}
{Guzm{\'a}n}, A.~E., {Guzm{\'a}n}, V.~V., {Garay}, G., {Bronfman}, L., \& {Hechenleitner}, F. 2018, \apjs, 236, 45

\bibitem[{{Hassel} {et~al.}(2008){Hassel}, {Herbst}, \& {Garrod}}]{hassel2008}
{Hassel}, G.~E., {Herbst}, E., \& {Garrod}, R.~T. 2008, \apj, 681, 1385

\bibitem[{{Hoare} {et~al.}(2012){Hoare}, {Purcell}, {Churchwell}, {Diamond}, {Cotton}, {Chandler}, {Smethurst}, {Kurtz}, {Mundy}, {Dougherty}, {Fender}, {Fuller}, {Jackson}, {Garrington}, {Gledhill}, {Goldsmith}, {Lumsden}, {Mart{\'\i}}, {Moore}, {Muxlow}, {Oudmaijer}, {Pandian}, {Paredes}, {Shepherd}, {Spencer}, {Thompson}, {Umana}, {Urquhart}, \& {Zijlstra}}]{Hoare+12}
{Hoare}, M.~G., {Purcell}, C.~R., {Churchwell}, E.~B., {et~al.} 2012, \pasp, 124, 939

\bibitem[{{Jim{\'e}nez-Serra} {et~al.}(2008){Jim{\'e}nez-Serra}, {Caselli}, {Mart{\'\i}n-Pintado}, \& {Hartquist}}]{jimenezserra2008}
{Jim{\'e}nez-Serra}, I., {Caselli}, P., {Mart{\'\i}n-Pintado}, J., \& {Hartquist}, T.~W. 2008, \aap, 482, 549

\bibitem[{{Jow} {et~al.}(2018){Jow}, {Hill}, {Scott}, {Soler}, {Martin}, {Devlin}, {Fissel}, \& {Poidevin}}]{Jow2018}
{Jow}, D.~L., {Hill}, R., {Scott}, D., {et~al.} 2018, \mnras, 474, 1018

\bibitem[{Kennicutt \& Evans(2012)}]{Kennicutt_2012}
Kennicutt, R.~C. \& Evans, N.~J. 2012, Annual Review of Astronomy and Astrophysics, 50, 531–608

\bibitem[{{K{\"o}nig} {et~al.}(2017){K{\"o}nig}, {Urquhart}, {Csengeri}, {Leurini}, {Wyrowski}, {Giannetti}, {Wienen}, {Pillai}, {Kauffmann}, {Menten}, \& {Schuller}}]{konig2017}
{K{\"o}nig}, C., {Urquhart}, J.~S., {Csengeri}, T., {et~al.} 2017, \aap, 599, A139

\bibitem[{{Le Teuff} {et~al.}(2000){Le Teuff}, {Millar}, \& {Markwick}}]{leteuff2000}
{Le Teuff}, Y.~H., {Millar}, T.~J., \& {Markwick}, A.~J. 2000, \aaps, 146, 157

\bibitem[{{Leonardis} {et~al.}(2006){Leonardis}, {Bischof}, \& {Pinz}}]{leonardis2006}
{Leonardis}, A., {Bischof}, H., \& {Pinz}, A. 2006, Berlin:Springer, Vol. 3951

\bibitem[{{Li} {et~al.}(2022{\natexlab{a}}){Li}, {Sanhueza}, {Lu}, {Lee}, {Zhang}, {Bovino}, {Sabatini}, {Liu}, {Kim}, {Morii}, {Tafoya}, {Tatematsu}, {Sakai}, {Wang}, {Li}, {Silva}, {Izumi}, \& {Allingham}}]{li2022crosscorrelation}
{Li}, S., {Sanhueza}, P., {Lu}, X., {et~al.} 2022{\natexlab{a}}, \apj, 939, 102

\bibitem[{{Li} {et~al.}(2022{\natexlab{b}}){Li}, {Sanhueza}, {Lu}, {Lee}, {Zhang}, {Bovino}, {Sabatini}, {Liu}, {Kim}, {Morii}, {Tafoya}, {Tatematsu}, {Sakai}, {Wang}, {Li}, {Silva}, {Izumi}, \& {Allingham}}]{li2022}
{Li}, S., {Sanhueza}, P., {Lu}, X., {et~al.} 2022{\natexlab{b}}, \apj, 939, 102

\bibitem[{{Li} {et~al.}(2020){Li}, {Sanhueza}, {Zhang}, {Nakamura}, {Lu}, {Wang}, {Liu}, {Tatematsu}, {Jackson}, {Silva}, {Guzm{\'a}n}, {Sakai}, {Izumi}, {Tafoya}, {Li}, {Contreras}, {Morii}, \& {Kim}}]{LiOutflows2020}
{Li}, S., {Sanhueza}, P., {Zhang}, Q., {et~al.} 2020, \apj, 903, 119

\bibitem[{{Liu} {et~al.}(2020){Liu}, {Evans}, {Kim}, {Goldsmith}, {Liu}, {Zhang}, {Tatematsu}, {Wang}, {Juvela}, {Bronfman}, {Cunningham}, {Garay}, {Hirota}, {Lee}, {Kang}, {Li}, {Li}, {Mardones}, {Qin}, {Ristorcelli}, {Tej}, {Toth}, {Wu}, {Wu}, {Yi}, {Yun}, {Liu}, {Peng}, {Li}, {Li}, {Lee}, {Shen}, {Baug}, {Wang}, {Zhang}, {Issac}, {Zhu}, {Luo}, {Soam}, {Liu}, {Xu}, {Wang}, {Zhang}, {Ren}, \& {Zhang}}]{Liu2020ATOMShog}
{Liu}, T., {Evans}, N.~J., {Kim}, K.-T., {et~al.} 2020, \mnras, 496, 2790

\bibitem[{{Louvet} {et~al.}(2016){Louvet}, {Motte}, {Gusdorf}, {Nguy{\^e}n Luong}, {Lesaffre}, {Duarte-Cabral}, {Maury}, {Schneider}, {Hill}, {Schilke}, \& {Gueth}}]{louvet2016}
{Louvet}, F., {Motte}, F., {Gusdorf}, A., {et~al.} 2016, \aap, 595, A122

\bibitem[{{Luna} {et~al.}(2017){Luna}, {Luna-Ferr{\'a}ndiz}, {Mill{\'a}n}, {Domingo}, {Mu{\~n}oz Caro}, {Santonja}, \& {Satorre}}]{luna2017}
{Luna}, R., {Luna-Ferr{\'a}ndiz}, R., {Mill{\'a}n}, C., {et~al.} 2017, \apj, 842, 51

\bibitem[{{Martin-Pintado} {et~al.}(1992){Martin-Pintado}, {Bachiller}, \& {Fuente}}]{martinpintado1992}
{Martin-Pintado}, J., {Bachiller}, R., \& {Fuente}, A. 1992, \aap, 254, 315

\bibitem[{{Millar} {et~al.}(1989){Millar}, {Bennett}, \& {Herbst}}]{millar1989}
{Millar}, T.~J., {Bennett}, A., \& {Herbst}, E. 1989, \apj, 340, 906

\bibitem[{{Molinari} {et~al.}(2019){Molinari}, {Baldeschi}, {Robitaille}, {Morales}, {Schisano}, {Traficante}, {Merello}, {Molinaro}, {Vitello}, {Sciacca}, \& {Liu}}]{molinari2019}
{Molinari}, S., {Baldeschi}, A., {Robitaille}, T.~P., {et~al.} 2019, \mnras, 486, 4508

\bibitem[{{Molinari} {et~al.}(2008){Molinari}, {Pezzuto}, {Cesaroni}, {Brand}, {Faustini}, \& {Testi}}]{molinari2008}
{Molinari}, S., {Pezzuto}, S., {Cesaroni}, R., {et~al.} 2008, \aap, 481, 345

\bibitem[{{Molinari} {et~al.}(2025){Molinari}, {Schilke}, {Battersby}, {Ho}, {Sanchez-Monge}, {Traficante}, {Jones}, {Beltran}, {Beuther}, {Fuller}, {Zhang}, {Klessen}, {Walch}, {Tang}, {Benedettini}, {Elia}, {Coletta}, {Mininni}, {Schisano}, {Avison}, {Law}, {Nucara}, {Soler}, {Stroud}, {Wallace}, {Wells}, {Ahmadi}, {Brogan}, {Hunter}, {Liu}, {Pezzuto}, {Su}, {Zimmermann}, {Zhang}, {Wyrowski}, {De Angelis}, {Liu}, {Clarke}, {Fontani}, {Klaassen}, {Koch}, {Johnston}, {Lebreuilly}, {Liu}, {Lumsden}, {Moeller}, {Moscadelli}, {Kuiper}, {Lis}, {Peretto}, {Pfalzner}, {Rigby}, {Sanhueza}, {Rygl}, {van der Tak}, {Zinnecker}, {Amaral}, {Bally}, {Bronfman}, {Cesaroni}, {Goh}, {Hoare}, {Hatchfield}, {Hennebelle}, {Henning}, {Kim}, {Kim}, {Maud}, {Merello}, {Nakamura}, {Plume}, {Qin}, {Svoboda}, {Testi}, {Veena}, \& {Walker}}]{ALMAGAL1}
{Molinari}, S., {Schilke}, P., {Battersby}, C., {et~al.} 2025, \aap, 696, A149

\bibitem[{{Molinari} {et~al.}(2016){Molinari}, {Schisano}, {Elia}, {Pestalozzi}, {Traficante}, {Pezzuto}, {Swinyard}, {Noriega-Crespo}, {Bally}, {Moore}, {Plume}, {Zavagno}, {di Giorgio}, {Liu}, {Pilbratt}, {Mottram}, {Russeil}, {Piazzo}, {Veneziani}, {Benedettini}, {Calzoletti}, {Faustini}, {Natoli}, {Piacentini}, {Merello}, {Palmese}, {Del Grande}, {Polychroni}, {Rygl}, {Polenta}, {Barlow}, {Bernard}, {Martin}, {Testi}, {Ali}, {Andr{\'e}}, {Beltr{\'a}n}, {Billot}, {Carey}, {Cesaroni}, {Compi{\`e}gne}, {Eden}, {Fukui}, {Garcia-Lario}, {Hoare}, {Huang}, {Joncas}, {Lim}, {Lord}, {Martinavarro-Armengol}, {Motte}, {Paladini}, {Paradis}, {Peretto}, {Robitaille}, {Schilke}, {Schneider}, {Schulz}, {Sibthorpe}, {Strafella}, {Thompson}, {Umana}, {Ward-Thompson}, \& {Wyrowski}}]{cutexmolinari2016}
{Molinari}, S., {Schisano}, E., {Elia}, D., {et~al.} 2016, \aap, 591, A149

\bibitem[{{Molinari} {et~al.}(2011){Molinari}, {Schisano}, {Faustini}, {Pestalozzi}, {di Giorgio}, \& {Liu}}]{cutexMolinari+11}
{Molinari}, S., {Schisano}, E., {Faustini}, F., {et~al.} 2011, \aap, 530, A133

\bibitem[{{Motte} {et~al.}(2007){Motte}, {Bontemps}, {Schilke}, {Schneider}, {Menten}, \& {Brogui{\`e}re}}]{motte2007SiOinCygnusX}
{Motte}, F., {Bontemps}, S., {Schilke}, P., {et~al.} 2007, \aap, 476, 1243

\bibitem[{{Purcell} {et~al.}(2008){Purcell}, {Hoare}, \& {Diamond}}]{Purcell+08}
{Purcell}, C.~R., {Hoare}, M.~G., \& {Diamond}, P. 2008, in Astronomical Society of the Pacific Conference Series, Vol. 387, Massive Star Formation: Observations Confront Theory, ed. H.~{Beuther}, H.~{Linz}, \& T.~{Henning}, 389

\bibitem[{{Roueff} {et~al.}(2007){Roueff}, {Parise}, \& {Herbst}}]{roueff2007}
{Roueff}, E., {Parise}, B., \& {Herbst}, E. 2007, \aap, 464, 245

\bibitem[{{Sakai} {et~al.}(2022){Sakai}, {Sanhueza}, {Furuya}, {Tatematsu}, {Li}, {Aikawa}, {Lu}, {Zhang}, {Morii}, {Nakamura}, {Takemura}, {Izumi}, {Hirota}, {Silva}, {Guzman}, {Sakai}, \& {Yamamoto}}]{SakaiASHES_DCNoutflow}
{Sakai}, T., {Sanhueza}, P., {Furuya}, K., {et~al.} 2022, \apj, 925, 144

\bibitem[{{S{\'a}nchez-Monge} {et~al.}(2025){S{\'a}nchez-Monge}, {Brogan}, {Hunter}, {Ahmadi}, {Avison}, {Beltr{\'a}n}, {Beuther}, {Coletta}, {Fuller}, {Johnston}, {Jones}, {Liu}, {Mininni}, {Molinari}, {Schilke}, {Schisano}, {Su}, {Traficante}, {Zhang}, {Battersby}, {Benedettini}, {Elia}, {Ho}, {Klaassen}, {Klessen}, {Law}, {Lis}, {Liu}, {Maud}, {M{\"o}ller}, {Moscadelli}, {Pezzuto}, {Rygl}, {Sanhueza}, {Soler}, {Stroud}, {Tang}, {van der Tak}, {Walker}, {Wallace}, {Walch}, {Wells}, {Wyrowski}, {Zhang}, {Allande}, {Bronfman}, {Dann}, {De Angelis}, {Fontani}, {Henning}, {Kim}, {Kuiper}, {Merello}, {Nakamura}, {Nucara}, \& {Rigby}}]{ALMAGAL2}
{S{\'a}nchez-Monge}, {\'A}., {Brogan}, C.~L., {Hunter}, T.~R., {et~al.} 2025, \aap, 696, A150

\bibitem[{{Sanhueza} {et~al.}(2013){Sanhueza}, {Jackson}, {Foster}, {Jimenez-Serra}, {Dirienzo}, \& {Pillai}}]{SanhuezaSiOnarrow}
{Sanhueza}, P., {Jackson}, J.~M., {Foster}, J.~B., {et~al.} 2013, \apj, 773, 123

\bibitem[{{Santos} {et~al.}(2022){Santos}, {Chuang}, {Lamberts}, {Fedoseev}, {Ioppolo}, \& {Linnartz}}]{santos2022}
{Santos}, J.~C., {Chuang}, K.-J., {Lamberts}, T., {et~al.} 2022, \apjl, 931, L33

\bibitem[{{Schilke} {et~al.}(1997){Schilke}, {Walmsley}, {Pineau des Forets}, \& {Flower}}]{schilke1997}
{Schilke}, P., {Walmsley}, C.~M., {Pineau des Forets}, G., \& {Flower}, D.~R. 1997, \aap, 321, 293

\bibitem[{{Shirley}(2015)}]{Shirley2015}
{Shirley}, Y.~L. 2015, \pasp, 127, 299

\bibitem[{{Simons} {et~al.}(2020){Simons}, {Lamberts}, \& {Cuppen}}]{simons2020}
{Simons}, M.~A.~J., {Lamberts}, T., \& {Cuppen}, H.~M. 2020, \aap, 634, A52

\bibitem[{{Soler} {et~al.}(2019){Soler}, {Beuther}, {Rugel}, {Wang}, {Clark}, {Glover}, {Goldsmith}, {Heyer}, {Anderson}, {Goodman}, {Henning}, {Kainulainen}, {Klessen}, {Longmore}, {McClure-Griffiths}, {Menten}, {Mottram}, {Ott}, {Ragan}, {Smith}, {Urquhart}, {Bigiel}, {Hennebelle}, {Roy}, \& {Schilke}}]{soler_astroHOG}
{Soler}, J.~D., {Beuther}, H., {Rugel}, M., {et~al.} 2019, \aap, 622, A166

\bibitem[{{Soler} {et~al.}(2013){Soler}, {Hennebelle}, {Martin}, {Miville-Desch{\^e}nes}, {Netterfield}, \& {Fissel}}]{soler2013}
{Soler}, J.~D., {Hennebelle}, P., {Martin}, P.~G., {et~al.} 2013, \apj, 774, 128

\bibitem[{{Soler} {et~al.}(2023){Soler}, {Zucker}, {Peek}, {Heyer}, {Goldsmith}, {Glover}, {Molinari}, {Klessen}, {Hennebelle}, {Testi}, {Colman}, {Benedettini}, {Elia}, {Mininni}, {Pezzuto}, {Schisano}, \& {Traficante}}]{soler2023}
{Soler}, J.~D., {Zucker}, C., {Peek}, J.~E.~G., {et~al.} 2023, \aap, 675, A206

\bibitem[{{Stahler} \& {Palla}(2004)}]{stahlerpalla}
{Stahler}, S.~W. \& {Palla}, F. 2004, {The Formation of Stars}

\bibitem[{{Syed} {et~al.}(2020){Syed}, {Wang}, {Beuther}, {Soler}, {Rugel}, {Ott}, {Brunthaler}, {Kerp}, {Heyer}, {Klessen}, {Henning}, {Glover}, {Goldsmith}, {Linz}, {Urquhart}, {Ragan}, {Johnston}, \& {Bigiel}}]{syed2020}
{Syed}, J., {Wang}, Y., {Beuther}, H., {et~al.} 2020, \aap, 642, A68

\bibitem[{{Taniguchi} {et~al.}(2019){Taniguchi}, {Saito}, {Sridharan}, \& {Minamidani}}]{taniguchi2019}
{Taniguchi}, K., {Saito}, M., {Sridharan}, T.~K., \& {Minamidani}, T. 2019, \apj, 872, 154

\bibitem[{{Towner} {et~al.}(2024){Towner}, {Ginsburg}, {Dell'Ova}, {Gusdorf}, {Bontemps}, {Csengeri}, {Galv{\'a}n-Madrid}, {Louvet}, {Motte}, {Sanhueza}, {Stutz}, {Bally}, {Baug}, {Chen}, {Cunningham}, {Fern{\'a}ndez-L{\'o}pez}, {Liu}, {Lu}, {Nony}, {Valeille-Manet}, {Wu}, {{\'A}lvarez-Guti{\'e}rrez}, {Bonfand}, {Di Francesco}, {Nguyen-Luong}, {Olguin}, \& {Whitworth}}]{TownerAlmaIMFOutflow}
{Towner}, A.~P.~M., {Ginsburg}, A., {Dell'Ova}, P., {et~al.} 2024, \apj, 960, 48

\bibitem[{{Turner}(2001)}]{turner2001}
{Turner}, B.~E. 2001, \apjs, 136, 579

\bibitem[{{Tychoniec} {et~al.}(2021){Tychoniec}, {van Dishoeck}, {van't Hoff}, {van Gelder}, {Tabone}, {Chen}, {Harsono}, {Hull}, {Hogerheijde}, {Murillo}, \& {Tobin}}]{tychoniecnew}
{Tychoniec}, {\L}., {van Dishoeck}, E.~F., {van't Hoff}, M. L.~R., {et~al.} 2021, \aap, 655, A65

\bibitem[{{van Dishoeck} \& {Blake}(1998)}]{vandischoekblake1998}
{van Dishoeck}, E.~F. \& {Blake}, G.~A. 1998, \araa, 36, 317

\bibitem[{{Wakelam} {et~al.}(2004){Wakelam}, {Caselli}, {Ceccarelli}, {Herbst}, \& {Castets}}]{wakelam2004}
{Wakelam}, V., {Caselli}, P., {Ceccarelli}, C., {Herbst}, E., \& {Castets}, A. 2004, \aap, 422, 159

\bibitem[{{Wakelam} {et~al.}(2011){Wakelam}, {Hersant}, \& {Herpin}}]{wakelam2011}
{Wakelam}, V., {Hersant}, F., \& {Herpin}, F. 2011, \aap, 529, A112

\bibitem[{{Watanabe} {et~al.}(2004){Watanabe}, {Nagaoka}, {Shiraki}, \& {Kouchi}}]{watanabe2004}
{Watanabe}, N., {Nagaoka}, A., {Shiraki}, T., \& {Kouchi}, A. 2004, \apj, 616, 638

\bibitem[{{Wells} {et~al.}(2024){Wells}, {Beuther}, {Molinari}, {Schilke}, {Battersby}, {Ho}, {S{\'a}nchez-Monge}, {Jones}, {Scheuck}, {Syed}, {Gieser}, {Kuiper}, {Elia}, {Coletta}, {Traficante}, {Wallace}, {Rigby}, {Klessen}, {Zhang}, {Walch}, {Beltr{\'a}n}, {Tang}, {Fuller}, {Lis}, {M{\"o}ller}, {van der Tak}, {Klaassen}, {Clarke}, {Moscadelli}, {Mininni}, {Zinnecker}, {Maruccia}, {Pezzuto}, {Benedettini}, {Soler}, {Brogan}, {Avison}, {Sanhueza}, {Schisano}, {Liu}, {Fontani}, {Rygl}, {Wyrowski}, {Bally}, {Walker}, {Ahmadi}, {Koch}, {Merello}, {Law}, \& {Testi}}]{molly2024}
{Wells}, M.~R.~A., {Beuther}, H., {Molinari}, S., {et~al.} 2024, \aap, 690, A185

\end{thebibliography}

\appendix

\section{Masks creation}
\begin{figure*}[h!]
    \centering
    \includegraphics[width=\linewidth]{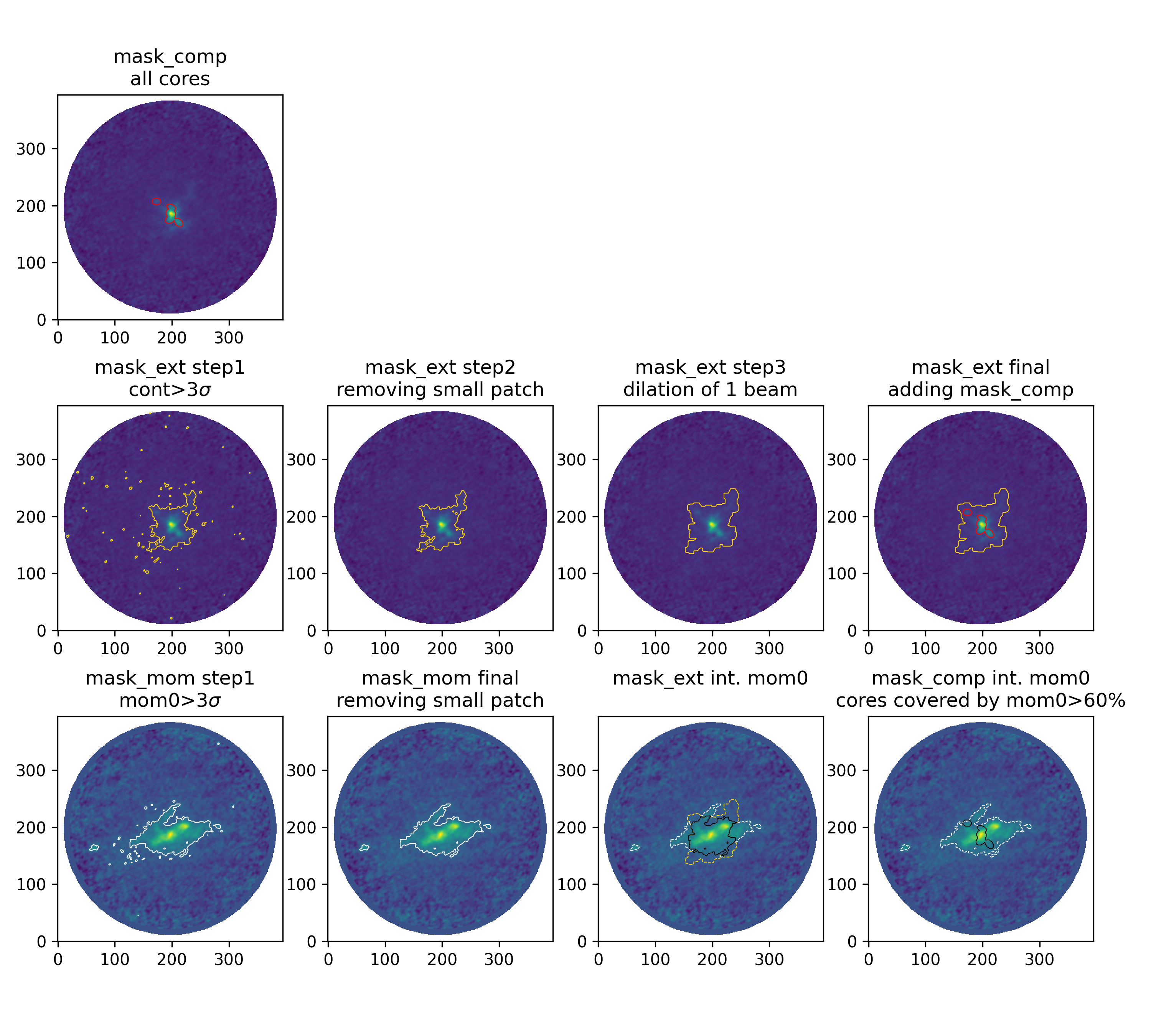}
    \caption{ Figure illustrating the steps defining the masks used in this paper for source AG024.4616+0.1981 for the continuum and for the moment-0 of H$_2$CO 3$_{0,3}$-2$_{0,2}$. Upper row: continuum mask$_{\rm comp.}$; middle row: steps for the creation of continuum mask$_{\rm ext.}$; bottom row first two panel from left: steps for the creation of the mask on moment-0; bottom row last two panel on the right: intersection masks used for the analysis.}
    \label{fig:masks all}
\end{figure*}
\label{appendix:continuummasks}
\subsection{Continuum masks}
In this paper, we created two different masks on the continuum maps.\\ \indent
The first mask we defined contains all the compact sources detected in the field of view by CuTEx, and we call it mask$_{\rm comp.}$. CuTEx \citep{cutexMolinari+11} individuates compact sources inside continuum maps building a curvature image by double-differentiation of the initial map in four different directions. This method allows an easier
identification (by applying a threshold on the curvatures maps)
of compact sources, which stand out in the curvature images, while diffuse and background emission is strongly dampened. Identified peaks in curvature maps are then fitted (on the original map) with 2D elliptical Gaussians plus a background that can be a constant background (zero order) or an inclined plane (first order) background\footnote{The best order for the backgound is automatically assessed by CuTEx.}, in order to obtain the position and dimensions of the sources (major axis $\sigma_A$, minor axis $\sigma_B$, and position angle) and the integrated flux background subtracted. As last step, we required that the background-subtracted peak
flux from photometry is higher than 5 times the map rms to confirm the compact source. CuTEx has been run on the continuum \textit{\sc 7m+tm2} maps using the same methodology of \citet{ALMAGAL3} for the \textit{\sc 7m+tm2+tm1} maps (the catalog for \textit{\sc 7m+tm2} will be presented in Coletta et al. in prep.). For the mask$_{\rm{com.}}$ we create ellipses for each core that took as semi-axis $2\sigma_A$ and $2\sigma_B$ from the 2D Gaussian fits, to recover the 86\% of the integral of the source\footnote{For 2D-Gaussian distribution centered around (0,0) the integral between $+\sigma$ and $-\sigma$ correspond to the 39\% of the total area, while between $+2\sigma$ and $-2\sigma$ to the 86\%.}. An example of mask$_{\rm{com.}}$ for the source AG024.4616+0.1981 is visible in red in the upper row of Fig. A.1
\\ \indent
The second mask, which includes the most diffuse and extended continuum emission, is defined as the continuum emission with a SNR higher than 3,  (mask$_{\rm{ext.}}$). In the ALMA continuum map, the noise level is not uniform but increases from the center of the map toward the edge, following the shape of the primary beam. Therefore, we build a map of noise by estimating the noise level of the continuum image not corrected for the primary beam, where the noise is uniform all over the map, with a sigma clipping method  and then we divided for the map of the primary beam. In the first step of the mask creation we select all of the pixels above a SNR of 3. In these first step a lot of noise peaks just above SNR 3 are included in the mask. Therefore, as second step we excluded from the first raw mask all the isolated regions with a number of pixels $<0.8\times N_{\rm{beam}}$, where $N_{\rm{beam}}$ is the number of pixels inside the area of a beam. As third step we dilated the mask by one beam \citep{dame2011}, to encompass possible emission just below the 3 SNR threshold, and as final step we include all of the pixel of mask$_{\rm{com.}}$ in mask$_{\rm{ext.}}$, if case not all of them where already selected. The steps for the creation of  mask$_{\rm{ext.}}$ are illustrated in orange in the second row of image A.1 for source AG024.4616+0.1981. The last step for this source is not needed, since all of the pixel of mask$_{\rm{com.}}$ were already in mask$_{\rm{ext.}}$ at step 3.\\

\subsection{Moment-0 masks}
For the moment-0 maps we derived a maps of the noise, using the same methodology described above for the continuum mask$_{\rm{ext.}}$. In the first step of the mask creation we select all of the pixels above a SNR of 3.  Therefore, in the second and final step we excluded the regions that do not have any pixel with SNR above 5 or that have a to-
tal number of pixels < 0.8$\times N_{\rm{beam}}$. The first requirement is
connected to the fact that integrated line intensity below 5 SNR
would result in spectral lines not identified above 3 SNR in the
spectrum, while the second is to avoid spurious noise fluctuation
peaks in the mask. As the last step, we required that at least one
of the regions remaining inside the mask has a number of pixels
> N$_{\rm{beam}}$, since this would be the size of emission of a point-like source. If this last condition is met, we considered the transition
as detected in the field. The two steps for the creation of the moment-0 mask for the transition of H$_2$CO 3$_{0,3}$-2$_{0,2}$ for source AG024.4616+0.1981 are visible in the first two columns of the bottom row of Fig. A.1 in white. 

\subsection{Intersection masks}
 We calculated the astroHOG V$_{\rm{N}}$ coefficient and the Spearman's coefficient on the intersection between the moment-0 mask and mask$_{\rm{ext.}}$. An example is visible in the third column of the bottom row of Figure A.1. For the compact emission, we run the analysis on the mask of the compact cores with detection of the selected transition, that we defined as the cores that are covered by more than 60\% by the mask of the moment-0 of the selected transitions. An example is visible in the last column of the bottom row of Figure A.1. In this case all the cores defined in the continuum met our condition.

 \section{Definition of the range of integration for moment-0 maps}
 In order to create the moment-0 maps for all the sources and all the selected transitions (1013 sources $\times$ 9 transitions), we created an automated routine that derives the best velocity range for integrating the cubes. The aim of the routine is to individuates the velocity range that encompass all the line (including high-velocity wings) in all the pixel of the source, to then integrate all the pixel over the same velocity range. \\ \indent
As first step, we resampled spatially the cube on the two positions axis by a factor 6, the average number of pixels in a beam size. For each source, we created an average spectrum from all the pixels with emission above 4$\sigma$ ($\sigma$ is the standard deviation of the cube along the frequency axis) in the resampled cube in an interval of 7 km/s around the systemic velocity of the clump, to avoid selecting a different line close in frequency, and we fit it with a Gaussian profile. 
We repeated the Gaussian fit for the spectrum of the pixel with the highest line intensity peak, since in some cases the line on the brightest pixel shows higher FWHM than the median spectrum defined before. We took the largest full width at half maximum (FWHM$^{\rm max}$) of the two fits and the central velocity, $\varv_{\rm{fit}}$, of the first fit to define as integration limits  $\varv_{\rm{fit}}-2{\rm FWHM}^{\rm max}$, and $\varv_{\rm{fit}}+2{\rm FWHM}^{\rm max}$. Moreover, considering the possible presence of non-Gaussian wings in strong lines, for pixels with emission above 5$\sigma$ we calculated the initial and final endpoints of the wings (i.e. where the emission of the wings goes to zero) as the two values $\varv_{\rm{i}}$ and $\varv_{\rm{f}}$ where the signal-to-noise ratio (SNR), starting from the center of the line, goes below 0.5. This ensures that the full wing is included inside the integration, since in many cases wings are very noisy and the signal has channels below 1 SNR before the end of the wing.
Finally, to compute the moment-0 maps, we adopt the starting point as
the minimum value between $\varv_{\rm{i}}$ and $\varv_{\rm{fit}}-2{\rm FWHM}^{\rm max}$, and the endpoint as the maximum value between $\varv_{\rm{f}}$ and $\varv_{\rm{fit}}+2{\rm FWHM}^{\rm max}$. The CH$_{3}$CN $12_{0}-11_{0}$ and $12_{1}-11_{1}$ lines are heavily blended in several sources. 
Therefore, for these two transitions, we created a single moment-0 map that includes the emission of both transitions, and then we fitted two Gaussian assuming they have the same FWHM.\\ \indent 
In Figure B.1 we show the different spectra used for defining the extreme of integration for the line of H$_2$CO 3$_{0,3}$-2$_{0,2}$ (upper row) and CH$_{3}$CN $12_{0,1}-11_{0,1}$ (bottom row) for source AG024.4616+0.1981. In the first panel (from left) of the upper row we can see the median spectrum of all the pixels above 4$\sigma$ and its best fit. The value of the FWHM of this fit define the first guess for the extreme of integration (dashed vertical magenta line) shown in all the panels. The second panel shows the spectrum of the pixel with the highest peak intensity with its best fit. The value of the FWHM of this fit define the second guess for the extreme of integration (dashed vertical green line) shown in all the panels. The third panel show the spectrum of the pixel with more extended blue-shifted wings, with the blue vertical line indicating the channel where the emission of the blue-wing goes below 0.5 SNR. The last panel show the spectrum of the pixel with more extended red-shifted wings, with the blue vertical line indicating the channel where the emission of the red-wing goes below 0.5 SNR. As defined before, we then selected the largest extreme of integration to create the moment-0 maps, in this case the value defined by the blue and red solid vertical lines.\\\indent
The bottom row of Fig. B.1 shows the same spectra as the upper row, but for the case of  CH$_{3}$CN, where only one $\rm{moment-0}$ map have been created integrating together the transitions $12_{0,1}-11_{0,1}$, since they are in most cases blended. For this reason the first two panels (from left) show a two-Gaussian fit, and the extreme of integration took into account the separation in veocity of the two peaks. For this source there are no pixels with blue-wing or red-wing above the range of integration already defined from the 2 two-Gaussian fit, so there is no spectra in the last two panels on the right. In this case the extreme of integration selected for the creation of the moment-0 map are those indicated by the vertical magenta dashed lines.
\\\indent
 All the maps for each field and transition were then visually inspected to validate them. This was done by looking at the line observed in the spectra averaged on each continuum compact source detected by the source extraction algorithm CuTEx (CUrvature Thresholding EXtractor \citealt{cutexMolinari+11, cutexmolinari2016} see Sect. 4.4; following the same methodology used in \citet{ALMAGAL3} for \textit{\sc 7m+tm2+tm1} data) in the clump, and checking that the velocity extremes used for the integration agree with the velocity ranges of the lines in these spectra, and no blending with other lines is present.
 \begin{figure*}
     \centering
     \includegraphics[width=0.9\linewidth, trim={4cm 0 4cm 0}, clip]{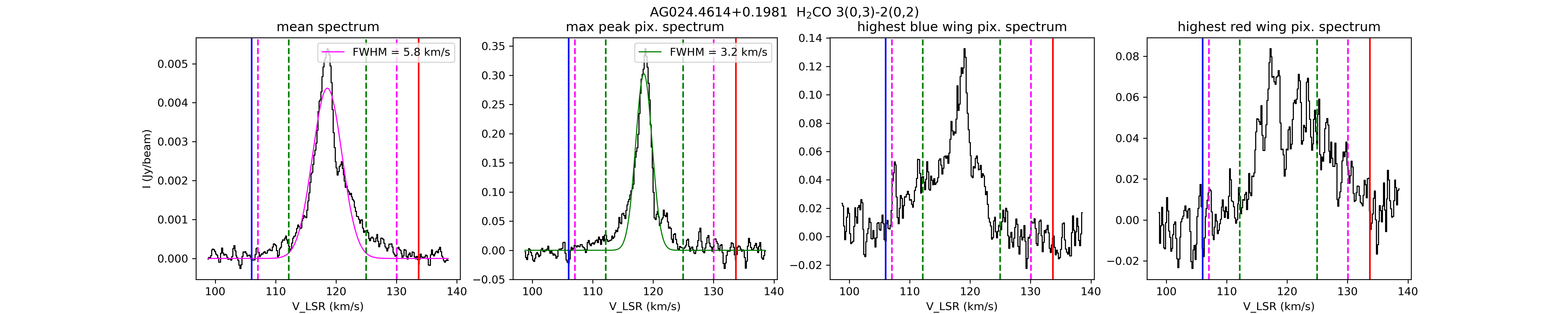}\\
     \includegraphics[width=0.9\linewidth, trim={4cm 0 4cm 0}, clip]{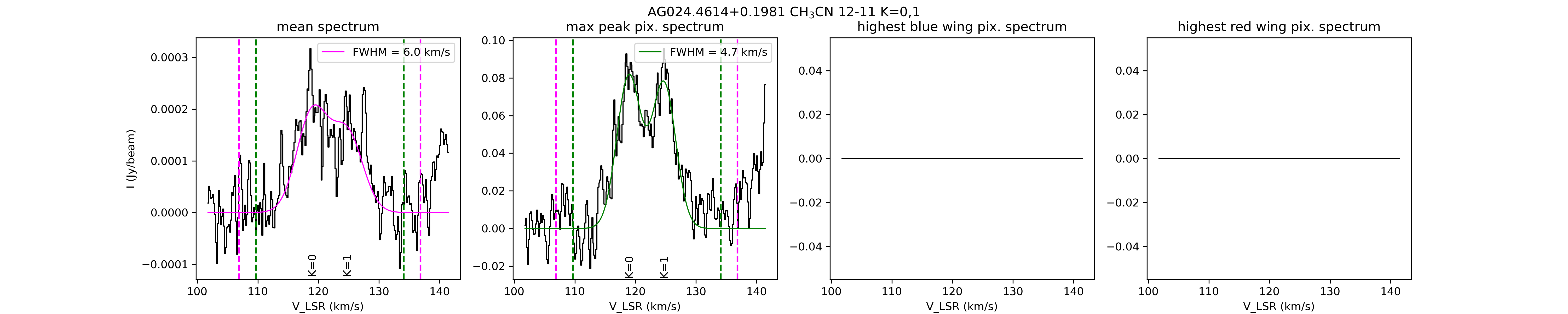}
     \caption{Steps of the determination of the range of integration for the moment-0 maps for the example source AG024.4616+0.1981 for H$_2$CO 3(0,3)-2(0,2) (top row) and CH$_{3}$CN $12_{0,1}-11_{0,1}$ (bottom row). Top row from left: 1) mean spectrum over all the pixel with peak emission above 4 SNR, with its best fit and derived range of integration in magenta; 2) pixel with highest peak spectrum, with its best fit and derived range of integration in green; 3) spectrum of the pixel with the blue-wing terminating at the lowest velocity, marked by the blue vertical line; 4) spectrum of the pixel with the red-wing terminating at the highest velocity, marked by the red vertical line. Bottom row: same as top row, but for CH$_{3}$CN $12_{0,1}-11_{0,1}$ for which a 2-Gaussian fit is performed to include both K=0 and K=1 line. No spectra are shown in the last two panel since no pixel was found with wings above the integration range defined by the first two panels. }
     \label{fig:integrationrange}
 \end{figure*}
 
  \section{Noise correlation in astroHOG}
  \label{appendix:astrohognoise}
\begin{figure}
      \centering
      \includegraphics[width=9cm]{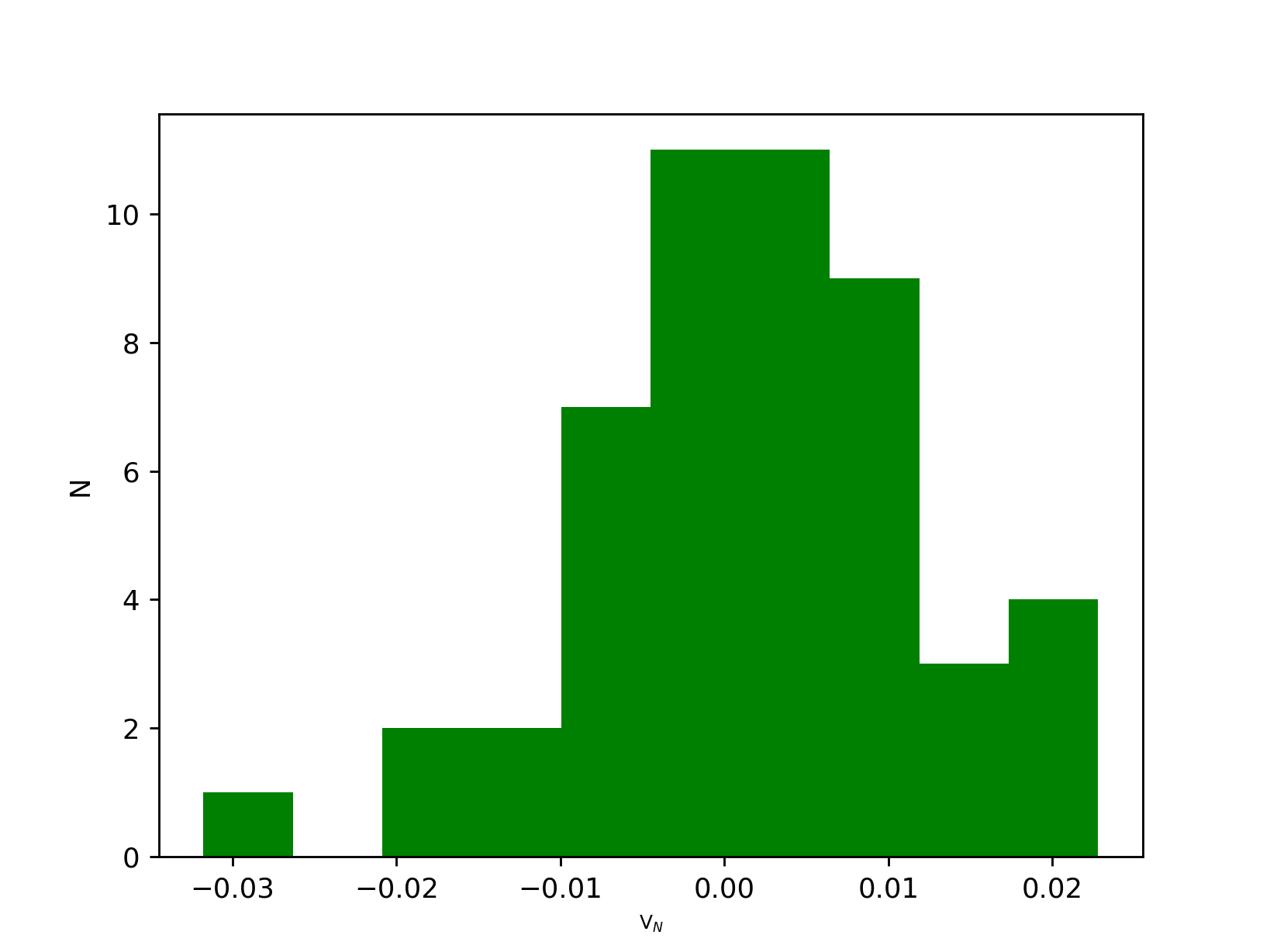}
      \caption{Distribution of the $V_{\rm{N}}$ parameter, calculated by \texttt{astroHOG} on pairs of noise-only moment-0 maps.}
      \label{fig:histonoise_noise}
  \end{figure}

\begin{figure*}
      \centering
      \includegraphics[width=18cm, trim={0.5cm 2.1cm 0.7cm 1.7cm}, clip]{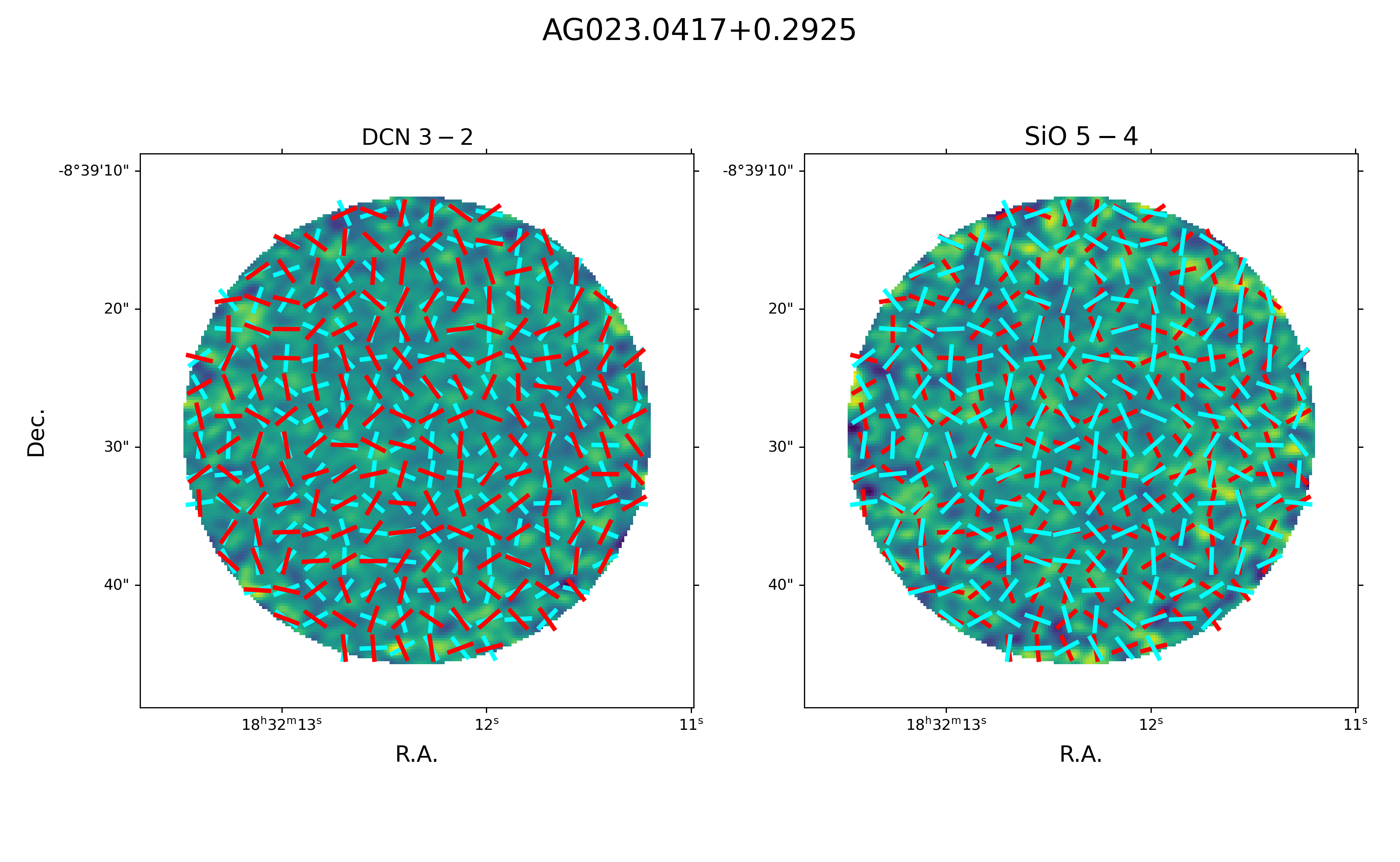}
      \caption{Gradients comparison on a pair of noise-only maps from Table \ref{tab:astroHOGnoise} for source AG023.0417+0.2925. The color scale units are Jy/beam km\,s$^{-1}$ in both panels. The cyan lines indicate the direction of the gradients of the noise map in the left panel, while the red lines indicate the direction of the gradient of the noise map in the right panel. For this pair of noise maps $V_{\rm{N}}$ is 0.013. }
      \label{fig:plot_noise_noise}
  \end{figure*}

  To estimate the significance of the projected Rayleigh statistics parameter $V_{\rm{N}}$, in output of \texttt{astroHOG}, we performed the morphological comparison on noise-only maps. We selected couples of moment-0 maps of the same sources of two non-detected lines, for a total of 50 couples of noise maps, and ran \texttt{astroHOG} on the total FOV of the maps. The list of moment-0 maps used and the results of \texttt{astroHOG} are given in Table \ref{tab:astroHOGnoise}. The histogram of the values of $V_{\rm{N}}$ from the correlation of noise-only maps is given in Fig. \ref{fig:histonoise_noise}, while in Fig. \ref{fig:plot_noise_noise} we plotted one couple of noise maps with over imposed the gradients directions in the two maps. The median absolute deviation of the correlation of noise-only maps is 0.6\%, with all the values being below 3\%.

  \begin{table*}[]
      \centering
       \caption{Results of \texttt{astroHOG} significance test on  moment-0 maps with non-detection. }
      \begin{tabular}{lll|c}
      \hline\hline
        ALMAGAL ID & noise-map 1& noise-map 2 & $V_{\rm{N}}$  \\
         \hline
AG022.7965-0.4814 &	DCN $3-2$ & SiO $5-4$&		0.00939	\\
AG022.8624+0.4213 &	DCN $3-2$ & SiO $5-4$&		-0.00727\\
AG022.9296-0.3401 &	DCN $3-2$ & SiO $5-4$&		0.01956	\\
AG023.0417+0.2925 &	DCN $3-2$ & SiO $5-4$&		0.01276	\\
AG023.2440+0.0062 &	DCN $3-2$ & SiO $5-4$&		0.02281	\\
AG023.2969-0.0711 &	DCN $3-2$ & SiO $5-4$&		-0.00187\\
AG023.3221-0.0621 &	DCN $3-2$ & SiO $5-4$&		0.00212	\\
AG023.4407+0.4334 &	DCN $3-2$ & SiO $5-4$&		-0.00007\\
AG023.5349-0.0590 &	DCN $3-2$ & SiO $5-4$&		-0.00727\\
AG023.5628+0.4862 &	DCN $3-2$ & SiO $5-4$&		0.00478	\\
AG024.4982+0.0543 &	HCCCN $24-23$ & CH$_3$OH $4_{2,3}-3_{1,2}$	&		0.00601	\\
AG024.5259-0.2868 &	HCCCN $24-23$ & CH$_3$OH $4_{2,3}-3_{1,2}$	&		-0.00527\\
AG024.5243+0.3190 &	HCCCN $24-23$ & CH$_3$OH $4_{2,3}-3_{1,2}$	&		-0.00180\\
AG024.5412-0.1309 &	HCCCN $24-23$ & CH$_3$OH $4_{2,3}-3_{1,2}$	&		0.00692	\\
AG024.5640+0.3385 &	HCCCN $24-23$ & CH$_3$OH $4_{2,3}-3_{1,2}$	&		-0.00322\\
AG024.7042-0.1263 &	HCCCN $24-23$ & CH$_3$OH $4_{2,3}-3_{1,2}$	&		0.00876	\\
AG024.8555+0.0051 &	HCCCN $24-23$ & CH$_3$OH $4_{2,3}-3_{1,2}$	&		0.00978	\\
AG024.9506-0.0552 &	HCCCN $24-23$ & CH$_3$OH $4_{2,3}-3_{1,2}$	&		0.00839	\\
AG025.0283-0.4634 &	HCCCN $24-23$ & CH$_3$OH $4_{2,3}-3_{1,2}$	&		0.00706	\\
AG025.1783+0.2107 &	HCCCN $24-23$ & CH$_3$OH $4_{2,3}-3_{1,2}$	&		-0.00448\\
AG026.3308-0.0457 &	H$_2$CO $3_{2,1} - 2_{2,0}$	 & DCN $3-2$		&		0.00029	\\
AG026.4959+0.3215 &	H$_2$CO $3_{2,1} - 2_{2,0}$	 & DCN $3-2$		&		-0.00553\\
AG026.6280-0.0647 &	H$_2$CO $3_{2,1} - 2_{2,0}$	 & DCN $3-2$		&		0.00403	\\
AG026.8474+0.1794 &	H$_2$CO $3_{2,1} - 2_{2,0}$	 & DCN $3-2$		&		-0.00264\\
AG026.9310+0.1783 &	H$_2$CO $3_{2,1} - 2_{2,0}$	 & DCN $3-2$		&		-0.00454\\
AG026.9567-0.0755 &	H$_2$CO $3_{2,1} - 2_{2,0}$	 & DCN $3-2$		&		0.00517	\\
AG027.0291+0.2804 &	H$_2$CO $3_{2,1} - 2_{2,0}$	 & DCN $3-2$		&		0.00237	\\
AG027.0854-0.5640 &	H$_2$CO $3_{2,1} - 2_{2,0}$	 & DCN $3-2$		&		0.00963	\\
AG027.6856+0.0977 &	H$_2$CO $3_{2,1} - 2_{2,0}$	 & DCN $3-2$		&		0.01119	\\
AG027.8819+0.2192 &	H$_2$CO $3_{2,1} - 2_{2,0}$	 & DCN $3-2$		&		0.00440	\\
AG253.4449-0.4146 &	CH$_3$CN $12-11$ \tiny{K=0,1}	 & SO $6_5-5_4$		&		-0.01323\\
AG256.6469-0.0506 &	CH$_3$CN $12-11$ \tiny{K=0,1}	 & SO $6_5-5_4$		&		0.00276	\\
AG256.9065-0.2753 &	CH$_3$CN $12-11$ \tiny{K=0,1}	 & SO $6_5-5_4$		&		-0.01729\\
AG259.5640-0.9212 &	CH$_3$CN $12-11$ \tiny{K=0,1}	 & SO $6_5-5_4$		&		0.00626	\\
AG259.5936-1.3007 &	CH$_3$CN $12-11$ \tiny{K=0,1}	 & SO $6_5-5_4$		&		-0.00018\\
AG259.6098-1.3001 &	CH$_3$CN $12-11$ \tiny{K=0,1}	 & SO $6_5-5_4$		&		-0.00112\\
AG260.6880-1.3930 &	CH$_3$CN $12-11$ \tiny{K=0,1}	 & SO $6_5-5_4$		&		0.00538	\\
AG012.9785-0.2712 &	CH$_3$CN $12-11$ \tiny{K=0,1}	 & SO $6_5-5_4$		&		-0.00344\\
AG013.0166-0.1797 &	CH$_3$CN $12-11$ \tiny{K=0,1}	 & SO $6_5-5_4$		&		-0.03177\\
AG013.2123+0.0404 &	CH$_3$CN $12-11$ \tiny{K=0,1}	 & SO $6_5-5_4$		&		-0.01261\\
AG326.8567-0.0574 &	H$_2$CO $3_{0,3}-2_{0,2}$	 & CH$_3$OCHO $17_{3,14}-16_{3,13}\,$E	&		-0.01646\\
AG326.8832-0.2663 &	H$_2$CO $3_{0,3}-2_{0,2}$	 & CH$_3$OCHO $17_{3,14}-16_{3,13}\,$E	&		0.02187	\\
AG327.0692-0.2884 &	H$_2$CO $3_{0,3}-2_{0,2}$	 & CH$_3$OCHO $17_{3,14}-16_{3,13}\,$E	&		0.01961	\\
AG327.0931-0.3027 &	H$_2$CO $3_{0,3}-2_{0,2}$	 & CH$_3$OCHO $17_{3,14}-16_{3,13}\,$E	&		0.00052	\\
AG327.0937+0.5558 &	H$_2$CO $3_{0,3}-2_{0,2}$	 & CH$_3$OCHO $17_{3,14}-16_{3,13}\,$E	&		0.01112	\\
AG327.2300-0.5027 &	H$_2$CO $3_{0,3}-2_{0,2}$	 & CH$_3$OCHO $17_{3,14}-16_{3,13}\,$E	&		-0.00714\\
AG327.2420-0.2282 &	H$_2$CO $3_{0,3}-2_{0,2}$	 & CH$_3$OCHO $17_{3,14}-16_{3,13}\,$E	&		0.01374	\\
AG327.2455-0.2176 &	H$_2$CO $3_{0,3}-2_{0,2}$	 & CH$_3$OCHO $17_{3,14}-16_{3,13}\,$E	&		-0.00352\\
AG327.6594-0.0974 &	H$_2$CO $3_{0,3}-2_{0,2}$	 & CH$_3$OCHO $17_{3,14}-16_{3,13}\,$E	&		0.01255	\\
AG328.1177+0.5564 &	H$_2$CO $3_{0,3}-2_{0,2}$	 & CH$_3$OCHO $17_{3,14}-16_{3,13}\,$E	&		0.00113	\\
\hline
      \end{tabular}
     
      \label{tab:astroHOGnoise}
  \end{table*}

\section{Table of Results}
In Tables D.1 and D.2 we give the results of astroHOG analysis, of Pearson's correlation coefficient, and Spearman's correlation coefficient, together with the number of pixels of the various masks adopted in the analysis for the first 37 sources for H$_2$CO 3$_{0,3}$-2$_{0,2}$ for mask$_{\rm{ext.}}$ and mask$_{\rm{comp.}}$, respectively. The complete tables, as well as the tables for all of the other transitions are given in Zenodo\footnote{https://doi.org/10.5281/zenodo.15236491}.

\onecolumn
\small
\setlength{\tabcolsep}{2pt}
\begin{landscape}
\begin{table*}[]
\centering
\caption{Results of \texttt{astroHOG}, Pearson's correlation, and Spearman's correaltion for the mask$_{\rm{ext.}}$ for H$_2$CO $3_{0,3}-2_{0,2}$. }
      \begin{tabular}{cccccccccccccccccccc}
      \hline\hline

ALMAGAL ID   &   Original ID   &    Npix$_{\rm{ext.}}$   &   Npix$_{\rm{com.}}$   &   flag$_{1}$   &   det.   &   Npix$_{\rm{mom0}}$   &   Npix$_{\rm{inext.}}$   &   Npix$_{\rm{incom.}}$   &   flag$_{2}$   &   Npix$_{\rm{maxreg}}$   &   flag$_{3}$   &   V   &   errV   &   V$_{N}$   &   errV$_{N}$   &   $\rho_{\rm{p}}$  &   err $\rho_{\rm{p}}$ &   $\rho_{\rm{s}}$  &   err $\rho_{\rm{s}}$\\
\hline

AG010.1313-0.7687	&	46677	&	2136	&	441 	&	T	&	Y	&	260	&	220	&	0	&	T	&	220	&	T	&	-0.60		&	0.50	&	-0.03		&	0.02	&	0.17		&	0.04	&	0.18	&	0.05	\\ 
AG010.1392-0.7831	&	46696	&	1134	&	127 	&	T	&	Y	&	706	&	322	&	127	&	T	&	322	&	T	&	16.50		&	0.70	&	0.65		&	0.03	&	0.80		&	0.01	&	0.65	&	0.02	\\ 
AG010.2283-0.2064	&	47184	&	5479	&	766 	&	T	&	Y	&	1501	&	1057	&	390	&	T	&	927	&	T	&	13.70		&	0.60	&	0.30		&	0.01	&	0.46		&	0.01	&	0.38	&	0.01	\\ 
AG010.6185-0.0314	&	48727	&	1992	&	379 	&	T	&	Y	&	2086	&	1084	&	295	&	T	&	1084	&	T	&	23.40		&	0.70	&	0.50		&	0.02	&	0.73		&	0.00	&	0.68	&	0.01	\\ 
AG010.6691-0.2196	&	49005	&	5524	&	910 	&	T	&	Y	&	1443	&	1180	&	326	&	T	&	1160	&	T	&	17.10		&	0.40	&	0.35		&	0.01	&	0.54		&	0.01	&	0.47	&	0.01	\\ 
AG010.6985-0.0994	&	49143	&	725	&	0 	&	T	&	Y	&	316	&	81	&	0	&	T	&	81	&	F	&	-		&	-	&	-		&	-	&	-		&	-	&-	&-	\\ 
AG010.7516-0.1987	&	49337	&	4434	&	249 	&	T	&	Y	&	921	&	436	&	249	&	F	&	434	&	T	&	-		&	-	&	-		&	-	&	-		&	-	&-	&	-	\\ 
AG011.0542-0.0419	&	50433	&	638	&	0 	&	T	&	Y	&	45	&	42	&	0	&	T	&	42	&	F	&	-		&	-	&	-		&	-	&	-		&	-	&	-	&	-\\ 
AG011.0829-0.5311	&	50555	&	3981	&	540 	&	T	&	Y	&	1739	&	924	&	108	&	T	&	572	&	T	&	2.10		&	0.40	&	0.05		&	0.01	&	-0.13		&	0.01	&	-0.20	&	0.01	\\ 
AG011.2019-0.0909	&	51020	&	296	&	0 	&	T	&	N	&	0	&	0	&	0	&	F	&	0	&	F	&	-		&	-	&	-		&	-	&	-		&	-	&	-	&	-	\\ 
AG011.3614+0.8017	&	51554	&	4595	&	0 	&	T	&	N	&	0	&	0	&	0	&	F	&	0	&	F	&	-		&	-	&	-		&	-	&	-		&	-	&	-	&	-	\\ 
AG011.6992-0.3011	&	52831	&	2068	&	141 	&	T	&	Y	&	414	&	414	&	141	&	T	&	317	&	T	&	21.50		&	0.60	&	0.75		&	0.02	&	0.93		&	0.00	&	0.82	&	0.01	\\ 
AG011.9039-0.1403	&	53795	&	8340	&	1009 	&	T	&	Y	&	2913	&	2889	&	781	&	T	&	2889	&	T	&	38.00		&	0.60	&	0.50		&	0.01	&	0.83		&	0.00	&	0.76	&	0.00	\\ 
AG011.9183-0.6122	&	53869	&	6340	&	920 	&	T	&	Y	&	4497	&	3202	&	681	&	T	&	2997	&	T	&	14.60		&	0.60	&	0.18		&	0.01	&	0.55		&	0.00	&	0.13	&	0.01	\\ 
AG011.9370-0.6160	&	54001	&	10786	&	1254 	&	T	&	Y	&	4010	&	3967	&	1028	&	T	&	2305	&	T	&	11.00		&	1.00	&	0.13		&	0.01	&	0.19		&	0.01	&	0.09	&	0.01	\\ 
AG012.1889-0.1334	&	55190	&	1113	&	0 	&	T	&	Y	&	46	&	46	&	0	&	T	&	46	&	F	&	-		&	-	&	-		&	-	&	-		&	-	&	-	&	-	\\ 
AG012.2087-0.1017	&	55288	&	5445	&	706 	&	T	&	Y	&	5605	&	4175	&	706	&	T	&	4175	&	T	&	46.20		&	0.50	&	0.51		&	0.01	&	0.84		&	0.00	&	0.76	&	0.00	\\ 
AG012.4040-0.4681	&	56184	&	7305	&	610 	&	T	&	Y	&	616	&	616	&	251	&	T	&	265	&	T	&	19.10		&	0.50	&	0.54		&	0.02	&	0.82		&	0.01	&	0.71	&	0.01	\\ 
AG012.4977-0.2233	&	56634	&	8256	&	716 	&	T	&	Y	&	876	&	821	&	251	&	T	&	361	&	T	&	2.00		&	0.70	&	0.05		&	0.02	&	0.25		&	0.02	&	0.18	&	0.02	\\ 
AG012.5571-0.3485	&	56859	&	1349	&	218 	&	T	&	Y	&	692	&	45	&	0	&	T	&	45	&	F	&	-		&	-	&	-		&	-	&	-		&	-	&	-	&	-	\\ 
AG012.6794-0.1830	&	57335	&	4822	&	968 	&	T	&	Y	&	2817	&	2529	&	721	&	T	&	2521	&	T	&	27.80		&	0.70	&	0.39		&	0.01	&	0.95		&	0.00	&	0.69	&	0.01	\\ 
AG012.7207-0.2175	&	57596	&	6322	&	496 	&	T	&	Y	&	3724	&	3111	&	496	&	T	&	3111	&	T	&	17.70		&	0.50	&	0.22		&	0.01	&	0.40		&	0.00	&	0.61	&	0.00	\\ 
AG012.7360-0.1031	&	57641	&	6258	&	866 	&	T	&	Y	&	1330	&	1231	&	527	&	T	&	912	&	T	&	5.60		&	0.70	&	0.12		&	0.01	&	0.15		&	0.01	&	0.17	&	0.01	\\ 
AG012.8535-0.2265	&	58312	&	9396	&	987 	&	T	&	Y	&	2962	&	2932	&	702	&	T	&	2690	&	T	&	9.50		&	0.70	&	0.12		&	0.01	&	0.39		&	0.00	&	0.32	&	0.01	\\ 
AG012.9008-0.2404	&	58516	&	7414	&	983 	&	T	&	Y	&	2321	&	2251	&	866	&	T	&	2109	&	T	&	16.30		&	0.30	&	0.24		&	0.00	&	0.13		&	0.00	&	0.32	&	0.00	\\ 
AG012.9048-0.0306	&	58529	&	14825	&	1044 	&	T	&	Y	&	7437	&	6422	&	614	&	T	&	5705	&	T	&	0.00		&	1.00	&	0.00		&	0.01	&	0.27		&	0.00	&	0.12	&	0.01	\\ 
AG012.9084-0.2604	&	58549	&	8286	&	761 	&	T	&	Y	&	4001	&	3860	&	761	&	T	&	3860	&	T	&	33.50		&	0.70	&	0.38		&	0.01	&	0.78		&	0.00	&	0.52	&	0.01	\\ 
AG012.9156-0.3341	&	58583	&	4989	&	804 	&	T	&	Y	&	2279	&	1747	&	661	&	T	&	1747	&	T	&	0.60		&	0.50	&	0.01		&	0.01	&	0.42		&	0.00	&	0.32	&	0.01	\\ 
AG013.0166-0.1797	&	58971	&	1743	&	159 	&	T	&	Y	&	554	&	553	&	159	&	T	&	553	&	T	&	5.50		&	0.70	&	0.16		&	0.02	&	0.39		&	0.02	&	0.26	&	0.02	\\ 
AG013.0965-0.1454	&	59262	&	6246	&	766 	&	T	&	Y	&	1194	&	733	&	190	&	T	&	356	&	T	&	6.80		&	0.40	&	0.20		&	0.01	&	0.61		&	0.01	&	0.40	&	0.02	\\ 
AG013.1319-0.1498	&	59390	&	6024	&	705 	&	T	&	Y	&	1859	&	1859	&	705	&	T	&	1859	&	T	&	23.90		&	0.30	&	0.39		&	0.00	&	0.50		&	0.00	&	0.56	&	0.00	\\ 
AG013.1787+0.0603	&	59542	&	9451	&	1142 	&	T	&	Y	&	3513	&	3157	&	332	&	T	&	2739	&	T	&	4.70		&	0.60	&	0.06		&	0.01	&	0.21		&	0.01	&	-0.00	&	0.01	\\ 
AG013.1845-0.1071	&	59561	&	5770	&	617 	&	T	&	Y	&	1693	&	1479	&	240	&	T	&	578	&	T	&	3.40		&	0.60	&	0.06		&	0.01	&	0.38		&	0.01	&	0.21	&	0.01	\\ 
AG013.2105-0.1440	&	59672	&	12273	&	2154 	&	T	&	Y	&	3467	&	3460	&	736	&	T	&	3345	&	T	&	9.50		&	0.70	&	0.11		&	0.01	&	0.06		&	0.01	&	0.03	&	0.01	\\ 
AG013.2123+0.0404	&	59679	&	6189	&	720 	&	T	&	Y	&	305	&	305	&	174	&	F	&	182	&	T	&	-		&	-	&	-		&	-	&	-		&	-	&	-	&	-	\\

\hline
      \end{tabular}
   
      \label{tab:astroHOGexampleH2CO}
      \small{ \textbf{Notes:} Npix$_{\rm{ext.}}$: number of pixels in the mask$_{\rm{ext.}}$ defined from the continuum; Npix$_{\rm{com.}}$: number of pixels in the mask$_{\rm{com.}}$ defined from the catalog of cores in the continuum; flag$_1$: "T" if Npix$_{\rm{ext.}}$>2\,Npix$_{\rm{com.}}$, it is the selection criteria to confirm that there is a diffuse emission for the selected source; det: "Y" if the molecule is detected from the moment-0 maps emission; Npix$_{\rm{mom0}}$: number of pixels in the mask of the emission of the moment-0 map; Npix$_{\rm{inext}}$: number of pixels in the intersection between the mask$_{\rm{ext.}}$ and the mask of the moment-0 emission; Npix$_{\rm{incom}}$: number of pixels in the continuum cores that are covered at least for the 60\% of their area by emission in the moment-0 map; flag$_{2}$: "T" if Npix$_{\rm{inext}}$>2\,Npix$_{\rm{incom}}$, it is the first of the two selection criteria to confirm that the molecular emission is tracing the diffuse emission; Npix$_{\rm{maxreg}}$: number of pixels of the larger region inside the intersection mask between the mask$_{\rm{ext.}}$ and the mask of the moment-0 emission; flag$_{2}$: "T" if Npix$_{\rm{maxreg}}$ is larger than 10 times the area of the beam in pixels, it is the second of the two selection criteria to confirm that the moleculare emission is tracing the diffuse emission; V: results from astroHOG; errV: error on the results from astroHOG based on MC method; V$_{N}$: normalized V, see Eq. (4); errV$_{N}$: error on the normalized V; $\rho_{\rm{p}}$: Pearson's correlation coefficient; err$\rho_{\rm{p}}$: error on the Pearson's correlation coefficient based on MC method; $\rho_{\rm{s}}$: Spearman's correlation coefficient; err$\rho_{\rm{s}}$: error on the Spearman's correlation coefficient based on MC method. The full version of this table is available through Zenodo (https://doi.org/10.5281/zenodo.15236491). The analogouses of this Table for all the other molecular transition analyzed in this paper are available through Zenodo. }
  \end{table*}
\end{landscape}

\newpage
\onecolumn
\small
\setlength{\tabcolsep}{4pt}
\begin{landscape}
\begin{table*}[]
\centering
\caption{Results of \texttt{astroHOG}, Pearson's correlation, and Spearman's correlation for the mask$_{\rm{comp}}$ for H$_2$CO $3_{0,3}-2_{0,2}$. }
      \begin{tabular}{ccccccccccccccc}
      \hline\hline

ALMAGAL ID   &   Original ID   &      Npix$_{\rm{com}}$    &   det.   &   Npix$_{\rm{mom0}}$   &    Npix$_{\rm{incom}}$   &  flag &  V   &   errV   &   V$_{N}$   &   errV$_{N}$   &   $\rho_{\rm{p}}$  &   err $\rho_{\rm{p}}$ &   $\rho_{\rm{s}}$  &   err $\rho_{\rm{s}}$\\
\hline
AG010.1313-0.7687	&	46677	&	441	&	Y	&	260	&	0	&	F	&	-		&	-	&	-		&	-	&	-		&	-	&	-	&	-	\\ 
AG010.1392-0.7831	&	46696	&	127	&	Y	&	706	&	127	&	True	&	12.20		&	0.20	&	0.77		&	0.01	&	0.76		&	0.02	&	0.72	&	0.02	\\ 
AG010.2283-0.2064	&	47184	&	766	&	Y	&	1501	&	390	&	True	&	13.90		&	0.50	&	0.50		&	0.02	&	0.67		&	0.01	&	0.65	&	0.02	\\ 
AG010.6185-0.0314	&	48727	&	379	&	Y	&	2086	&	295	&	True	&	11.60		&	0.50	&	0.48		&	0.02	&	0.72		&	0.01	&	0.78	&	0.01	\\ 
AG010.6691-0.2196	&	49005	&	910	&	Y	&	1443	&	326	&	True	&	11.80		&	0.30	&	0.46		&	0.01	&	0.21		&	0.01	&	0.19	&	0.01	\\ 
AG010.6985-0.0994	&	49143	&	0	&	Y	&	316	&	0	&	F	&	-		&	-	&	-		&	-	&	-		&	-	&	-	&	-	\\ 
AG010.7516-0.1987	&	49337	&	249	&	Y	&	921	&	249	&	True	&	2.80		&	0.30	&	0.13		&	0.01	&	0.44		&	0.02	&	0.47	&	0.02	\\ 
AG011.0542-0.0419	&	50433	&	0	&	Y	&	45	&	0	&	F	&	-		&	-	&	-		&	-	&	-		&	-	&	-	&	-	\\ 
AG011.0829-0.5311	&	50555	&	540	&	Y	&	1739	&	108	&	True	&	4.20		&	0.20	&	0.29		&	0.01	&	0.54		&	0.02	&	0.57	&	0.02	\\ 
AG011.2019-0.0909	&	51020	&	0	&	N	&	0	&	0	&	F	&	-		&	-	&	-		&	-	&	-		&	-	&	-	&	-	\\ 
AG011.3614+0.8017	&	51554	&	0	&	N	&	0	&	0	&	F	&	-		&	-	&	-		&	-	&	-		&	-	&	-	&	-	\\ 
AG011.6992-0.3011	&	52831	&	141	&	Y	&	414	&	141	&	True	&	14.50		&	0.10	&	0.86		&	0.01	&	0.89		&	0.01	&	0.87	&	0.01	\\ 
AG011.9039-0.1403	&	53795	&	1009	&	Y	&	2913	&	781	&	True	&	20.20		&	0.30	&	0.51		&	0.01	&	0.75		&	0.00	&	0.83	&	0.00	\\ 
AG011.9183-0.6122	&	53869	&	920	&	Y	&	4497	&	681	&	True	&	15.20		&	0.50	&	0.41		&	0.01	&	0.67		&	0.00	&	0.27	&	0.01	\\ 
AG011.9370-0.6160	&	54001	&	1254	&	Y	&	4010	&	1028	&	True	&	1.30		&	0.70	&	0.03		&	0.02	&	-0.01		&	0.01	&	-0.10	&	0.01	\\ 
AG012.1889-0.1334	&	55190	&	0	&	Y	&	46	&	0	&	F	&	-		&	-	&	-		&	-	&	-		&	-	&	-	&	-	\\ 
AG012.2087-0.1017	&	55288	&	706	&	Y	&	5605	&	706	&	True	&	16.70		&	0.50	&	0.44		&	0.01	&	0.88		&	0.00	&	0.90	&	0.00	\\ 
AG012.4040-0.4681	&	56184	&	610	&	Y	&	616	&	251	&	True	&	19.60		&	0.20	&	0.88		&	0.01	&	0.76		&	0.01	&	0.72	&	0.01	\\ 
AG012.4977-0.2233	&	56634	&	716	&	Y	&	876	&	251	&	True	&	7.70		&	0.40	&	0.34		&	0.02	&	0.48		&	0.02	&	0.48	&	0.02	\\ 
AG012.5571-0.3485	&	56859	&	218	&	Y	&	692	&	0	&	F	&	-		&	-	&	-		&	-	&	-		&	-	&	-	&	-	\\ 
AG012.6794-0.1830	&	57335	&	968	&	Y	&	2817	&	721	&	True	&	17.30		&	0.50	&	0.46		&	0.01	&	0.96		&	0.00	&	0.83	&	0.00	\\ 
AG012.7207-0.2175	&	57596	&	496	&	Y	&	3724	&	496	&	True	&	-2.00		&	0.20	&	-0.06		&	0.01	&	0.13		&	0.00	&	0.16	&	0.00	\\ 
AG012.7360-0.1031	&	57641	&	866	&	Y	&	1330	&	527	&	True	&	8.10		&	0.50	&	0.25		&	0.01	&	0.17		&	0.01	&	0.33	&	0.02	\\ 
AG012.8535-0.2265	&	58312	&	987	&	Y	&	2962	&	702	&	True	&	7.80		&	0.60	&	0.21		&	0.02	&	0.31		&	0.01	&	0.41	&	0.01	\\ 
AG012.9008-0.2404	&	58516	&	983	&	Y	&	2321	&	866	&	True	&	8.60		&	0.30	&	0.21		&	0.01	&	0.43		&	0.01	&	0.58	&	0.01	\\ 
AG012.9048-0.0306	&	58529	&	1044	&	Y	&	7437	&	614	&	True	&	5.50		&	0.60	&	0.16		&	0.02	&	0.38		&	0.01	&	0.32	&	0.01	\\ 
AG012.9084-0.2604	&	58549	&	761	&	Y	&	4001	&	761	&	True	&	15.80		&	0.50	&	0.41		&	0.01	&	0.80		&	0.00	&	0.83	&	0.00	\\ 
AG012.9156-0.3341	&	58583	&	804	&	Y	&	2279	&	661	&	True	&	6.60		&	0.50	&	0.18		&	0.01	&	0.71		&	0.00	&	0.70	&	0.00	\\ 
AG013.0166-0.1797	&	58971	&	159	&	Y	&	554	&	159	&	True	&	4.00		&	0.60	&	0.22		&	0.03	&	0.49		&	0.03	&	0.49	&	0.03	\\ 
AG013.0965-0.1454	&	59262	&	766	&	Y	&	1194	&	190	&	True	&	6.80		&	0.30	&	0.35		&	0.01	&	0.75		&	0.01	&	0.75	&	0.01	\\ 
AG013.1319-0.1498	&	59390	&	705	&	Y	&	1859	&	705	&	True	&	9.80		&	0.20	&	0.26		&	0.01	&	0.57		&	0.01	&	0.64	&	0.01	\\ 
AG013.1787+0.0603	&	59542	&	1142	&	Y	&	3513	&	332	&	True	&	6.90		&	0.30	&	0.27		&	0.01	&	0.27		&	0.02	&	0.21	&	0.01	\\ 
AG013.1845-0.1071	&	59561	&	617	&	Y	&	1693	&	240	&	True	&	10.00		&	0.50	&	0.46		&	0.02	&	0.70		&	0.01	&	0.69	&	0.01	\\ 
AG013.2105-0.1440	&	59672	&	2154	&	Y	&	3467	&	736	&	True	&	1.20		&	0.70	&	0.03		&	0.02	&	0.10		&	0.02	&	0.07	&	0.01	\\ 
AG013.2123+0.0404	&	59679	&	720	&	Y	&	305	&	174	&	True	&	11.80		&	0.80	&	0.63		&	0.04	&	0.65		&	0.03	&	0.65	&	0.03	\\

\hline
      \end{tabular}
   
      \label{tab:astroHOGexampleH2CO}
      \small{\textbf{Notes:}  Npix$_{\rm{com}}$: number of pixels in the mask$_{\rm{com.}}$ defined from the catalog of cores in the continuum; det: "Y" if the molecule is detected from the moment-0 maps emission; Npix$_{\rm{mom0}}$: number of pixels in the mask of the emission of the moment-0 map; Npix$_{\rm{incom}}$: number of pixels in the continuum cores that are covered at least for the 60\% of their area by emission in the moment-0 map; flag: "T" if Npix$_{\rm{incom}}$  is larger than 3 times the area of the beam in pixels, in order to have reliable results; V: results from astroHOG; errV: error on the results from astroHOG based on MC method; V$_{N}$: normalized V, see Eq. (4); errV$_{N}$: error on the normalized V; $\rho_{\rm{p}}$: Pearson's correlation coefficient; err$\rho_{\rm{p}}$: error on the Pearson's correlation coefficient based on MC method; $\rho_{\rm{s}}$: Spearman's correlation coefficient; err$\rho_{\rm{s}}$: error on the Spearmans's correlation coefficient based on MC method. The full version of this table is available through Zenodo. The analogouses of this Table for all the other molecular transition analyzed in this paper are available through Zenodo (https://doi.org/10.5281/zenodo.15236491). }
  \end{table*}

\end{landscape}
\twocolumn
\section{HOG results excluding failed automatic routine sources for moment maps and sources with CORNISH counterparts}
\begin{figure*}[h!]
    \centering
    \includegraphics[width=9cm, trim={0.5cm 3.4cm 1cm 0.7cm}, clip]{figures/recapV_all.png}
        \includegraphics[width=9cm, trim={0.5cm 3.4cm 1cm 0.7cm}, clip]{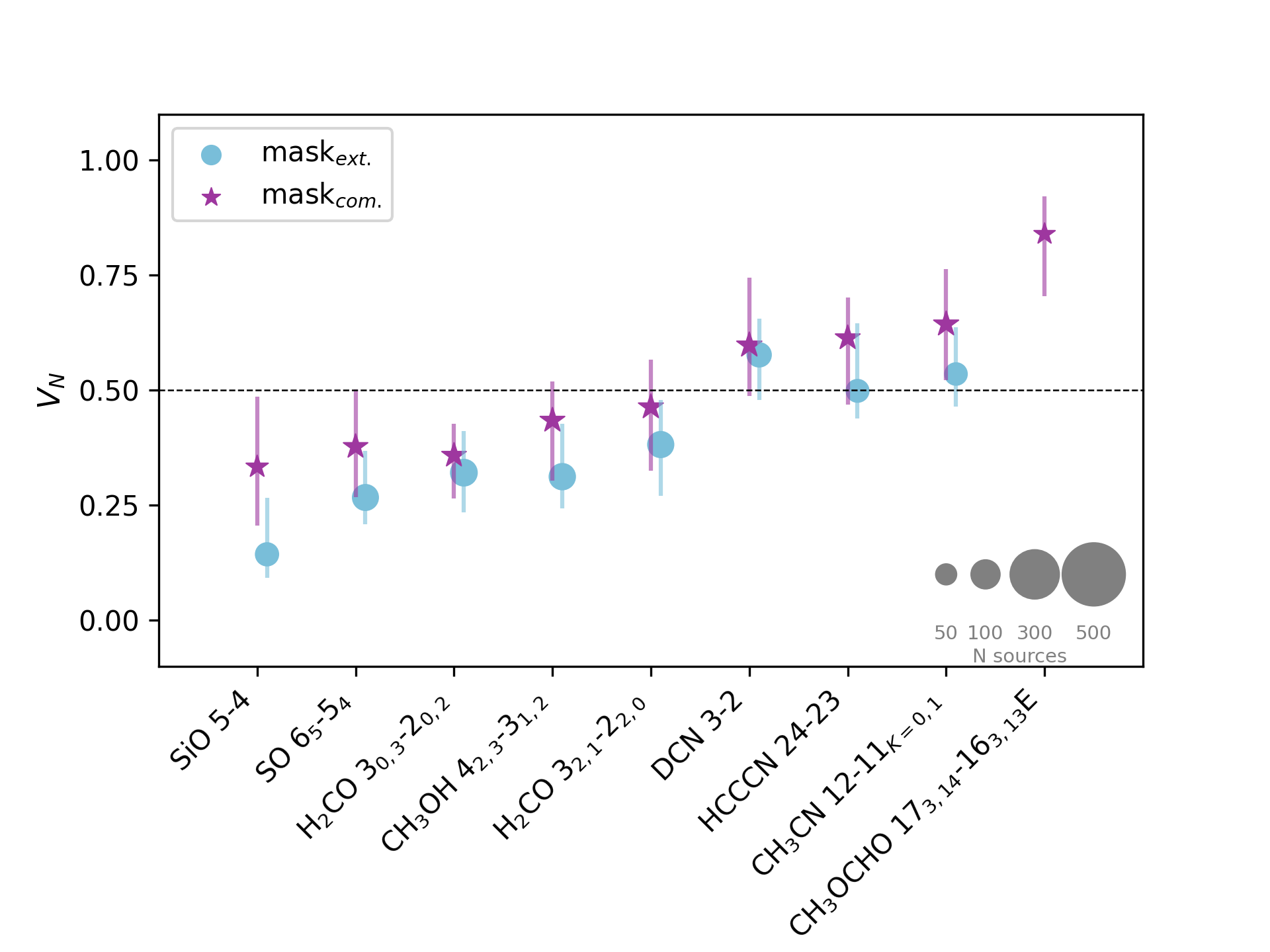 }\\
        \includegraphics[width=9cm, trim={0.5cm 0 1cm 0.7cm}, clip]{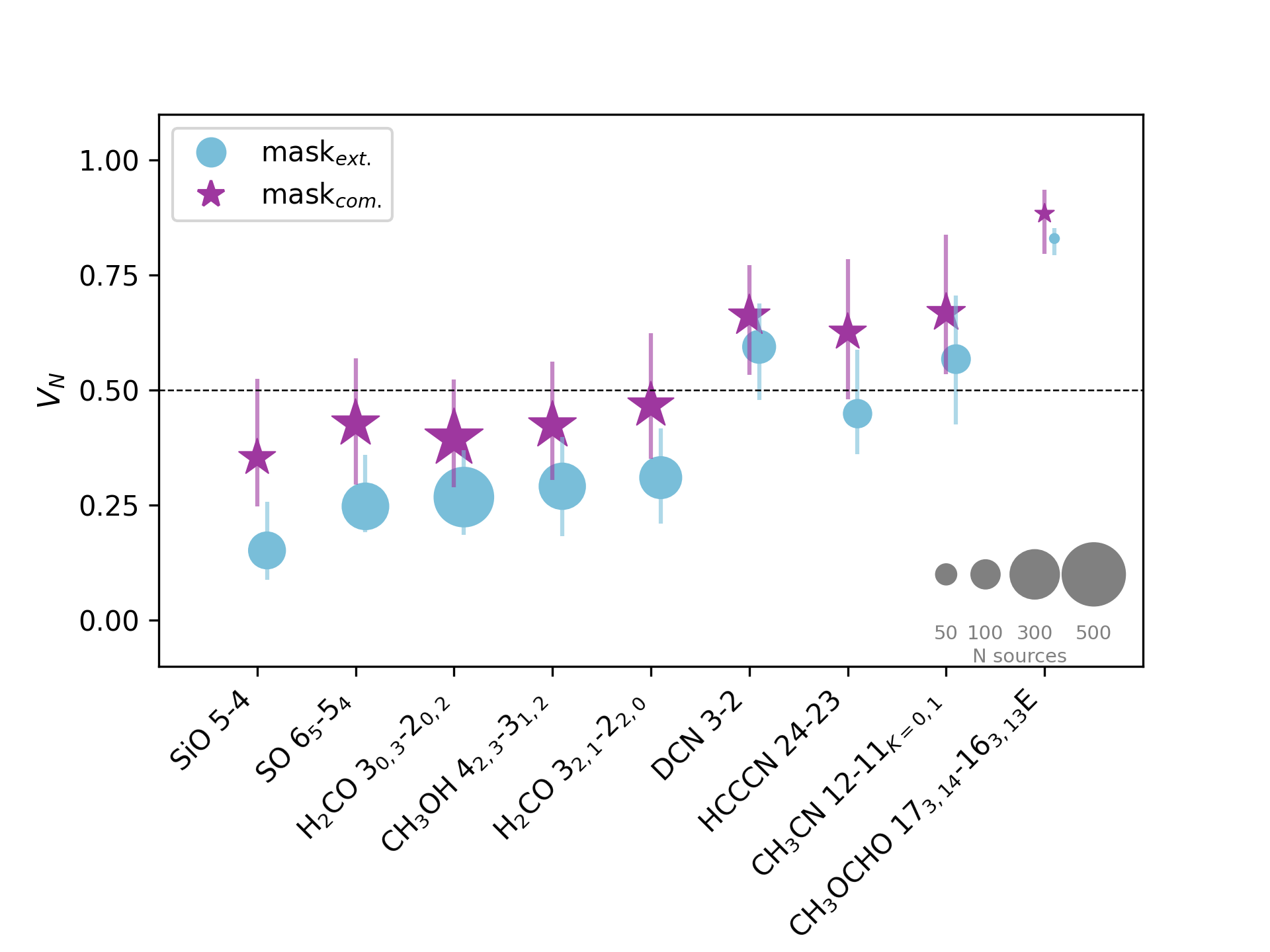 }\\

    \caption{Top left panel: same of Fig. 9, i.e. results of the astroHOG analysis on all the sample; top right panel: same plot only for the sources for which the automated routine for the creation of moment-0 maps failed; bottom panel: same plot excluding from all the sources those with a CORNISH counterpart.}
    \label{fig:NORADIO}
\end{figure*}

In the main text we have presented the results of the analysis of the histogram of oriented gradients (HOG) in Fig. 9 on all the sources. However, for 97 sources we created the moment-0 maps using a fixed velocity range of $+/-5$\,km/s, since the automated routine failed due mostly to blendings in line-rich sources. The moment-maps of these sources therefore do not encompass the entire line in the pixels of the line-rich cores to avoid line blendings, and are not integrating possible high velocity wings also in less line-rich regions of the clump. In Figure E.1 we show that the inclusion of these sources, with moment-0 maps defined in a different way, do not alter the main result shown in the main text, since the differences between the results on all the sources (upper-left panel) and the results only on the 97 sources (upper-right panel) are negligible and well within the range of the distribution.\\ Moreover, for 112 sources we found a counterpart in emission from CORNISH. In these sources a part of the continuum emission could be affected by free-free emission. Since the importance of this contamination, both in the area of the maps affected, and in the relative importance in the flux measured, is not known, we show in the bottom row of Figure E.1 that removing these contaminated sources does not alter the main results shown in the text (and in the upper-left panel of Figure E.1). 

%% This command is needed to show the entire author+affiliation list when
%% the collaboration and author truncation commands are used.  It has to
%% go at the end of the manuscript.
%\allauthors

%% Include this line if you are using the \added, \replaced, \deleted
%% commands to see a summary list of all changes at the end of the article.
%\listofchanges

\end{document}